\documentclass[pra,aps,floatfix,floats,superscriptaddress,notitlepage]{revtex4-1}

\pdfoutput=1

\usepackage{epsfig}
\usepackage{enumerate}
\usepackage{amsmath}
\usepackage{amssymb}
\usepackage{hyperref}
\usepackage{color}
\usepackage{theorem}
\usepackage{braket}
\usepackage{mathtools}
\usepackage{comment}
\usepackage{tensor}
\usepackage{tikz}
\usepackage{empheq}
\usepackage{hyperref}
\usetikzlibrary{decorations.markings}

\newcommand{\be}{\begin{equation}}
\newcommand{\ee}{\end{equation}}
\newcommand{\bea}{\begin{eqnarray}}
\newcommand{\eea}{\end{eqnarray}}

\definecolor{violet}{rgb}{0.62,0,1}
\definecolor{lightblue}{rgb}{0.12,0.56,1}
\definecolor{green}{rgb}{0.13,0.55,0.13}
\def\nn{\nonumber\\}

\def\tr{\textrm{tr}}

\definecolor{violet}{rgb}{0.62,0,1}
\definecolor{lightblue}{rgb}{0.62,0,1}
\definecolor{green}{rgb}{0.13,0.55,0.13}

\def\doi{http://dx.doi.org/}

\newcommand{\xt}{\zeta}
\newcommand{\limth}{{\textstyle\lim_ {\rm th}}}

\newcommand{\ak}{a}

\def\eqref#1{(\ref{#1})}

\def\doi{http://dx.doi.org/}

\begin{document}

\title{Low-Temperature Transport in Out-of-Equilibrium XXZ Chains}

\author{Bruno Bertini}
\address{SISSA and INFN, via Bonomea 265, 34136, Trieste, Italy}
\address{Department of physics, FMF, University of Ljubljana, Jadranska 19, SI-1000 Ljubljana, Slovenia}
\author{Lorenzo Piroli}
\address{SISSA and INFN, via Bonomea 265, 34136, Trieste, Italy}
\begin{abstract}
We study the low-temperature transport properties of out-of-equilibrium XXZ spin-$1/2$ chains. We consider the protocol where two semi-infinite chains are prepared in two thermal states at small but different temperatures and suddenly joined together. We focus on the qualitative and quantitative features of the profiles of local observables, which at large times $t$ and distances $x$ from the junction become functions of the ratio $\zeta=x/t$. By means of the generalized hydrodynamic equations, we analyse the rich phenomenology arising by considering different regimes of the phase diagram. In the gapped phases, variations of the profiles are found to be exponentially small in the temperatures but described by non-trivial functions of $\zeta$. We provide analytical formulae for the latter, which give accurate results also for small but finite temperatures. In the gapless regime, we show how the three-step conformal predictions for the profiles of energy density and energy current are naturally recovered from the hydrodynamic equations. Moreover, we also recover the recent non-linear Luttinger liquid predictions for low-temperature transport: universal peaks of width $\Delta\zeta\propto T$ emerge at the edges of the light cone in the profiles of generic observables. Such peaks are described by the same function of $\zeta$ for all local observables.
\end{abstract}

\maketitle

\section{Introduction}
Transport problems provide a natural setting to investigate several fascinating aspects of many-body quantum physics. They show many celebrated examples of emergent collective phenomena, \emph{i.e.} effects that can not be explained by separately considering the behaviour of the system's microscopic components. Indisputably, the collective nature of these problems is also at the basis of many difficulties in their theoretical understanding and a full characterisation of transport problems in ``real" three dimensional systems is currently out of reach. A simplified framework where such problems can be more effectively addressed has been identified in one-dimensional quantum systems~\cite{mazur,suzuki,czp-95,znp-97,nma-98,zotos-99,ag-02,h-m-03,bfks-05,hhb-07, 
spa-09,steinigeweg-09,steinigeweg-11,steinigeweg2-11,znidaric-11,kbm-12,khlh-13,
KaIM13,steinigeweg-14,KaKM14,KaMH14,VaKM15,SHZB16, Karr17,AbWi05,JTKA11,LLMM17}.
Besides providing several useful toy models, one-dimensional systems display unusual effects and allow one to investigate important questions such as, the implication of conservation laws for the transport dynamics  \cite{mazur,suzuki,czp-95,znp-97} or the existence of universal features \cite{SoCa08, CaHD08, BeDo12, DoHB14, BeDo15, BeDo16, ADSV16, BeDo16Review, LLMM17, DuSC17, BPC:short,EiME16,Korm17,PeGa17,SDCV17,EiBa17}. Furthermore, while for a long time the interest in these models has been purely academic, one-dimensional systems out-of-equilibrium are now routinely engineered in ultracold atomic laboratories ~\cite{bdz-08,ccgo-11,pssv-11,LaGS16,kww-06,HLFS07,gklk-12,fse-13,lgkr-13,glms-14,langen-15}, making several existing theoretical studies experimentally testable.

The standard physical setting to study transport phenomena is realised by connecting the system to two different external baths. This configuration, however, has the disadvantage of requiring an \emph{ad hoc} specification of the system-bath dynamics, complicating the analysis of the results. It is indeed difficult to disentangle generic features from those generated by the specific system-bath interaction considered. More definite predictions can be obtained by embedding system and bath in a larger closed system. A natural setting to observe non-trivial transport in this framework is a quantum quench from a piecewise homogeneous initial state. One studies the dynamics of the system after the sudden junction of two semi-infinite halves prepared in two different macroscopic states, for example two thermal states at different temperatures \cite{BGMS13,SpLe77}. Despite its conceptual simplicity, this problem hides the full intrinsic complexity of many-body quantum physics, and until recently analytical understanding was limited to the case of free systems~\cite{ARRS99,AsPi03,AsBa06,PlKa07,LaMi10,EiRz13,DVBD13,Bert17, CoKa14,EiZi14,CoMa14,DeMV15,DLSB15,VSDH16,KoZi-1-17, Mint11,MiSo13}. The effects of non-trivial interactions were considered only through numerical studies~\cite{GKSS05,SaMi13,AlHe14,BDVR16,ViIR17,PBPD17} or the formulation of \emph{ad hoc} conjectures \cite{DVMR14,CCDH14,Zoto16}.  As an important exception, analytic predictions could be obtained for the low-temperature transport of gapless systems. Indeed, in this limit the systems are described by appropriate universal effective theories~\cite{SoCa08,CaHD08,BeDo12,DoHB14,BeDo15,BeDo16,ADSV16,BeDo16Review,LLMM17,DuSC17, BPC:short}.

The last year has witnessed a major breakthrough within the framework of transport in one-dimensional closed systems, as the so called ``generalised hydrodynamic approach" was introduced in the works~\cite{CaDY16,BCDF16}. This approach allows for an analytical treatment of transport in integrable models, even in the presence of non-trivial interactions. The essence of the generalised hydrodynamic approach is to describe the large time behaviour of an inhomogeneous system by a family of space- and time- dependent stationary states which are determined by solving an infinite system of continuity equations for the densities of conserved charges. This description exploits several results previously obtained in the study of homogeneous quantum quenches in integrable models, such as the possibility of characterising the large time expectations of local observables in terms of a generalised Gibbs ensemble \cite{RDYO07} (see also the recent reviews~\cite{CaEM16,CaCa16,EsFa16,ViRi16,IMPZ16,LaGS16}). As a result of the generalised hydrodynamic approach, we are now able to provide a complete asymptotic characterisation of the local properties of integrable systems in non-homogeneous settings. For the protocol described above, namely two semi-infinite subsystems suddenly joined together, this statement can be made more precise. In that case the generalised hydrodynamics gives the exact characterisation of the system in the scaling limit of infinite distances  $x$ from the junction and times $t$ with fixed $\zeta=x/t$; observables on different ``rays" $\zeta$ relax to different stationary states $\boldsymbol{\rho}({\zeta})$, called locally quasi-stationary states~\cite{BeFa16}.  The method introduced in \cite{CaDY16,BCDF16} also allows one to study other protocols, from the non-equilibrium dynamics in the presence of localised defects \cite{BeFa16,DeLucaBastianello} to the release of quantum gases from a confining trap \cite{DDKY17}. Moreover, similar ideas have been recently employed to compute exact entanglement dynamics after a quantum quench \cite{AlCa16,Alba17,MBPC17}. Finally, these developments have also led to important results in the framework of linear transport \cite{BVKM17,BVKM17-2,IlDe17,DoSp17-2,IlDe2-17}. In fact, many studies already appeared, investigating further aspects of the generalised hydrodynamic approach in several integrable models \cite{DDKY17,BVKM17,BVKM17-2, DoYo16,DoSY17,DoSp17,DoYC17,Spoh17,PDCB17,CDV:analyticDW,F:higherhydro,Bulc17,CaBM17,Fago17-charges}.

The generalised hydrodynamic approach is conveniently implemented by using a thermodynamic Bethe ansatz (TBA) formalism. One represents each GGE at fixed $x$ and $t$ by the corresponding distributions of quasi-momenta (or rapidities) of quasi-particle excitations. The family of $x$- and $t$- dependent quasi-momenta distributions is fixed by solving a compact set of integro-differential equations. In the vast majority of the cases studied up to now such equations have been solved numerically using different schemes~\cite{CaDY16,BCDF16,BVKM17-2} (see however \cite{CDV:analyticDW}, where fully analytic results were obtained in the XXZ chain in the case of a domain-wall initial configuration). Even if such numerical solutions are shown to converge very well and give an excellent description of the profiles of local observables~\cite{BCDF16,BVKM17-2}, to gain a deeper insight into the physics of the problem it is useful to obtain some fully analytical solutions. These solutions can be used, for example, to compare the predictions of generalised hydrodynamics to that of the previously mentioned universal theories~\cite{SoCa08,CaHD08,BeDo12,DoHB14,BeDo15,BeDo16,ADSV16,BeDo16Review,LLMM17,DuSC17, BPC:short}. In this work we follow this logic: focussing on the prototypical case of the XXZ spin-$1/2$ chain we consider the transport dynamics originated by joining two thermal states at small but different temperatures and we develop a fully analytic low-temperature expansion of the profiles of local observables.

The XXZ chain is particularly interesting from the point of view of low-temperature transport, as it displays a non-trivial ground-state phase diagram depending on the anisotropy parameter and the external magnetic field. Different phases are characterised by different structures of elementary excitations; this generates qualitative differences in the transport dynamics. In fact, qualitative differences in the transport dynamics between the different phases emerge even for finite temperatures and can be understood in terms of a different structure of the conservations laws~\cite{IDWC15,PiVC16,PoVW17,IlQC17}. The most prominent one is the emergence of sub-ballistic spin transport in the gapped phase~\cite{LjZP17,MiMK17, PDCB17}.

The main result of our analysis can be summarised as follows. In the gapped phases, variations of the profiles are found to be exponentially small in the temperatures and are described by non-trivial functions of $\zeta$. We derive analytic formulae for these functions and show that they give accurate results also for small but finite temperatures. In the gapless regime, the leading order contributions for the profiles of energy density and current are $\propto T^2$ and coincide with the conformal field theory predictions of Ref.~\cite{BeDo12}: they assume a three step form identifying an effective light cone. For generic observables, however, we also find the region of width $\propto T$ at the edges of the light cone which has been recently predicted in Ref.~\cite{BPC:short} using a non-linear Luttinger liquid description. In such region the leading corrections are $\propto T$ and they are proportional to the same smooth function of $\zeta$ for all local observables.

The paper is laid out as follows. In Section~\ref{sec:the_model} we introduce the model considered and summarise its thermodynamic Bethe ansatz description. In Section~\ref{sec:ground} we review the structure of the ground state of the model. In Section~\ref{sec:protocol} we discuss our protocol to generate out-of-equilibrium dynamics and briefly review the TBA treatment of generalised hydrodynamics. Low-temperature expansions in the three different phases of the model are constructed in the subsequent sections. In Section~\ref{sec:lowtgapless} we consider the gapless phase, while the two gapped phases are examined in Section~\ref{sec:lowtgapped}. Section~\ref{sec:conclusions} contains our conclusions. Two appendices complement the main text with technical details.      

\section{The model}\label{sec:the_model}

We consider a XXZ spin-$1/2$ chain in an external magnetic field $h$, as described by the following Hamiltonian
\begin{equation}
	\boldsymbol H = \frac{J}{4}\sum_{j=-L/2}^{L/2-1}\left[\boldsymbol\sigma_{j}^{x}\boldsymbol\sigma_{j+1}^{x}+\boldsymbol\sigma_{j}^{y}\boldsymbol\sigma_{j+1}^{y}+\Delta ( \boldsymbol\sigma_{j}^{z}\boldsymbol\sigma_{j+1}^{z}-1)\right]- h \sum_{j=-L/2}^{L/2-1} (\boldsymbol \sigma_j^z-1)\,.
\label{eq:Hamiltonian_XXZ}
\end{equation}
Here we denote by $\boldsymbol\sigma_j^\alpha$ ($\alpha=x,y,z$) the Pauli matrices representing the spin-$1/2$ degrees of freedom at lattice sites $j=1,2,\ldots, L$ and we assume periodic boundary conditions, namely $\boldsymbol\sigma_{L/2}^\alpha = \boldsymbol\sigma_{-L/2}^\alpha$. The real parameter $\Delta$ corresponds to an interaction anisotropy along the $z$-direction. Note that the Hamiltonian \eqref{eq:Hamiltonian_XXZ} commutes with the $z$-component of the total spin, so that its eigenstates have a well-defined number of down spins. 

\begin{figure}
	\includegraphics[width=0.65\textwidth]{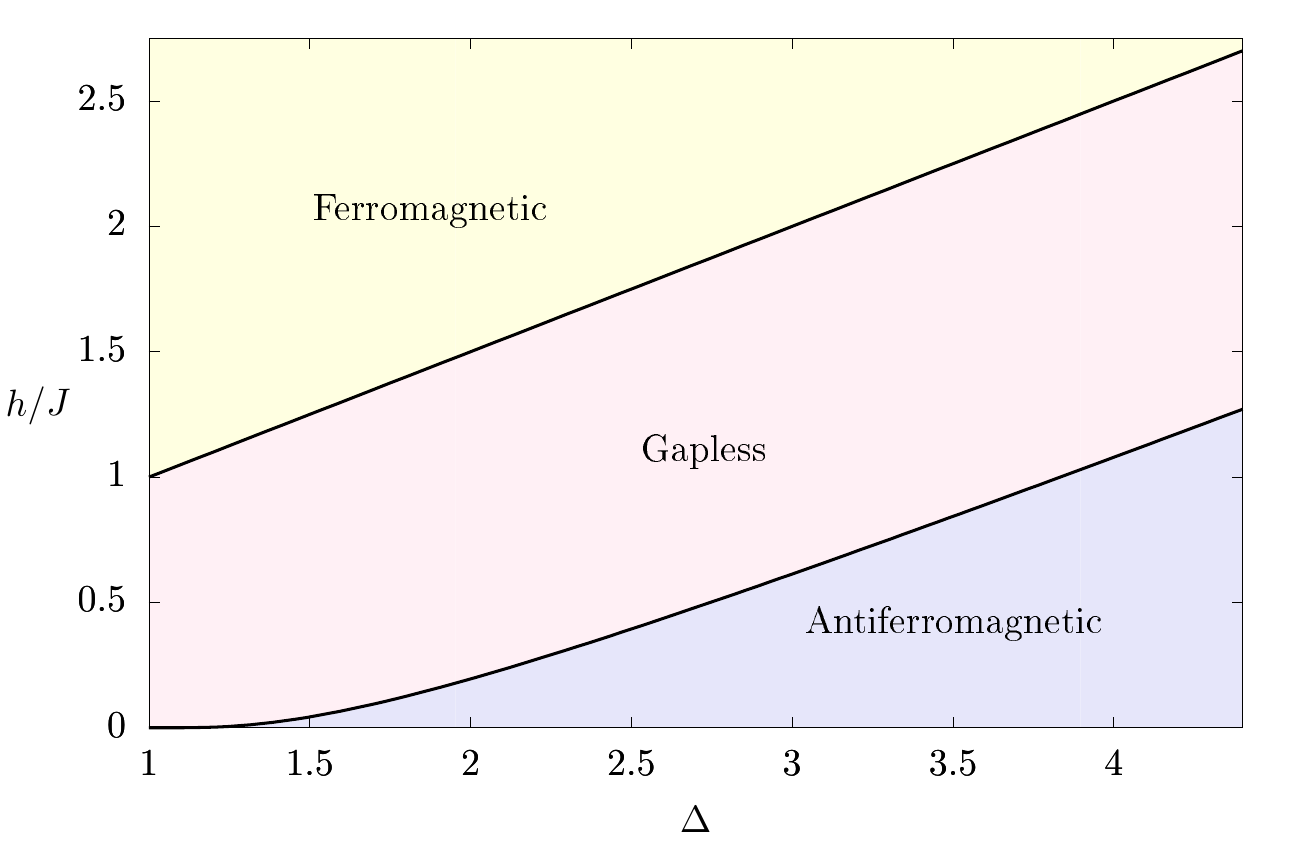}
	\caption{The phase diagram of the XXZ spin-$1/2$ chain in an external magnetic field $h$. For fixed values of the anisotropy $\Delta>1$ all the phases are explored by increasing the value of the field $h$. The lowest line corresponds to the curve parametrized in Eq.~\eqref{eq:lower_critical_field}, while the upper one corresponds to $h=J(1+\Delta)/2$.}
	\label{fig:phasediagram}
\end{figure}

Throughout this work we consider the case of $J,h>0$ and  $\Delta \geq1$. For the purposes of our paper this is not a restriction as this choice of the parameters allows one to study  all the three different phases of the model at low temperatures, see Fig.~\ref{fig:phasediagram}.

In the following, we make use of the parametrisation
\be
\Delta=\cosh\eta\,, \qquad\qquad\eta \geq 0\,.\label{eq:eta}
\ee
The eigenvalues of the Hamiltonian \eqref{eq:Hamiltonian_XXZ} for a finite number $L$ of lattice sites can be constructed exactly, by means of the Bethe ansatz method~\cite{korepin, takahashi, gaudin}. A nice physical interpretation of these eigenstates can be obtained in the limit of large $L$, when the string hypothesis applies~\cite{takahashi}. In this limit the eigenstates become scattering states of quasi-particles called magnons which represent propagating spin flips. Due to the non-trivial interactions, the magnons can also form bound states, called strings, which are considered as independent quasi-particles. Each string is labelled by an integer $j$, running from $1$ to $\infty$; ${j=1}$ corresponds to free magnons. 

Here we are interested in the thermodynamic limit 
\be
L,N\rightarrow \infty \qquad \qquad \text{with fixed}\qquad \qquad n\equiv\frac{N}{L}\,.
\ee
In this limit, denoted by $\limth$, the rapidities become dense and the states are conveniently described by their densities; one needs a density of quasi-momenta $\rho_j(\lambda)$ for each different type of bound states. The densities of quasi-momenta are dual to the densities of un-occupied quasi-momenta $\rho^h_j(\lambda)$, called holes. These two sets of functions are related by the so-called thermodynamic Bethe equations \cite{takahashi}
 \be
\rho_{j}^t(\lambda)\equiv\rho^h_j(\lambda)  +\rho_j(\lambda)  = \ak_j(\lambda) -\Big[ \sum_{k} \,T_{jk} \ast \rho_k \Big] (\lambda) \,.
\label{eq:tba}
\ee
Here we introduced the total root density $\rho_{t,n}(\lambda)$ and the function
\begin{equation}
\ak_j(\lambda)=\frac{1}{\pi} \frac{\sinh\left( j\eta\right)}{\cosh (j \eta) - \cos( 2 \lambda)}\,,
\label{eq:afunction}
\end{equation}
which is related to $p_j(\lambda)$, the momentum of a bound state $j$ with rapidity $\lambda$, as follows 
\be
a_j(\lambda)=\frac{1}{2\pi}p_j'(\lambda)\,.
\ee
We also introduced the function $T_{jk}(\lambda)$, which is the scattering kernel between bound states. In terms of $\ak_{j}(\lambda)$ it reads as 
\be
T_{jk}(\lambda)=(1-\delta_{j k}) \ak_{|j-k|}(\lambda)+2\ak_{|j-k|+2}(\lambda)+\cdots +2\ak_{j+k-2}(\lambda)+\ak_{j+k}(\lambda)\,.
\ee
Finally, we defined the convolution as 
\be
[f \ast g](\lambda)= \int_{-\pi/2}^{\pi/2}{\rm d} \mu\, f(\lambda - \mu) g(\mu)\,.
\label{eq:convolution}
\ee

In this paper we always consider cases where the state, in the thermodynamic limit, is completely specified by the root densities $\{\rho_n(\lambda),\rho^h_n(\lambda)\}$ (see, however, Ref.~\cite{PDCB17}). Since $\{\rho_n(\lambda)\}$ and $\{\rho^h_n(\lambda)\}$ are connected via Eq.~\eqref{eq:tba}, a single one of these sets is enough to specify the state; a usual parametrisation is to use the ratio of the two 
\begin{equation}
\label{eq:etan}
\eta_n (\lambda) = \frac{\rho_n^h(\lambda)}{\rho_n(\lambda)}\,,
\end{equation} 
or alternatively the ``filling functions" $\vartheta_n(\lambda)$ 
\begin{equation}
\vartheta_n(\lambda) = \frac{\rho_n(\lambda)}{\rho_n(\lambda) + \rho^h_n(\lambda)} \equiv \frac{1}{1 + \eta_n(\lambda)}\,,
\label{eq:fillingfunctions}
\end{equation}
which give the fraction of occupied momenta in the state. 

\subsection{Conserved charges}
\label{sec:charges}
Since $\{\rho_j(\lambda)\}$ completely specifies the thermodynamic state, it determines all the values of the conserved charges of the system. For example the energy density is given by
\begin{equation}
\limth \frac{\langle \boldsymbol{H} \rangle}{L}  =    \sum_{j} \int_{-\pi/2}^{\pi/2} d \lambda\,\, e_j (\lambda) \rho_j(\lambda)\,.
\label{eq:energy_density_gapped}
\end{equation} 
Here the functions $e_j(\lambda)$ are called single particle eigenvalues of the energy and read as 
\be
e_j (\lambda)=  -\pi J \sinh\eta\, a_j(\lambda)+2 h j \,,\qquad\qquad\qquad j=1,\ldots\,. 
\ee 
More generally, we can define an entire set of local conserved charges $\{\boldsymbol{Q}_m\}_{m=1}^\infty$ \cite{Grabowski94, Grabowski95}, whose densities have an expectation value of the form \eqref{eq:energy_density_gapped} on the state specified by $\{\rho_n(\lambda)\}$. The single-particle eigenvalues of the $m$-th charge are given by  
\begin{equation}
q^{(m+1)}_j(\lambda)   = i (-1)^m\frac{(\sinh \eta)^m}{2 ^m} \partial_\lambda^m \log\left[\frac{\sin\left(\lambda+i\eta j/2\right)}{\sin\left(\lambda-i\eta j/2\right)}\right]\,,\qquad\qquad j=1,\ldots\,,\qquad m\geq1\,.
\end{equation}
The magnetisation density in the state specified by $\{\rho_j(\lambda)\}$  can instead be written as 
\begin{equation}
\limth \frac{\braket{\boldsymbol{S^z}}}{L}  = \limth  \frac{\braket{\boldsymbol\sigma^z_i}}{2}= \frac{1}{2}  - \sum_j j \int_{-\pi/2}^{\pi/2}  d\lambda\,\, \rho_j(\lambda)\,,\qquad \forall\, i\,,
\label{eq:magnetization_gapped}
\end{equation}
where $\boldsymbol{S^z}=\frac{1}{2}\sum_i \boldsymbol\sigma^z_i$ is the total spin in the $z$ direction. Note that it does not appear in the set $\{\boldsymbol{Q}_m\}_{m=1}^\infty$. 

\subsection{Dressing equations and velocities}
Let us consider the system in a large finite volume $L$, in an eigenstate of the Hamiltonian specified by a set of rapidities $\{\lambda_\alpha^n\}$, which are distributed according to $\{\rho_n(\lambda)\}$ in the thermodynamic limit. Elementary excitations on this eigenstate can be constructed by injecting an extra string of type $n$ with rapidity $\lambda$. This operation induces a change in the expectation value of the energy as well as those of all other conserved charges $\boldsymbol{Q}$ of the system, namely we have  
\begin{align}
\braket{\boldsymbol{Q}} \to \braket{\boldsymbol{Q}} + q_n^d(\lambda)\,.
\end{align}
Here we introduced the \emph{dressed charge} $q_n^d(\lambda)$, which is a deformation  $O(L^0)$ of the expectation value of the charge $\boldsymbol{Q}$. The dressed charge encodes non-trivial information about all the particles in the system, as adding the extra string of type $n$ we forced all the other particles in the state to rearrange their rapidities. The derivative of $q^{d}_n(\lambda)$ with respect to $\lambda$, can be expressed as a linear integral equation that takes a universal form for any conserved charge of the system
\be
q_n^{d\,\prime} (\lambda)=q_n^{\prime}  (\lambda) - \Big[ \sum_{k} \,T_{nk} \ast ( q_k^{d\,\prime}  \vartheta_k) \Big]  (\lambda)\,.
\label{eq:dressing}
\ee
Comparing \eqref{eq:dressing} to \eqref{eq:tba} we see that the total root density 
\be
\rho^t_{n}(\lambda) \equiv \rho_n(\lambda) + \rho^h_n(\lambda)\,,
\ee
is proportional to the derivative of the dressed momentum $p_n^{d\, \prime}(\lambda)$, namely 
\begin{equation}
\rho^t_n(\lambda) =\frac{1}{2 \pi}  p_n^{d\,\prime} (\lambda)\,.
\label{eq:dressedmomentumderivative}
\end{equation}
The dressed energy $\varepsilon_n(\lambda)$ instead fulfils 
\be
\varepsilon_n^{\prime} (\lambda)=e_n^{\prime}  (\lambda) - \Big[ \sum_{k} \,T_{nk} \ast ( \varepsilon^{\prime}  \vartheta_k) \Big]  (\lambda)\,.
\label{eq:dressedenderivative}
\ee
Using $p_n^{d}(\lambda)$ and $\varepsilon_n(\lambda)$ we can find the group velocity of an elementary excitation of type $n$ and rapidity $\lambda$ as \cite{korepin,BoEL14}
\begin{equation}
v_n (\lambda) = \frac{\partial \varepsilon_n (\lambda)}{\partial p_n^d(\lambda)} = \frac{\varepsilon_n^{\prime} (\lambda)}{2 \pi \rho_n^t (\lambda)}\,.
\label{eq:velocity}
\end{equation}

\section{The thermal equilibrium state and the ground state}
\label{sec:ground}


When the system is in thermal equilibrium at temperature $T$, its state is described by the following density matrix
\be
\boldsymbol \rho = \frac{1}{Z}e^{-\beta \boldsymbol H}\,,\qquad\qquad\qquad Z=\tr\left[e^{-\beta \boldsymbol H}\right]\,.
\label{eq:thermalstate}
\ee 
The root densities characterising this state in the context of TBA are determined in terms of a set of integral equations~\cite{takahashi}; it is convenient to express them using the ratio $\eta_n(\lambda)$ (\emph{cf}. Eq.~\eqref{eq:etan})
\begin{equation}
\label{eq:therm}
T \log \eta_j(\lambda)=  e_j(\lambda)+ T \sum_{k}\,  \left[T_{jk}\ast \log \left( 1 + \eta_{k}^{-1}\right)\right](\lambda)\,.
\end{equation}
It is useful to rewrite these equations in the following ``partially decoupled" form~\cite{takahashi}
\begin{align}
&\log\eta_n(x) = -\frac{J \pi \sinh \eta}{ T}s(x)\delta_{n,1}+s\ast \log[(1+\eta_{n-1})(1+\eta_{n+1})](x)\,, \quad n\geq 1\,,\label{eq:TBA}\\
&\lim_{n\rightarrow\infty}\frac{\log\eta_{n}(x)}{n} = \frac{2 h}{T}\,, \label{eq:lastthermaldecoupledD>1}
\end{align}
where we adopted the convention $\eta_0(x)\equiv 0$ and we introduced 
\begin{equation}
s(x) = \frac{1}{ 2\pi} \sum_{j=-\infty}^\infty  \frac{e^{2 i j x}}{ \cosh (j \eta)} =\frac{K(w)}{\pi^2} \,{\rm dn}\Bigl(\frac{2K(w)}{\pi}x\Big |w\Bigr)\,.
\label{eq:defKernelS}
\end{equation}
Here ${\rm dn}(x|w)$ is a Jacobi elliptic function, $K(w)$ is the complete elliptic integral of the first kind, and $w$ is the unique solution to~\cite{gradstein}
\be\label{eq:w}
K(1-w)=\frac{\eta}{\pi}K(w)\, .
\ee

The thermal TBA equations \eqref{eq:TBA} and \eqref{eq:lastthermaldecoupledD>1} can be used to study the ground-state properties of the system by taking the $T\rightarrow0^+$ limit. As ground-state properties encode crucial information on the transport behaviour at small temperatures, it is useful to review them before analysing the transport scenario. We also refer to \cite{takahashi} for further detail.

\begin{figure}
	\includegraphics[width=0.97\textwidth]{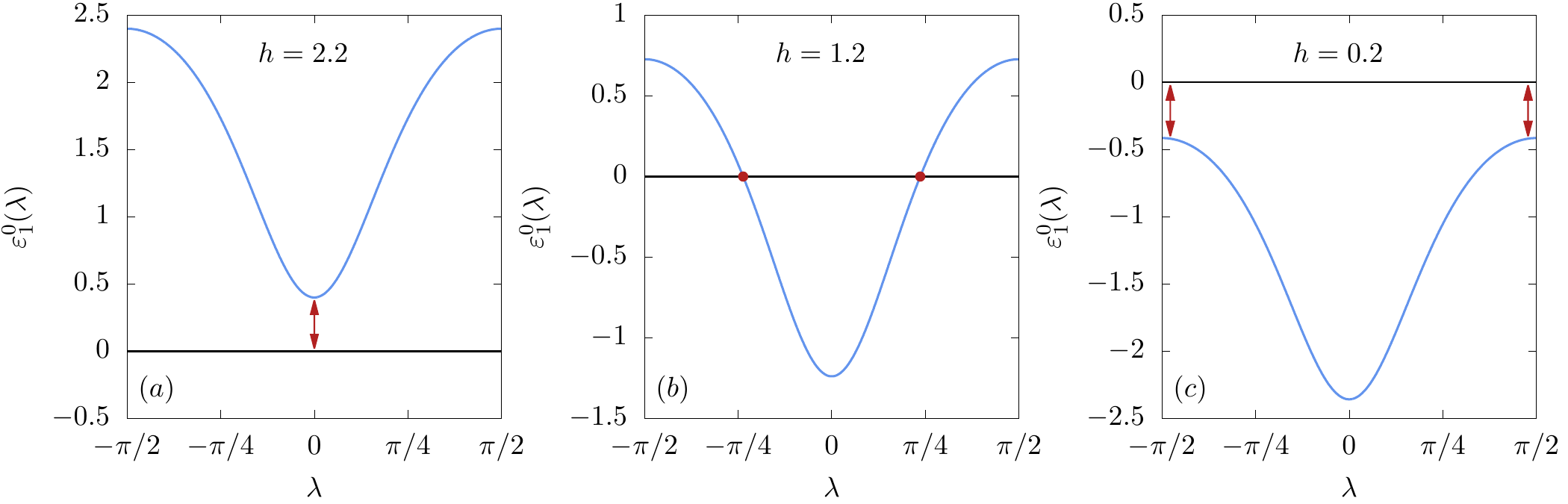}
	\caption{Dressed energies of the elementary excitations above the ground state for different phases of the Hamiltonian. The anisotropy is chosen to be $\Delta=3$. Figures $(a)$ and $(c)$  correspond to the ferromagnetic and antiferromagnetic phases. They display a finite gap in the spectrum of excitations (red arrows). Figure $(b)$ shows the dressed energy in the gapless phase. Two zeros appear in correspondence to the Fermi quasi-momenta (red points), allowing for excitations for which no energy cost is required. }
	\label{fig:dressed_energies}
\end{figure}
In the limit ${T\rightarrow0^+}$ , the thermal functions $\eta_n(\lambda)$ fulfilling~\eqref{eq:therm} diverge exponentially in $1/T$ and it is useful to introduce the \emph{thermal dressed energies}, defined as follows 
\be
\varepsilon^{\rm th}_n(\lambda)\equiv T\log\eta_n(\lambda)\,.
\label{eq:thermalde}
\ee
The thermal dressed energies remain finite in the zero-temperature limit. Moreover, by taking the derivative of \eqref{eq:therm} and integrating by parts it is easy to verify that $\varepsilon^{\rm th\,\prime}_n(\lambda)$ satisfies \eqref{eq:dressedenderivative}. The functions \eqref{eq:thermalde} are then nothing but the dressed energies of elementary excitations on the thermal state and from now on we denote them by $\varepsilon_n(\lambda)$.

Considering the decoupled TBA equations \eqref{eq:TBA} and \,\eqref{eq:lastthermaldecoupledD>1} we see that if $h>0$
\be
\varepsilon_j(\lambda)>0\qquad{\rm for}\qquad j\geq 2\,.
\label{eq:property}
\ee
Using this property, we can simplify the thermal TBA equations \eqref{eq:therm} in the zero temperature limit. The result reads as 
\be
\varepsilon^{0}_n(\lambda) =  e_n(\lambda)-\left[T_{n1}\ast \varepsilon^{0-}_1\right] (\lambda)\,, \qquad \qquad \qquad n=1,\ldots,+\infty\,,
\label{eq:limitepsilonn}
\ee
where we denoted by $\varepsilon^{0}_n(\mu)$ the thermal dressed energies for zero temperature and we defined 
\be
\varepsilon^{0\,-}_1(\mu)=\frac{1}{2}\left(\varepsilon^{0}_1(\mu)-|\varepsilon^{0}_1(\mu)|\right)\,.
\ee
Note that all the dressed energies are determined once $\varepsilon^{0}_1(\lambda)$ is known. 

The zero temperature dressed energy $\varepsilon^{0}_1(\mu)$ gives direct information on the critical regions of the model: if it has a zero, excitations can be produced with no energy cost and the system is gapless while it is gapped otherwise. We note that $\varepsilon^{0}_1(\lambda)$ is a continuous monotonic function of $|\lambda|$, which allows us to identify three separate phases, see Fig.~\ref{fig:dressed_energies}. Two of them are gapped and correspond to the dressed energy having a definite sign: they are respectively characterised by $\varepsilon^{0}_1(0)>0$ and $\varepsilon^{0}_1(\pi/2)<0$. The third regime is instead gapless and is realised when ${\varepsilon^{0}_1(0)<0}$ and ${\varepsilon^{0}_1(\pi/2)>0}$.

\subsection{Ferromagnetic regime}
Let us first determine the regime where the dressed energy ${\varepsilon^{0}_1(\lambda)}$ is positive for all $\lambda$. We note that $e_1(\lambda)\geq e_1(0)=2h-J (1+\Delta)$. As a consequence, if ${h>J(1+\Delta)/2}$, then $e_1(\lambda)>0$ and, using $T_{11}(\lambda)=a_2(\lambda)>0$, from \eqref{eq:limitepsilonn} we conclude $\varepsilon^{0}_1(\lambda)>0$. Instead, for $h<J(1+\Delta)/2$ the driving term $e_1(\lambda)$ becomes negative for small enough $\lambda$, implying that also $\varepsilon^{0}_1(\lambda)$ changes sign. This can be easily proven by \emph{reductio ad absurdum}. If $\varepsilon^{0}_1(\lambda)$ is never negative we have 
\be
\varepsilon^{0}_1(\lambda)=e_1(\lambda)\,.
\ee
This is impossible as $e_1(\lambda)$ is negative for small enough $\lambda$. The statement is then proved.  

Focussing on $h>J(1+\Delta)/2$ and using again \eqref{eq:limitepsilonn} we have   
\be
\varepsilon^{0}_n (\lambda) = e_n(\lambda)>0\,,\qquad\qquad\qquad n\geq1\,.
\label{eq:ferromagneticenergy}
\ee
Since in this regime all the dressed energies are positive for $T=0$, all the root densities are zero. For $h\geq 0^+$, the ground state corresponds to the reference state where all the spins are up. For $h=0$ (and $\Delta\leq -1$) one of the two reference states (all spins up or all spins down) is selected in the thermodynamic limit as the ground state by spontaneous symmetry breaking. This regime is known as the ``ferromagnetic" phase of the XXZ spin-$1/2$ chain~\cite{takahashi}, and is reported in the upper part of  Fig.~\ref{fig:phasediagram}. In this phase, since all the root densities are zero, all the dressed quantities are equal to the bare ones. 

\subsection{Gapless regime}
\label{sec:groundgapless}

For $h<J(1+\Delta)/2$ and $h$ close enough to $J(1+\Delta)/2$, there must exist a single ($\varepsilon^{0}_1(|\mu|)$ is monotonic) point $B>0$ such that  
\be
\varepsilon^{0}_1(B)=0\,,
\ee
meaning that the system becomes gapless; this phase is depicted in the central part of Fig.~\ref{fig:phasediagram}. In this regime we can rewrite Eq.~\eqref{eq:limitepsilonn} for $\varepsilon^{0}_1(\lambda)$ as 
\be
\varepsilon^{0}_1(\lambda) = e_1(\lambda)- \, \int_{-B}^{B}\mathrm \!\!d\mu \,\,\, \ak_{2}(\lambda-\mu) \varepsilon^{0}_1(\mu)\,.
\label{eq:limitepsilon}
\ee

\subsection{Antiferromagnetic regime}
A second transition occurs when the maximum of $\varepsilon^{0}_1(\lambda)$, namely $\varepsilon^{0}_1(\pi/2)$, becomes zero; this means $B=\pi/2$. The solution to \eqref{eq:limitepsilon} when $B=\pi/2$ is explicitly computed in terms of Jacobi elliptic functions as follows~\cite{takahashi, gradstein}
\be
\varepsilon^{0}_1(\lambda) = -J\sqrt{\Delta^2-1} s(\lambda)+ h\,,
\label{eq:leading_eps1_anti}
\ee
where $s(\lambda)$ is defined in \eqref{eq:defKernelS}. The condition $\varepsilon^{0}_1(\pi/2)<0$ is then written in terms of the parameters as 
\be
h<h_c(J,\Delta)\equiv J \sqrt{\Delta^2-1}\, s(\pi/2)  \,.
\label{eq:lower_critical_field}
\ee
For $h<h_c(J,\Delta)$ the system becomes again gapped. In this phase, however, $\varepsilon^{0}_1(\lambda)$ is negative for all $\lambda$ and the root density of the $1$-strings becomes equal to $\rho_{1}^{t\,0}(\lambda)$, with
\be
\rho^0_{1}(\lambda)=\rho^{t\,0}_{1}(\lambda)=s(\lambda)\,.
\ee

\section{Quench protocol and Locally Quasi-Stationary States}
\label{sec:protocol}
 
We consider the non-equilibrium dynamics resulting from joining two XXZ chains in two thermal states at different temperatures. More precisely, we focus on the initial state
\begin{equation}
\boldsymbol\rho_0  = \frac{e^{- \beta_L \boldsymbol{H}_L}}{Z_L} \bigotimes \frac{e^{- \beta_R \boldsymbol{H}_R}}{Z_R}\,,
\label{eq:initialstate}
\end{equation}
where the left density matrix applies on the negative sites of the chain and the right one on the positive sites. 
As shown in Ref.~\cite{BeFa16,BCDF16, CaDY16} at late times the system is described by a locally quasi-stationary state (LQSS) around each ray $\zeta = x/t$ between the two maximal velocities of particle propagation in the left and the right state $-v_L<\zeta<v_R$.  The LQSS are macrostates specified by a set of particle densities $\rho_{j , \xt}(\lambda)$ which satisfy a local continuity equation of the form 
\begin{equation}
\partial_t \rho_{j , \xt}(\lambda) + \partial_x \left( v_{j,\xt}(\lambda ) \rho_{j, \xt} (\lambda) \right)=0\,.
\label{eq:continuityrho}
\end{equation}
This equation follows from the infinite set of continuity equations satisfied by the densities of local and quasi-local conserved charges of the system~\cite{BCDF16, CaDY16} and is the fundamental equation in the TBA treatment of generalised hydrodynamics. Equation~\eqref{eq:continuityrho} is most easily solved in terms of the filling functions $\vartheta_{j ,\xt}(\lambda)$ (\emph{cf}. Eq.~\eqref{eq:fillingfunctions}); the solution reads as 
\be
\label{eq:transportsolution}
\vartheta_{j ,\xt}(\lambda) = \vartheta^R_{j}(\lambda) H(\xt - v_{j, \xt}(\lambda))+ \vartheta^ L_{j } (\lambda) H(v_{j, \xt}(\lambda)-\xt)\,.
\ee
Here $H(x)$ is the step function---it is non-zero and equal to one only if $x>0$---and 
\be
\vartheta^{R/L}_{n}(\lambda)=\frac{1}{1+\eta^{R/L}_{n}(\lambda)}\,,
\ee
are the filling functions of the left and right thermal states. Here $\eta^{R/L}_{n}(\lambda)$ are the solutions to the thermal TBA equations~\eqref{eq:therm} respectively with $T=T_L$ and $T=T_R$. 
As the velocity of excitations $ v_{j, \xt}(\lambda)$ depends on the filling functions $\vartheta_{j ,\xt}(\lambda)$, Equation \eqref{eq:transportsolution} must be solved iteratively. Once the macrostates $\vartheta_{j ,\xt}(\lambda)$ at each ray $\xt$ are known, one can compute the expectation values of observables as a function of the ray $\xt$. For example, charge densities' profiles are computed as follows (\emph{cf}. Sec~\ref{sec:charges})
\be
\braket{\boldsymbol q_\ell}_\xt  =    \sum_{j} \int_{-\pi/2}^{\pi/2} d \lambda\,\, q_j (\lambda) \rho_{j,\xt}(\lambda)\,,
\label{eq:chargeprofile}
\ee
where $\boldsymbol q_\ell$ is a generic conserved charge density characterised by the bare charge $q_j(\lambda)$. Here $\ell$ labels the position on the lattice.
We note that Eq.~\eqref{eq:chargeprofile} implies that $\braket{\boldsymbol q_\ell}_\xt $ does not depend on $\ell$: this is due to the translational invariance of the LQSS at fixed ray. In the following we will use this property to lighten the notation: we will not report the explicit $\ell$ dependence of expectation values of local operators at fixed ray. For example, we will write the l.h.s. of \eqref{eq:chargeprofile} as $\braket{\boldsymbol q}_\xt$. One can also consider conserved charge currents $\boldsymbol J_{\boldsymbol q, \ell}$, defined in terms of $\boldsymbol q_\ell$ via the following ``discrete" continuity equation
\be
\boldsymbol J_{{\boldsymbol q},\ell+1}-\boldsymbol J_{{\boldsymbol q},\ell}=i[\boldsymbol q_\ell,\boldsymbol H]\,. 
\label{eq:discretecontinuityeq}
\ee
The current is completely specified up to operators with zero expectation value in any translationally invariant state if one requires it to have zero expectation value on the state with zero root densities. Current profiles of local charges can be computed as follows~\cite{BCDF16,CaDY16}
\be
\braket{\boldsymbol J_{\boldsymbol q}}_\xt  =    \sum_{j} \int_{-\pi/2}^{\pi/2} d \lambda\,\, q_j (\lambda)v_{j,\xt}(\lambda )  \rho_{j,\xt}(\lambda)\,.
\label{eq:currentprofile}
\ee
Finally, knowing the macrostate root densities $\rho_{j,\xt}(\lambda)$ one can also explicitly compute correlation functions on a fixed ray~\cite{PDCB17}, by employing analytic formulae recently derived in~\cite{PM:correlations}.

Our main goal here is to construct the low-temperature expansion of the solution \eqref{eq:transportsolution} and in turn of the profiles \eqref{eq:chargeprofile} and \eqref{eq:currentprofile}. As the structure of the expansion and the form of the corrections strongly depend on the phase of the model, it is convenient to consider the different phases separately. This is done in the following sections: in Sec.~\ref{sec:lowtgapless} we consider the gapless phase, in Sec.~\ref{sec:lowtferro} the ferromagnetic phase, and finally in Sec.~\ref{sec:lowtantiferro} we study the antiferromagnetic phase.

\section{Transport in the critical phase}
\label{sec:lowtgapless}

In this section we focus on the transport properties arising at small temperatures in the gapless phase. In particular, we compute analytically the profiles of charges and currents in the limit $T\to 0$. Concretely, we fix 
\be
r\equiv\frac{T_R}{T_L}\,,
\label{eq:defr}
\ee
and perform an expansion for small $T_L\equiv T$. We focus on the region in parameter space determined by
\be
 \Delta>1\qquad\text{and}\qquad h_c(J,\Delta)<h<J(1+\Delta)/2\,.
\ee 
All our results, however, will be applicable in the entire gapless phase of the XXZ spin chain, \emph{i.e.}, also for $-1<\Delta<1$ and $h<J(1+\Delta)/2$ if the appropriate definitions for the string momenta $p_j(\lambda)$ are used.  

Due to the absence of an inherent energy scale, at small temperatures the expectation values of observables deviate from their ground states by some power-law corrections in temperature. These corrections are those generating non-trivial transport dynamics. 

To find the low temperature corrections to the profiles of local observables,  we need to determine the corrections to the relevant TBA quantities, specifically to $\varepsilon_n(\lambda)$, $\rho^t_n(\lambda)$, and $v_n(\lambda)$. Let us begin by considering a warm-up example and find the low-temperature corrections acquired by the latter quantities in a homogeneous thermal state~\eqref{eq:thermalstate}. In Sec.~\ref{sec:inhomotbaexp} we adapt the procedure to inhomogeneous states of the form \eqref{eq:transportsolution} and present our results in Sec.~\ref{sec:profiles}.  
 
\subsection{Low-temperature expansion of the thermal TBA equations}
\label{sec:homotbaexp}

Let us start by following Ref.~\cite{TakalowT} and compute the first correction to the dressed energy $\varepsilon_1(\lambda)$ with respect to its ground state value $\varepsilon^0_1(\lambda)$, defined as the solution to \eqref{eq:limitepsilon}. Since the dressed energies $\varepsilon_n(\lambda)$ for $n>2$ are strictly positive (\emph{cf}. Eq.~\eqref{eq:property}), for small temperatures, Eq.~\eqref{eq:therm} for $n=1$ can be written as 
\begin{align}
\varepsilon_1(\lambda) &= e_1(\lambda)+  T \int_{-\pi/2}^{\pi/2} \!\!\!\!\mathrm d\mu \,\,\, \ak_{2}(\lambda-\mu) \ln \left( 1 + e^{-\frac{\varepsilon_1(\mu)}{T}}\right)+O(e^{-1/T})\,.
\label{eq:generatingcorrections}
\end{align}
Note that here and in the following we have implicitly set the energy scale $J=1$, so that small temperatures correspond to $T\ll 1$. If $T$ is small enough, the dressed energy $\varepsilon_1(\lambda)$ continues to have two zeroes, which we call $\pm B'$. Since the dressed energy is a symmetric function of $\lambda$, the two zeroes continue to be symmetrically disposed around 0. From \eqref{eq:generatingcorrections} and \eqref{eq:limitepsilon} we find 
\bea
 \quad\quad\varepsilon_1(\lambda) -\varepsilon_1^{0}(\lambda)&=& T \int_{-\pi/2}^{\pi/2} \!\!\mathrm d\mu  \,\,\, \ak_{2}(\lambda-\mu) \left(\ln \left[ 1 + e^{-\frac{\varepsilon_1(\mu)}{T}} \right]+\frac{\varepsilon^{0\,-}_1(\mu)}{T}\right)\nn
&=& - \int_{-B}^{B}\!\!\mathrm d\mu \,\,\, \ak_{2}(\lambda-\mu) \left(\varepsilon_1(\mu) -\varepsilon^{0}_1(\mu)\right) - \int_{B}^{B'}\!\!\mathrm d\mu \,\,\, \ak_{2}(\lambda-\mu) \varepsilon_1(\mu)\nn
&& - \int_{-B'}^{-B}\!\!\mathrm d\mu \,\,\, \ak_{2}(\lambda-\mu) \varepsilon_1(\mu) +T \int_{-\pi/2}^{\pi/2} \!\!\mathrm d\mu  \,\,\, \ak_{2}(\lambda-\mu) \ln \left[ 1 + e^{-\frac{|\varepsilon_1(\mu)|}{T}}\right]\,.\label{eq:firstorder}
\eea
The last term in the r.h.s. of \eqref{eq:firstorder} can be simplified by expanding the integrand around the points $\pm B'$, where $\varepsilon_1(\lambda)$  vanishes 
\bea
\int_{-\pi/2}^{\pi/2} \!\!\mathrm d\mu \,\, \ak_{2}(\lambda-\mu) \,\ln\left[1+e^{-\frac{|\varepsilon_1(\mu)|}{T}} \right]&=& \frac{2T}{\varepsilon'_1(B')} \left(\ak_{2}(\lambda-B') + \ak_{2}(\lambda+B') \right) \int_0^{\infty}{\rm d}x\,\ln\left[1+e^{-x} \right]\nn
&=& \frac{\pi^2 T}{6 \varepsilon'_1(B')} \left(\ak_{2}(\lambda-B') + \ak_{2}(\lambda+B') \right)\,.
\label{eq:driving}
\eea
Here we have neglected $O(T^3)$. Then, we see that the fourth contribution in \eqref{eq:firstorder} is $O(T^2)$, while the second and third are $O((B-B')^2)$. Here we assume 
\be
B-B'=O(T^\alpha)\qquad\qquad \alpha>0\,.
\ee
There are then two cases; $(i)$ $\alpha\leq 1$, $(ii)$ $\alpha>1$. Let us show that the case $(i)$ is impossible, we will do it by \emph{reductio ad absurdum}. In the case $(i)$ the fourth contribution to \eqref{eq:firstorder} can be neglected and we find 
\be
\varepsilon_1(\lambda)-\varepsilon_1^{0}(\lambda)= T^{2\alpha} F(\lambda) + O(T^2)\,,
\label{eq:DEespwrong}
\ee
where $F(\lambda)$ is a $T$-independent function. Computing \eqref{eq:DEespwrong} in $\lambda=B'$ and expanding in $B-B'$ we have 
\be
-\varepsilon_1^{0\,\prime}(B)(B-B')+O((B-B')^2)=  T^{2\alpha} F(B) + O(T^2)
\ee
which is inconsistent for $\alpha\leq1$, as it requires $F(B)=\varepsilon_1^{0\,\prime}(B)=0$. In the case $(ii)$ instead, we can neglect the second and third contributions in \eqref{eq:firstorder}. Doing that we find
\be
\delta \varepsilon_1(\lambda)=\frac{\pi^2 T^2}{6 \varepsilon_1^{0\prime}(B)} U(\lambda) + O(T^{2\alpha})\,,
\label{eq:DEesp}
\ee
where we introduced the short-hand notation 
\be
\delta f(\lambda)\equiv f(\lambda)-f_0(\lambda)\,,
\label{eq:shorthandnotation}
\ee
to denote the difference between a quantity and its ground state value. The function $U(\lambda)$ appearing in \eqref{eq:DEesp} is defined as the solution of
\be
U(\lambda) = \ak_{2}(\lambda-B) + \ak_{2}(\lambda+B) - \int_{-B}^{B}\mathrm \!\!d\mu \,\,\, \ak_{2}(\lambda-\mu)U(\mu)\,.
\label{eq:inteqU}
\ee
Note that from Eq.~\eqref{eq:inteqU} it follows that the function $U(\lambda)$ must be an even function of $\lambda$, as the kernel and the driving term are both even.

By computing \eqref{eq:DEesp} in $\lambda=B'$ we find $\alpha=2$. Equation \eqref{eq:DEesp} gives the desired first correction to the dressed energy $\varepsilon_1(\lambda)$ for finite temperatures. Note that $\varepsilon_1(\lambda)$ must have at least one zero to produce power law corrections in $T$; when the dressed energy is non-zero for all $\lambda$s the integral \eqref{eq:driving} is bounded by a term $\propto e^{-\beta \min_\lambda|\varepsilon_1(\lambda)| }$, leading to exponentially suppressed corrections. All this has a very natural physical interpretation: in the gapped case, when the temperature is low enough, the thermal excitation energy becomes smaller that the gap and no excitation can be produced; accordingly the corrections have a characteristic energy scale. In the gapless case, however, for any finite $T$ the thermal excitation energy is sufficient to create some excitations and, accordingly, the corrections have no energy scale. 

Let us now turn to consider the total root density and the dressed velocity of low energy quasi-particle excitations. The equations describing these quantities up to exponential corrections in $1/T$ are obtained by neglecting higher strings contributions from Eqs.~\eqref{eq:tba} and  \eqref{eq:velocity} for $n=1$ and read as 
\begin{align}
&\rho_{1}^t(\lambda)=\ak_1(\lambda)-\int_{-\pi/2}^{\pi/2} \!\!\mathrm d\mu \,\ak_2(\lambda-\mu)\vartheta_1(\mu)\rho^t_{1}(\mu)+O(e^{-1/T})\,,
\label{eq:tbalt}\\
&v_{1}(\lambda)\rho^t_{1}(\lambda)=\frac{1}{2\pi}e_1'(\lambda)-\int_{-\pi/2}^{\pi/2} \!\!\mathrm d\mu \,\ak_2(\lambda-\mu)\vartheta_1(\mu)v_{1}(\mu)\rho^t_{1}(\mu)+O(e^{-1/T})\,.
\label{eq:velocitieslt}
\end{align}
The zero temperature limit of these equations reads as 
 \begin{align}
&\rho_{1}^{t\,0}(\lambda)=\ak_1(\lambda)-\int_{-B}^{B} \!\!\mathrm d\mu \,\ak_2(\lambda-\mu)\rho^{t\,0}_{1}(\mu)\,,
\label{eq:rhot0}\\
&v_{1}^0(\lambda)\rho^{t\,0}_{1}(\lambda)=\frac{1}{2\pi}e_1'(\lambda)-\int_{-B}^{B} \!\!\mathrm d\mu \,\ak_2(\lambda-\mu)v^0_{1}(\mu)\rho^{t\,0}_{1}(\mu)\,.
\label{eq:velocitiest0}
\end{align}
From these expressions we see that finding the first corrections to $\rho^t_{1}(\lambda)$ and $v_1(\lambda)$ involves the expansion of Sommerfeld-like integrals   
\be
I(\beta)=\int_{-\pi/2}^{\pi/2}\!\!\!\!{\rm d}\mu\,\,\,\vartheta_1(\mu) f(\mu)
\label{eq:integralbeta}
\ee
where $f(\mu)$ is a smooth function and 
\be
\vartheta_1(\lambda)=\frac{1}{1+e^{\beta {\varepsilon_1(\lambda)}}}\,.
\ee
It is convenient to construct an expansion of this integral once and for all. This task is carried out in Appendix~\ref{app:sommerfeldoneT}, the result reads as 
\bea
 I(\beta)=\int^{B}_{-B}\!\!\!\!{\rm d}\mu\, f(\mu)+\frac{\pi^2 T^2}{6 (\varepsilon_1^{0\,\prime}(B))^2}\left[f'(B)-f'(-B)-\left(\frac{\varepsilon_1^{0\,\prime\prime}(B)}{\varepsilon_1^{0\,\prime}(B)}+U(B)\right)(f(B)+f(-B))\right]+O(T^4)\,.
\label{eq:expansion}
\eea
Note that this behaviour is found only when the dressed energy has a zero. When $\varepsilon_1(\lambda)\neq0$ for all $\lambda$s the corrections are once again exponential. Using this expansion in the equation for $v_1(\lambda)\rho^t_{1}(\lambda)$ together with \eqref{eq:velocitiest0} we find 
\bea
\delta \left(v_1\rho^t_{1}\right)(\lambda)\equiv v_1(\lambda)\rho^t_{1}(\lambda)-v_1^0(\lambda)\rho^{t\,0}_{1}(\lambda)&=&-\int^{\pi/2}_{0}\!\!{\rm d}\mu\,\left(\ak_2(\lambda-\mu)-\ak_2(\lambda+\mu)\right)\vartheta_1(\mu)v_1(\mu)\rho^t_{1}(\mu)\nn
&&+\int_{0}^{B}\!\!{\rm d}\mu \,\,\left(\ak_2(\lambda-\mu)-\ak_2(\lambda+\mu)\right)v_1^0(\mu)\rho_{1}^{t\,0}(\mu)\nn
&=&-\int^{B}_{0}\!\!\!\!{\rm d}\mu\,\left(\ak_2(\lambda-\mu)-\ak_2(\lambda+\mu)\right)\delta \left(v_1\rho^t_{1}\right)(\mu)\nn
&&+\frac{\pi T^2}{12 \varepsilon_1^{0\,\prime}(B)}\left[\ak_2'(\lambda+B)+\ak_2'(\lambda-B)\right]\nn
&&+\frac{\pi T^2}{12 \varepsilon_1^{0\,\prime}(B)}\left[\left(\ak_2(\lambda-B)-\ak_2(\lambda+B)\right)U(B)\right]+O(T^4)\,,
\eea
where in the second equality we explicitly used $\rho^t_1(\mu)v_1(\mu)=\varepsilon^{\prime}_1(\mu)/2\pi$. This expression can be rewritten as 
\be
\delta \left(v_1\rho^t_{1}\right)(\lambda)=\frac{\pi T^2}{12 \varepsilon_1^{0\,\prime}(B)} W(\lambda) + O(T^4)\,,
\label{eq:DDEesp}
\ee
where $W(\lambda)$ is defined as the solution of 
\bea
 W(\lambda) = - \int_{-B}^{B}\mathrm \!\!\!\!d\mu \,\, \ak_{2}(\lambda-\mu)W(\mu) + \ak_{2}'(\lambda-B) + \ak_{2}'(\lambda+B) + U(B)(\ak_{2}(\lambda-B) - \ak_{2}(\lambda+B)).
 \label{eq:W}
\eea
From this equation it follows that $W(\lambda)$ is an odd function: this can be seen by noting that the kernel is even and the driving term is odd. 
Proceeding analogously, and using $\rho^t_1(\mu)=\varepsilon^{\prime}_1(\mu)/(2v_1(\mu)\pi)$, we find the correction to $\rho^t_{1}(\lambda)$
\be
\delta \rho^t_{1}(\lambda)\equiv\rho^t_{1}(\lambda)-\rho^{t\,0}_{1}(\lambda)=\frac{\pi T^2}{12 \varepsilon_1^{0\,\prime}(B) v^0_1(B)} R(\lambda)+O(T^4)\,,
\ee
where $R(\lambda)$ solves
\bea
R(\lambda) =  -\int_{-B}^{B}\mathrm \!\!\!{\rm d}\mu \, \ak_{2}(\lambda-\mu)R(\mu)+\left(\ak_2(\lambda-B)+\ak_2(\lambda+B)\right)\left[U(B)+\frac{v_{1}^{0\,\prime}(B)}{v_{1}^{0}(B)}\right]+\left(\ak_2'(\lambda-B)-\ak_2'(\lambda+B)\right)\,.
\label{eq:R}
\eea
Note that this equation implies that $R(\lambda)$ is even. 

These equations allow, \emph{e.g.}, to find the first finite temperature correction to the energy density of the state. This can be done as follows 
\begin{align}
e&=    \sum_{j} \int_{-\pi/2}^{\pi/2} d \lambda\,\, e_j (\lambda) \rho_j(\lambda)= \int_{-\pi/2}^{\pi/2} d \lambda\,\, e_1 (\lambda) \vartheta_1(\lambda) \rho^t_1(\lambda)+O(e^{-\beta})\notag\\
&=e_0+\frac{\pi T^2}{12 \varepsilon_1^{0\,\prime}(B) v^0_1(B)} \left(\int_{-B}^{B} d \lambda\,\, e_1 (\lambda)R(\lambda)+2 e_1'(B)-2\left(\frac{v_1^{\prime}(B)}{v_1(B)}+U(B)\right)e_1(B)\right)+O(T^4)\notag\\
&=e_0+\frac{\pi T^2}{6 v^0_1(B)}+O(T^4)\,.
\end{align}
Here we defined 
\be
e_0= \int^B_{-B}{\rm d}\mu \,e_1(\mu) \rho^{t\,0}_{1}(\mu)\,.
\ee
In the last step we used 
\be
\int_{-B}^{B} d \lambda\,\, e_1 (\lambda)R(\lambda)+2 e_1'(B)-2\left(\frac{v_1^{\prime}(B)}{v_1(B)}+U(B)\right)e_1(B) =2\varepsilon_1^{0\,\prime}(B)\,.
\ee
This relation can be easily proven by using the identities \eqref{eq:identity1} and \eqref{eq:identity2} of Appendix~\ref{app:tbaidentitieslowt}. Note that the finite temperature correction agrees with the CFT result~\cite{affleckCFT, BCNCFT} for a theory of central charge equal to one and velocity of light equal to $v^0_1(B)$, the velocity of excitations in the ground state calculated at the ``Fermi point'' $B$.

\subsection{Low-temperature expansion in the inhomogeneous case}
\label{sec:inhomotbaexp}

Having settled the homogeneous case, let us now move on and undertake our main goal: developing a low temperature expansion of the late-time profiles determined by \eqref{eq:transportsolution}.  Considering the thermal filling functions in \eqref{eq:transportsolution} we immediately see that we can restrict to $\vartheta_{1,\xt}(\lambda)$; all the others are exponentially suppressed as $\varepsilon_{n,R/L}(\lambda)>0$ for $n\geq2$. To solve the problem, Equation~\eqref{eq:transportsolution} for $\vartheta_{1,\xt}(\lambda)$ must be complemented with the two equations \eqref{eq:tbalt} and \eqref{eq:velocitieslt} for the total root density $\rho^t_{1,\xt}(\lambda)$ and for the velocity $v_{1,\xt}(\lambda)$. At the lowest order in $T_L=T$ and $T_R= r T$ (\emph{cf.}~\eqref{eq:defr}), these quantities are the ground state ones, denoted by $\rho^{t\,0}_{1}(\lambda)$ and $v_1^{0}(\lambda)$, and are determined by the equations \eqref{eq:rhot0} and \eqref{eq:velocitiest0}. Importantly, they are constant in $\xt$. 

To find the first non trivial corrections to $\rho^t_{1,\xt}(\lambda)$ and $v_{1,\xt}(\lambda)$ for small but finite $T_L$ and $T_R$, and in turn some non-trivial dependence on the ray $\xt$, it is again convenient to construct the low temperature expansion of a Sommerfeld-like integral
\be
 I(\beta,r,\xt)=\int_{-\pi/2}^{\pi/2}\!\!\!{\rm d}\lambda\,\,\vartheta_{1,\xt}(\lambda)f(\lambda)\,.
 \label{eq:sommerfeldtwot}
 \ee
The calculation is thoroughly carried out in Appendix~\ref{app:integraltwoT}, retaining up to orders $O(T^2)$. As the result is quite cumbersome we do not report its full expression here. It is, however, instructive to consider its form for two different regimes of rays $\xt$. 

For rays $\xt$ which are $O(T^0)$ away from $\pm v^{0}_{1}(B)$, namely
\be
\lim_{T\rightarrow0}|\xt\pm v^{0}_{1}(B)|\neq0\,,
\label{eq:CFTregime}
\ee
we find 
\begin{align}
 I(\beta,r,\xt)=&\int^{B}_{-B}\!\!{\rm d}\lambda\, f(\lambda)+\frac{\pi^2 T^2}{6 (\varepsilon_1^{0\,\prime}(B))^2}\left(f^{\prime}(B)-\frac{\varepsilon_1^{0\,\prime\prime}(B)}{\varepsilon_1^{0\,\prime}(B)}f(B)-U(B)f(B)\right)H(v^{0}_{1}(B)-\xt)\nn
&+\frac{\pi^2 T^2}{6 (\varepsilon_1^{0\,\prime}(B))^2}\left(-f^{\prime}(-B)-\frac{\varepsilon_1^{0\,\prime\prime}(B)}{\varepsilon_1^{0\prime}(B)}f(-B)-U(B)f(-B)\right)H(-v^{0}_{1}(B)-\xt)\nn
&+\frac{\pi^2 T^2 r^2}{6 (\varepsilon_1^{\prime}(B))^2}\left(f^{\prime}(B)-\frac{\varepsilon_1^{\prime\prime}(B)}{\varepsilon_1^{\prime}(B)}f(B)-U(B)f(B)\right)H(\xt-v^{0}_{1}(B))\nn
&+\frac{\pi^2 T^2 r^2}{6 (\varepsilon_1^{\prime}(B))^2}\left(-f^{\prime}(-B)-\frac{\varepsilon_1^{\prime\prime}(B)}{\varepsilon_1^{\prime}(B)}f(-B)-U(B)f(-B)\right)H(\xt+v^{0}_{1}(B))+O(T^3)\,.
\label{eq:stepregion}
\end{align}
In this region, $I(\beta,r,\xt)$ takes different constant values depending on whether the ray is greater than $v^{0}_{1}(B)$, between $v^{0}_{1}(B)$ and $-v^{0}_{1}(B)$ or smaller than $-v^{0}_{1}(B)$. If $|\xt|>v^{0}_{1}(B)$, the result coincides with that reported in Eq.~\eqref{eq:expansion} for the low temperature corrections in a single thermal state. The region $-v^{0}_{1}(B)<\xt<v^{0}_{1}(B)$ is instead the non-equilibrium one, where half corrections come from the left thermal state and the other half from the right one. 

As our model has a non-linear dispersion, however, the expansion \eqref{eq:stepregion} does not hold close enough to the transition regions. Specifically, when $\xt$ is order $T$ close to the transition velocities, namely
\be
 \xt \pm v^{0}_{1}(B)\sim O(T)\,.
\label{eq:transition}
\ee
In this region there is a smooth dependence on the ray and, most importantly, the corrections are $O(T)$. They explicitly read as 
\be
I(\beta,r,\xt)=\int_{-B}^{B}{\rm d}\lambda\,\, f(\lambda)+\frac{\pi^2 T (1-r^2)v^{0\,\prime}_1(B)}{6\varepsilon_1'(B)|v^{0\,\prime}_1(B)|}\left[f(B) {\cal D}_r \left({\varepsilon^{0\,\prime}_1(B)}\frac{\xt-v^0_1(B)}{T |v_1^{0\,\prime}(B)|}\right)-f(-B) {\cal D}_r \left({\varepsilon^{0\,\prime}_1(B)}\frac{\xt+v^0_1(B)}{T |v_1^{0\,\prime}(B)|} \right)\right].
\label{eq:transitionregion}
\ee
Here to simplify the formulae we have neglected $O(T^2)$ contributions (their explicit form is reported in Appendix~\ref{app:integraltwoT}), and defined
\be
{\cal D}_r (z)\equiv\frac{6}{\pi^2(1-r^2)} \log(1+e^{z})-\frac{6r}{\pi^2(1-r^2)}\log(1+e^{z/r})\,.
\ee
The function ${\cal D}_r (z)$ is strongly peaked around zero, in particular we have
\be
\lim_{T\rightarrow0}\frac{1}{T}{\cal D}_r \left(\frac{z}{T}\right)=\delta(z)\,.
\label{eq:Dlimit}
\ee

Using the expansion of \eqref{eq:sommerfeldtwot} one can determine all the leading finite temperature corrections to the TBA quantities similarly to what we did in the homogeneous case in Sec.~\ref{sec:homotbaexp}.  This task is explicitly carried out in Appendix~\ref{app:tbaidentitieslowt} for $\rho^t_{1,\xt}(\lambda)$ and $v_{1,\xt}(\lambda)\rho^t_{1,\xt}(\lambda)$ (\emph{cf}. Eqs.~\eqref{eq:fullexpansionrhot} and \eqref{eq:fullexpansionvelrhot}).

\subsection{Low-temperature profiles}
\label{sec:profiles}

Let us now consider the profiles of charges~\eqref{eq:chargeprofile} and currents~\eqref{eq:currentprofile}. At low temperatures, all the contributions from higher strings is exponentially suppressed  in $1/T$ and can be safely neglected. Then, the expressions for the profiles can be simplified as follows 
\begin{align}
\braket{{\boldsymbol q}}_\xt&=\int_{-\pi/2}^{\pi/2} \!\!\!{\rm d}\lambda \,\, q(\lambda) \vartheta_{1,\xt}(\lambda) \rho^t_{1,\xt}(\lambda)+O(e^{-{1/T}})\,,\label{eq:chargedensityprofilelt}\\
\braket{\boldsymbol J_{\boldsymbol q}}_\xt &=\int_{-\pi/2}^{\pi/2} \!\!\!{\rm d}\lambda \,\, q(\lambda) \vartheta_{1,\xt}(\lambda) v_{1,\xt}(\lambda) \rho^t_{1,\xt}(\lambda)+O(e^{-1/T})\,,\label{eq:currentprofilelt}
\end{align}
 where $q(\lambda)$ is the bare charge for the first string---the only relevant one. For low enough temperatures, this expression can be further simplified, neglecting $O(T^3)$ we have 
 \begin{align}
\braket{{\boldsymbol q}}_\xt =& \int_{-\pi/2}^{\pi/2} \!\!\!{\rm d}\lambda \,\, q(\lambda) \vartheta_{1,\xt}(\lambda) \rho^{t\,0}_{1,\xt}(\lambda)+\int_{-B}^{B} \!\!\!{\rm d}\lambda \,\, q(\lambda) \delta \rho^t_{1,\xt}(\lambda)\notag\\
&+\frac{\pi^2 T (1-r^2)v^{0\,\prime}_1(B)}{6\varepsilon_1'(B)|v^{0\,\prime}_1(B)|}\int_{-B}^{B} \!\!\!{\rm d}\lambda \frac{\rm d}{{\rm d}\lambda}\left(q(\lambda)\delta\rho^{t}_{1,\xt}(\lambda) {\cal D}_r \left({\varepsilon^{0\,\prime}_1(\lambda)}\frac{\xt-v^0_1(\lambda)}{T v_1^{0\,\prime}(\lambda)}\right)\right)+O(T^3)\,,\label{eq:chargedensityprofilelt1}\\
\braket{\boldsymbol J_{\boldsymbol q}}_\xt =&\int_{-\pi/2}^{\pi/2} \!\!\!{\rm d}\lambda \,\, q(\lambda) \vartheta_{1,\xt}(\lambda) v^0_{1,\xt}(\lambda) \rho^{t\,0}_{1,\xt}(\lambda)+\int_{-B}^{B} \!\!\!{\rm d}\lambda \,\, q(\lambda) \delta\left(v_{1,\xt} \rho^{t}_{1,\xt}\right)(\lambda)\notag\\
&+\frac{\pi^2 T (1-r^2)v^{0\,\prime}_1(B)}{6\varepsilon_1'(B)|v^{0\,\prime}_1(B)|}\int_{-B}^{B} \!\!\!{\rm d}\lambda \frac{\rm d}{{\rm d}\lambda}\left(q(\lambda)\delta(v_{1,\xt}\rho^{t}_{1,\xt})(\lambda) {\cal D}_r \left({\varepsilon^{0\,\prime}_1(\lambda)}\frac{\xt-v^0_1(\lambda)}{T v_1^{0\,\prime}(\lambda)}\right)\right)+O(T^3)\,.\label{eq:currentprofilelt1}
\end{align}
Using the expansion of the integral \eqref{eq:sommerfeldtwot}, reported in Eq.~\eqref{eq:nonnaiveexpansionfinal}  of Appendix~\ref{app:integraltwoT}, together with the finite temperature corrections to $ \rho^t_{1, \xt}(\lambda)$ and $ v_{1,\xt}(\lambda) \rho^t_{1,\xt}(\lambda)$, presented in  \eqref{eq:fullexpansionrhot} and \eqref{eq:fullexpansionvelrhot} of Appendix \ref{app:tbaidentitieslowt}, we can explicitly write down the form of low-temperature profiles. This task is explicitly carried out in Appendix~\ref{app:tbaidentitieslowt}, retaining all the contributions up to $O(T^2)$. The results are presented in \eqref{eq:fullexpansioncharge} and \eqref{eq:fullexpansioncurrent}.  

Rather than reporting the full expression of the result, let us once again focus on the two regimes  \eqref{eq:CFTregime} and \eqref{eq:transition}. Away from the transition region---for rays satisfying \eqref{eq:CFTregime}---we find     
 \be
\braket{{\boldsymbol q}}_\xt = \frak d_0[q]+  T^2\left[ r^2 \frak d[q,1]\,H(\xt-v^{0}_{1}(B))+ \frak d[q,1] \,H(-v^{0}_{1}(B)-\xt)+\frak d[q,r] \,H(v^{0}_{1}(B)-|\xt|)\right]+O(T^3)\,,\label{eq:chargeslowtstep}
\ee
for the charge profile and 
\be
\braket{\boldsymbol J_{\boldsymbol q}}_\xt = \frak j_0[q]+ T^2\left[r^2  \frak j[q,1]\, H(\xt-v^{0}_{1}(B))+  \frak j[q,1]\, H(-v^{0}_{1}(B)-\xt)+ \frak j[q,r]\, H(v^{0}_{1}(B)-|\xt|)\right]+O(T^3)\,,\label{eq:currentslowtstep}
\ee
for the current. In the transition region, $\xt\pm v_1^0(B)\sim T$, up to $O(T)$ we find
\begin{align}
\braket{{\boldsymbol q}}_\xt&= \frak d_0[q]+ T\left[  \frak d_{-}[q,r]\,  {\cal D}_r \left({\varepsilon^{0\,\prime}_1(B)}\frac{\xt-v^0_1(B)}{T v_1^{0\,\prime}(B)} \right)+ \frak d_{+}[q,r]\, {\cal D}_r \left({\varepsilon^{0\,\prime}_1(B)}\frac{\xt+v^0_1(B)}{T v_1^{0\,\prime}(B)} \right)\right]+O(T^2)\,,\label{eq:chargeslowttransition}\\
\braket{\boldsymbol J_{\boldsymbol q}}_\xt&=\frak j_0[q]+ T\left[ \frak j_{-}[q,r] \,{\cal D}_r \left({\varepsilon^{0\,\prime}_1(B)}\frac{\xt-v^0_1(B)}{T v_1^{0\,\prime}(B)} \right)+\frak j_{+}[q,r]\,{\cal D}_r \left({\varepsilon^{0\,\prime}_1(B)}\frac{\xt+v^0_1(B)}{T v_1^{0\,\prime}(B)} \right)\right]+O(T^2)\,.\label{eq:currentslowttransition}
\end{align}
Here we introduced the zero temperature expectation values of the charge described by $q(\lambda)$ and the associated current  $\frak d_{0}[q]$ and $\frak j_{0}[q]$   
\be
\frak d_{0}[q] = \int^B_{-B}{\rm d}\mu \,q(\mu) \rho^{t\,0}_{1}(\mu)\,,\qquad\qquad  \frak j_0[q] = \int^B_{-B}{\rm d}\mu \,q(\mu) v^{0}_1(\mu) \rho^{t\,0}_{1}(\mu)\,.
\label{eq:gsvalues}
\ee
We also introduced the coefficients $\frak d_{\pm}[q,r]$, $\frak j_{\pm}[q,r]$, $ \frak d[q,r]$, and $\frak j[q,r]$, which read as    
\begin{align}
 \frak d[q,r]&=\frac{\pi}{12} \left[\frac{f_q'(B)-r^2 f'_q(-B) - \left( f_q(B)+r^2 f_q(-B)\right)\left(U(B)+\frac{v_1^{0\prime}(B)}{v_1^{0}(B)}\right)}{{ \varepsilon_1^{0\,\prime}\!(B)} v_1^0(B)} \right] \,,\label{eq:dcoeff}\\
 \frak j[q,r]&=\frac{\pi }{12 } \left[\frac{f_q'(B)+r^2 f_q'(-B)- \left(f_q(B)-r^2 f_q(-B)\right)U(B)}{{ \varepsilon_1^{0\,\prime}\!(B)}}\right]\,,\label{eq:jcoeff}\\
\frak j_{\pm}[q,r]&=\mp v^{0}_{1}(B)\frak d_{\pm}[q,r]=\frac{\pi \text{sgn}(v^{0\,\prime}_1(B))}{12}(1-r^2)f_q(\mp B)\,.\label{eq:dpmcoeff}
\end{align}
The function $f_q(B)$ appearing in these expressions is defined via the following integral equation  
\be
f_q(\lambda) = q(\lambda) - \int_{-B}^{B}\mathrm \!\!d\mu \,\,\, \ak_2(\lambda-\mu)f_q(\mu)\,. 
\label{eq:fqmain}
\ee
From this definition we immediately see that $f_e(\lambda)=\varepsilon_1^{0}(\lambda)$. In general, however, the function $f_q(\lambda)$ is different from the dressed charge $q^d(\lambda)$ (\emph{cf}. Eq.~\eqref{eq:dressing}). The latter, in the low temperature limit, is defined (up to a constant) by the following integral equation for its derivative 
\be
q^{d\,\prime}(\lambda) = q^{\prime}(\lambda) - \int_{-B}^{B}\mathrm \!\!d\mu \,\,\, \ak_2(\lambda-\mu)q^{\prime}(\mu)\,. 
\ee
Taking the derivative of \eqref{eq:fqmain} and integrating by parts we find 
\be
f_q'(\lambda)=q^{d\,\prime}(\lambda)+K_-(\lambda)f_q(B) -K_+(\lambda) f_q(B)\,,
\label{eq:derivativesfq}
\ee
where the functions $K_{\pm}(\lambda)$ are defined in \eqref{eq:defK}. 

\begin{figure}[h]
\includegraphics[width=0.9\textwidth]{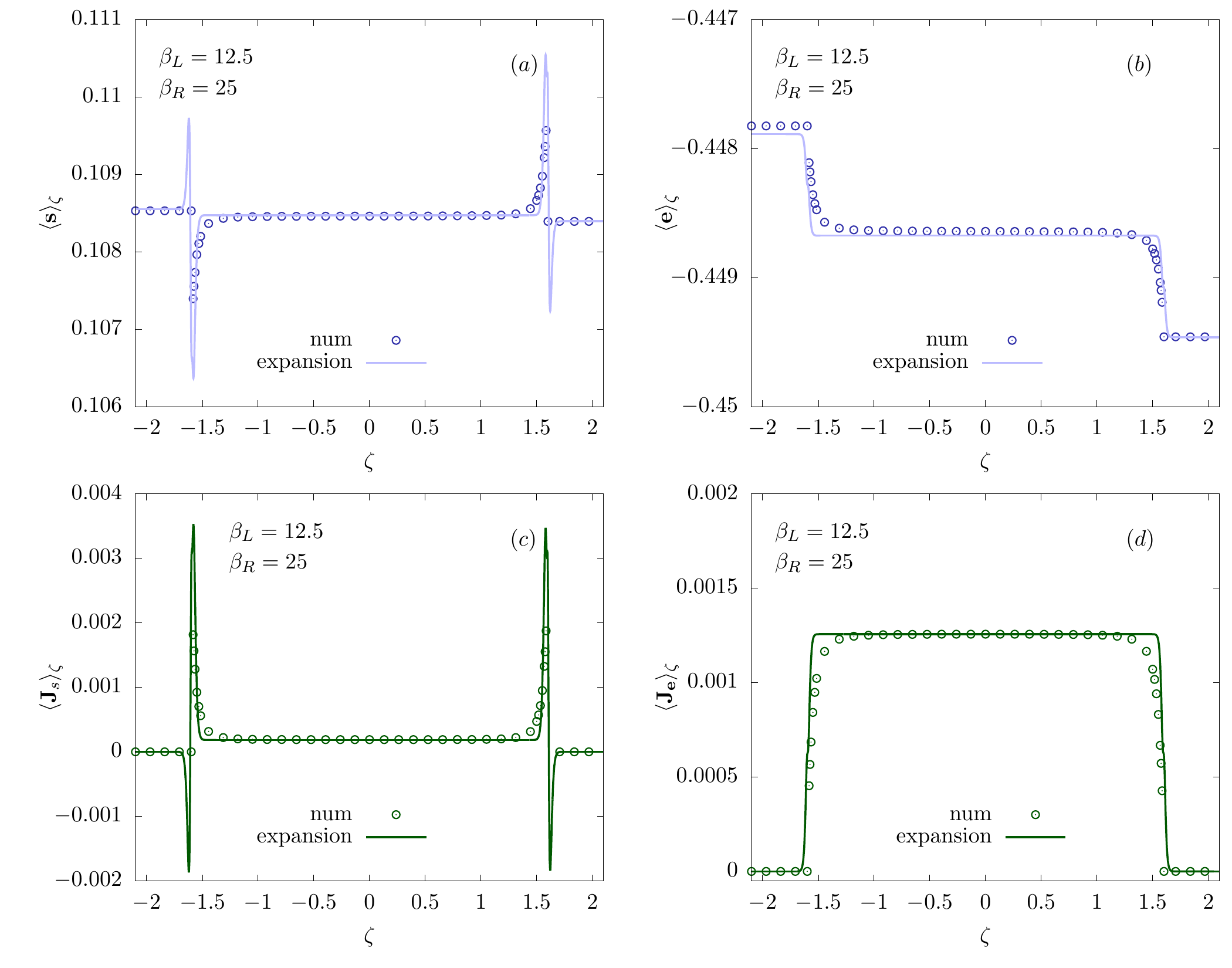}
\caption{Low-temperature profiles of spin and energy currents in the gapless phase. The parameters of the quench are chosen to be $\Delta=3$, $h=1.2$ while the two temperatures of the thermal states at the boundaries are $\beta_L=12.5$, $\beta_R=25$. The figure shows the comparison between the explicit numerical solution of the continuity equations~\eqref{eq:continuityrho} (circles) and the $O(T^2)$ analytic expansions \eqref{eq:fullexpansioncharge} and \eqref{eq:fullexpansioncurrent} (solid line).}
\label{fig:profiles_gapless12}
\end{figure}
\begin{figure}[h]
\includegraphics[width=0.9\textwidth]{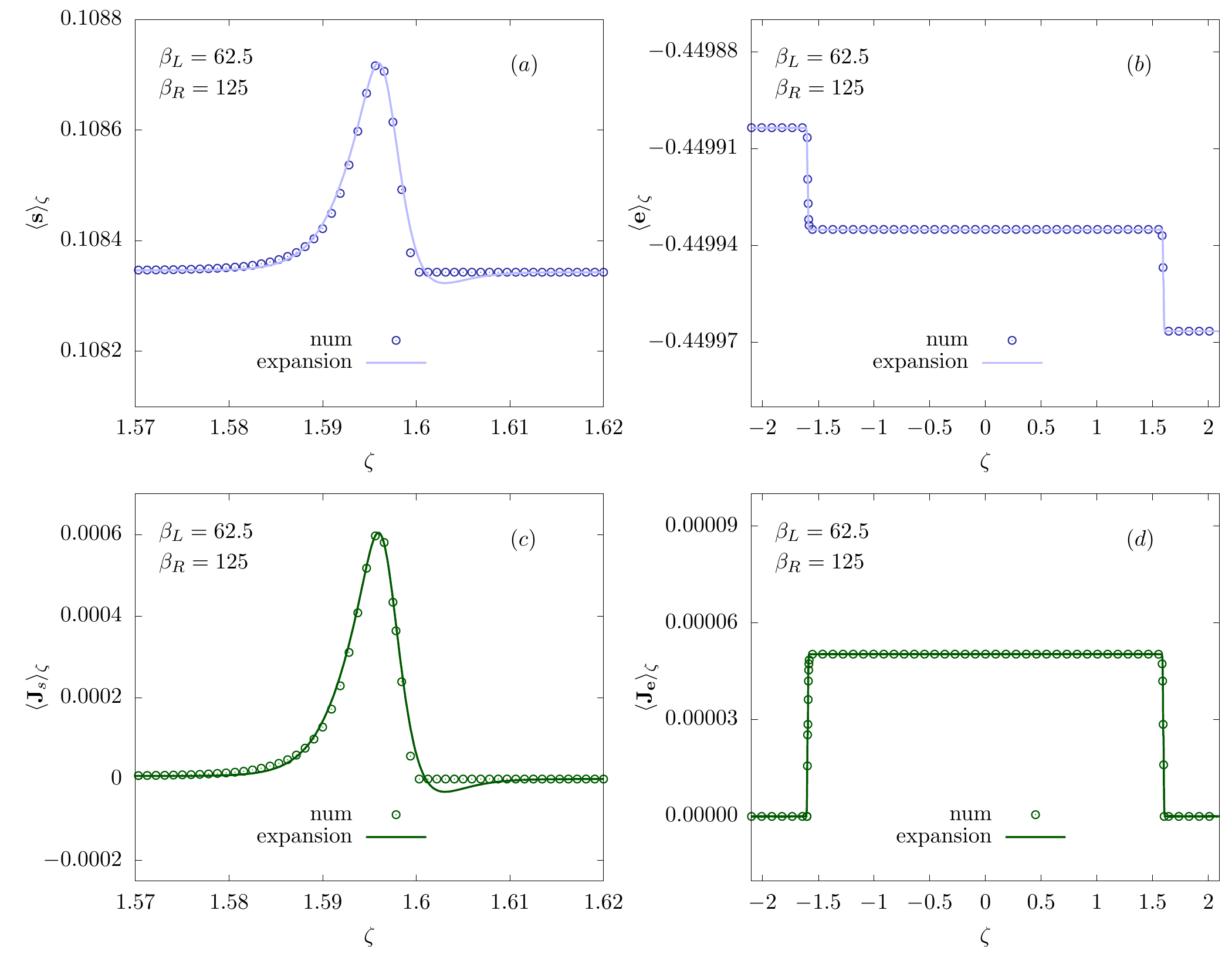}
\caption{Low-temperature profiles of spin and energy currents in the gapless phase. The parameters of the quench are chosen to be $\Delta=3$, $h=1.2$ while the two temperatures of the thermal states at the boundaries are $\beta_L=62.5$, $\beta_R=125$. The figure shows the comparison between the explicit numerical solution of the continuity equations~\eqref{eq:continuityrho}  (circles)  and the $O(T^2)$ analytic expansions \eqref{eq:fullexpansioncharge} and \eqref{eq:fullexpansioncurrent} (solid line). In the profiles of spin density and spin current the figure is zoomed around the ``non-CFT'' region (\emph{cf.} the main text).}
\label{fig:profiles_gapless62}
\end{figure}
\begin{figure}[h]
\includegraphics[width=0.9\textwidth]{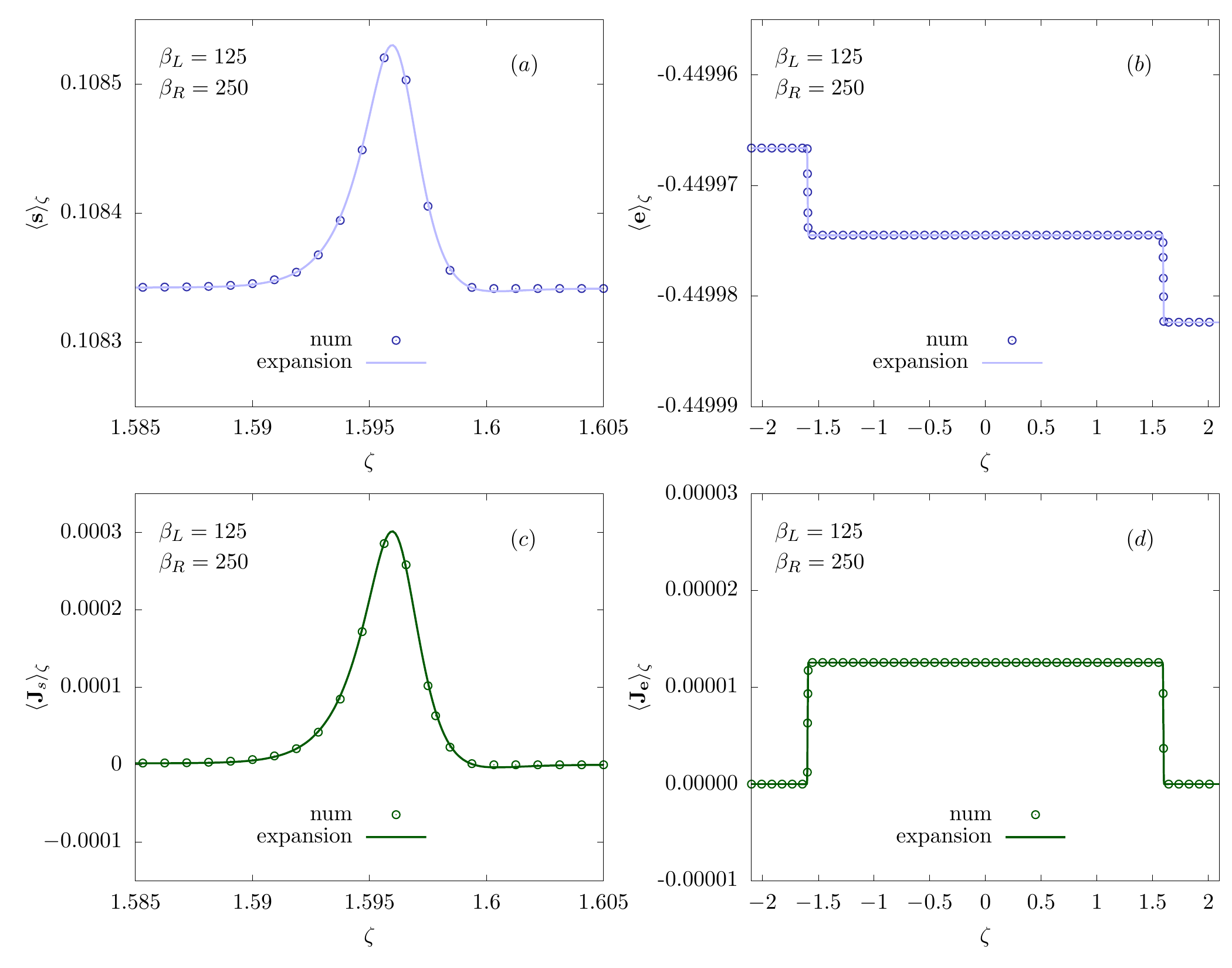}
\caption{Low-temperature profiles of spin and energy currents in the gapless phase. The parameters of the quench are chosen to be $\Delta=3$, $h=1.2$ while the two temperatures of the thermal states at the boundaries are $\beta_L=125$, $\beta_R=250$. The figure shows the comparison between the explicit numerical solution of the continuity equations~\eqref{eq:continuityrho}  (circles)  and the $O(T^2)$ analytic expansions \eqref{eq:fullexpansioncharge} and \eqref{eq:fullexpansioncurrent} (solid line). As in Fig.~\ref{fig:profiles_gapless62}, the profiles of spin density and spin current are zoomed around the ``non-CFT" region.}
\label{fig:profiles_gapless125}
\end{figure}
\begin{figure}[h]
\includegraphics[width=0.9\textwidth]{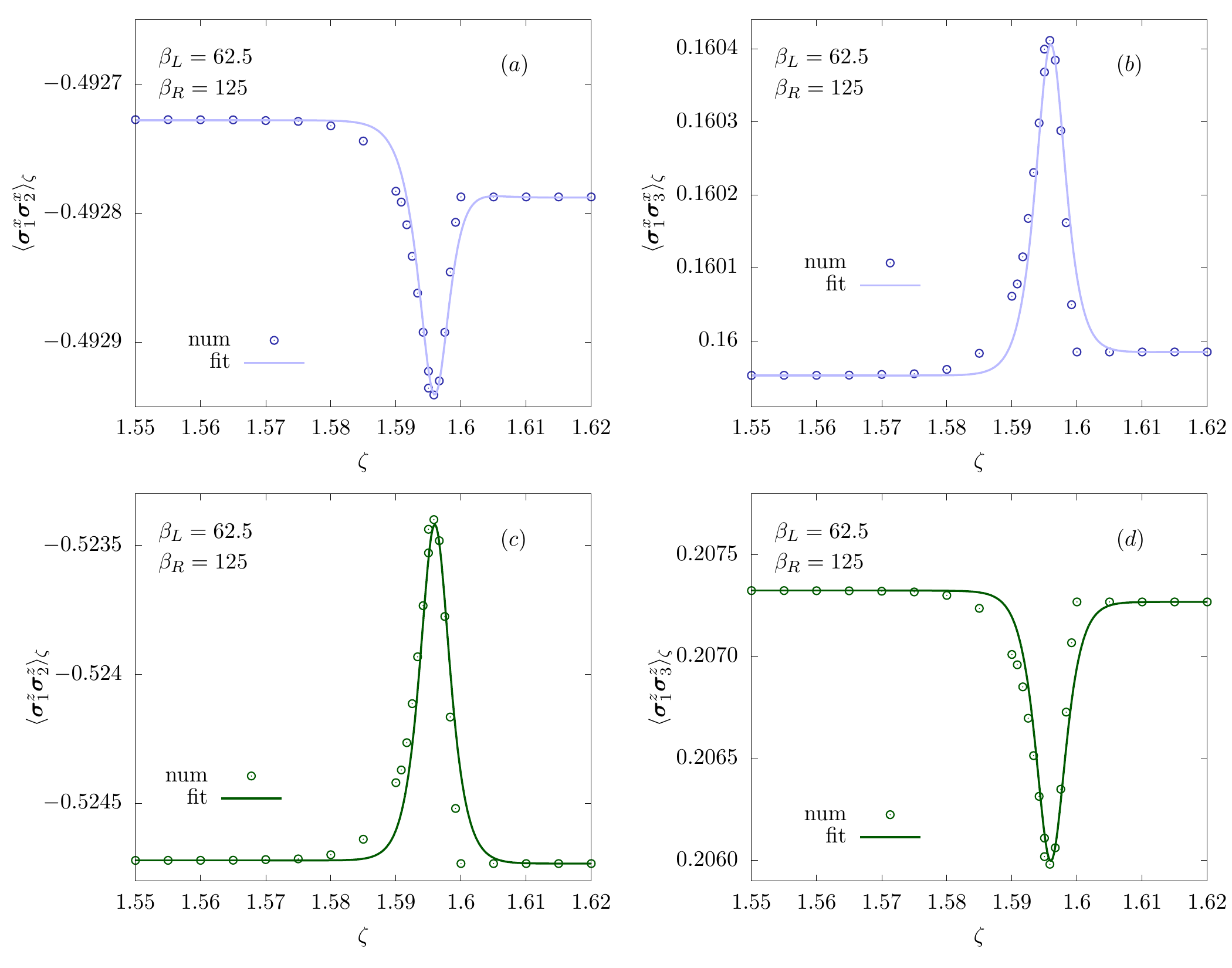}
\caption{Low-temperature profiles of local correlations. The parameters of the quench are chosen to be ${\Delta=3}$, ${h=1.2}$ while the two temperatures of the thermal states at the boundaries are ${\beta_L=62.5}$, ${\beta_R=125}$. Circles represent the exact result obtained using the explicit numerical solution of the continuity equations~\eqref{eq:continuityrho}  and the closed formulae of Ref.~\cite{PM:correlations}; the solid lines are instead the result of a single-parameter fit (see the main text).}
\label{fig:correlators}
\end{figure}
The expressions \eqref{eq:chargeslowtstep}--\eqref{eq:currentslowttransition} show that, at low temperatures, the light-cone structure emerging in the profiles of charge densities and currents is determined by the velocity $v^{0}_1(B)$ of gapless excitations. Away from the transition region, $\xt\pm v_1^0(B)\sim T$ the profiles have the structure observed in the framework of inhomogeneous conformal field theory~\cite{BeDo12,BeDo15,BeDo16,BeDo16Review}:  the ray dependence of the LQSS becomes trivial---the state is equal to the non-equilibrium steady state for $|\xt|<v^{0}_1(B)$ and respectively to the left and right thermal state for $\xt < -v^{0}_1(B)$ and $\xt > v^{0}_1(B)$. A non trivial ray dependence is recovered in the transition region $\xt\pm v_1^0(B)\sim T$, where the leading contribution in $T$ to the profiles of \emph{all} charges and currents are proportional to the function 
\be
{\cal D}_r \left(\frac{{\varepsilon^{0\,\prime}_1(B)}(\xt\pm v^0_1(B))}{T v_1^{0\,\prime}(B)}\right)\,.
\label{eq:functionD}
\ee
This result is exactly described by the non-linear Luttinger liquid prediction of Ref.~\cite{BPC:short}. Indeed, computing the effective mass of the dispersion $\varepsilon_1^{0}(\lambda)$ we have 
\be
(m^*)^{-1}= \frac{\partial^2 \varepsilon^0_1(\lambda)}{\partial p_1^{d\,0}(\lambda)^2}\bigg|_{p_1^{d\,0}(\lambda)=p_1^{d\,0}(B)} = \frac{v^{0\prime}_1(B)v^0_1(B)}{\varepsilon^{0\prime}_1(B)}\,.
\ee
Where $p_1^{d\,0}(\lambda)$ is the ground-state dressed momentum of the first string and we used $p_1^{d\,\prime}(\lambda)=2\pi \rho^{t}_1(\lambda)$ (\emph{cf}. Eq.~\eqref{eq:dressedmomentumderivative}). Plugging this into \eqref{eq:chargeslowttransition} and \eqref{eq:currentslowttransition} we see that they agree with the non-linear Luttinger liquid result by choosing a sound velocity equal to $v^{0}_1(B)$. 

The low-temperature profiles of some relevant observables are presented in Figures~\ref{fig:profiles_gapless12}, \ref{fig:profiles_gapless62}, and \ref{fig:profiles_gapless125}, where we compare the results of the low-temperature expansion with the exact ones, found by numerical iterations of \eqref{eq:transportsolution}. We see that for small enough temperatures the agreement becomes quantitatively excellent, see Fig.~ \ref{fig:profiles_gapless125}, while when increasing the temperature some spurious contributions of higher order in $T$ start to arise at the border of the transition region. See, e.g., the ``lower peaks" close to $\xt=v^{0}_1(B)$ in Figures~\ref{fig:profiles_gapless12} (a), \ref{fig:profiles_gapless12}(c), \ref{fig:profiles_gapless62}(a), and \ref{fig:profiles_gapless62}(c).

Note that in the profiles of the energy density and energy current the transition region behaviour is not observed: this is because the coefficients $\frak j_{\pm}[e_1,r]$ and $\frak d_{\pm}[e_1,r]$ vanish, as $f_e(B)=\varepsilon_1^{0}(B)=0$. Moreover, using \eqref{eq:dcoeff} and \eqref{eq:jcoeff} we find 
\be
\frak d[e_1,r]=\frac{\pi}{12v^0_1(B)}(1+r^2)\,,\qquad\qquad \frak j[e_1,r]=\frac{\pi }{12}(1-r^2) \,.\label{eq:jencoeff}
\ee
Plugging these expressions in the profiles \eqref{eq:chargeslowtstep} and \eqref{eq:currentslowtstep} we recover the inhomogeneous conformal field theory predictions~\cite{BeDo12,BeDo15,BeDo16,BeDo16Review} for a theory of central charge equal to one. At first this result might appear surprising: in the limit that we are considering, $x,t\rightarrow\infty$ with fixed $\zeta=x/t$, the time is much larger than the inverse curvature of the dispersion and it is natural to expect non-conformal effects. The conundrum is solved 
by taking into account the particular structure of the observables under exam
. Indeed by construction energy density and current are sensitive only to ``linear" modes at low temperatures~\cite{BeDo16Review}, as opposed to generic observables. In other words at low energies one has a mapping of the form
\be
\boldsymbol{e}_j\mapsto{\boldsymbol{e}_{\rm CFT}}_j\,,\qquad\qquad
\boldsymbol{J_{\boldsymbol e}}_j\mapsto{\boldsymbol{J_{\boldsymbol e}}_{\rm CFT}}_j\,,
\ee
this can, \emph{e.g.}, be thought as the result of a bosonisation procedure. Generic charges of the XXZ model, however, do not directly correspond to conserved charges of the underlying conformal field theory: this explains the non-conformal behaviour \eqref{eq:chargeslowttransition} and \eqref{eq:currentslowttransition} in the transition region.
Note that the profiles of charge densities and currents are related by the following continuity equation
\be
\partial_\xt \braket{\boldsymbol J_{\boldsymbol q}}_\xt -\xt \partial_\xt \braket{{\boldsymbol q}}_\xt  =0\,.
\ee
Such equation is obtained considering the expectation value of the continuity equation \eqref{eq:discretecontinuityeq} and taking the limit of infinite time $t$ and position $\ell$ with fixed ray $\xt=\ell/t$. Integrating we have 
\be
\braket{\boldsymbol J_{\boldsymbol q}}_\xt=\braket{\boldsymbol J_{\boldsymbol q}}_{\xt_0}+ \xt \braket{{\boldsymbol q}}_{\xt} -\xt_0 \braket{{\boldsymbol q}}_{\xt_0}  -\int_{\xt_0}^\xt \!\!{\rm d}s\,  \braket{{\boldsymbol q}}_s\,.
\ee
Using this equation we can obtain $\braket{\boldsymbol J_{\boldsymbol q}}_\xt$ from $\braket{{\boldsymbol q}}_\xt$. In particular, the integral of $\braket{{\boldsymbol q}}_\xt$ over the interior the transition region gives a contribution $\sim T^2$ as that over the exterior: this is a consequence of Eq.~\eqref{eq:Dlimit}. The ``accumulation" of charge in the transition region explains the asymmetry of the expressions \eqref{eq:dcoeff} and \eqref{eq:jcoeff}, namely the presence of the term involving $v_1^{0\,\prime}(B)$ in \eqref{eq:dcoeff}.

In the two regimes \eqref{eq:CFTregime} and \eqref{eq:transition}, the first non-trivial contributions in $T$ to the expectation values of all charge densities are proportional to the same functions. We can then use the reasoning of Ref.~\cite{BPC:short} to argue that the first non-trivial contribution in $T$ to all local observables in the two regimes \eqref{eq:CFTregime} and \eqref{eq:transition} must show the same $\xt$ dependence. This is demonstrated in Figure~\ref{fig:correlators} which reports the profile of some non-trivial local correlations compared with the following fitting function 
\be
{f_{\mathcal O}(\xt)=\braket{\mathcal O}_0+\braket{\mathcal O}_1 T^2+\braket{\mathcal O}_1T^2 \left ( r^2- 1\right )\mathcal G \left( \frac{\varepsilon^{0\,\prime}_1(B)(v^{0}_1(B)-\xt)}{T |v^{0\,\prime}_1(B)|}\right)+ a T {\cal D}_{r} \left(\frac{{\varepsilon^{0\,\prime}_1(B)}(\xt- v^0_1(B))}{T v_1^{0\,\prime}(B)}\right)}\,.
\ee 
Here $\mathcal G(-x/T)$ gives a smooth approximation of the step function for small $T$ (its precise definition is reported in Eq.~\eqref{eq:defGfun} of Appendix~\ref{app:detailslowtemperature}), the coefficients $\braket{\mathcal O}_0$ and $\braket{\mathcal O}_1$ are those of the first two orders in the low-$T$ expansion of the thermal expectation value of $\mathcal O$, and $a$ is the fitting parameter. The fit describes well the centre of the transition region and the conformal regions, while in the tails of the transition region there are discrepancies. This is expected since $f_{\mathcal O}(\xt)$ correctly reproduces only the $O(T)$ contribution in the transition region.

\section{Transport in the Gapped Phase}
\label{sec:lowtgapped}

As mentioned in the introduction, qualitative differences arise in the transport properties of the gapped phase of the model at all temperatures. These differences are even more marked at small temperatures. In this case, corrections to the zero-temperature limit are exponentially small in $T$, and profile functions of local observables are non-trivial, as we illustrate in detail in this section. In particular, we derive analytic formulae for the profile of local conserved charges and currents at the leading order. Comparison with numerical solution to the hydrodynamic equations reveals that the latter provide good qualitative predictions also at small but finite (\emph{i.e.} not infinitesimal) temperatures. In the gapped phase, the relevant small temperature limit is defined as follows. We set
\bea
\beta_L&=&\beta\,,\\
\beta_R&=&\beta+\Delta\beta\,,
\eea
and consider the limit $\beta\to\infty$, while we keep $\Delta\beta$ constant. Note that this is different from the procedure carried out in the critical phase, where the ratio of the temperatures was kept fixed. Exactly at $\beta=\infty$, the whole chain is in its ground state and one has clearly no transport dynamics, the profile of a generic observable $\boldsymbol{\mathcal{O}}$ being a constant function. We start by considering the ferromagnetic regime, while the antiferromagnetic phase will be analysed in the next subsection. 

\subsection{Ferromagnetic order}
\label{sec:lowtferro}
In the ferromagnetic regime the expansion is particularly simple. To construct it, it is convenient to consider the functions $\eta_{n,\zeta}(\lambda)$, defined in Eq.~\eqref{eq:etan}. It follows from \eqref{eq:transportsolution} that these functions display a simple dependence on the ray $\zeta$, namely they are Heaviside step functions of $\zeta$ (keeping $\lambda$ fixed). Moreover, using the properties of ground-state dressed energies, determined in Sec.~\ref{sec:ground}, one can immediately write down the leading behaviour of the functions $\eta_n(\lambda)$ outside of the light cone
\bea
\eta^{L/R}_n(\lambda)&\simeq & e^{+2nh\beta_{L/R} -a_n(\lambda) \beta J \pi \sinh\eta} \,,\qquad\qquad n\geq 1\,.
\eea
We have 
\bea
\eta_{n,\zeta}(\lambda)&\simeq &e^{\beta_R\left[2h n-J\pi a_n(\lambda)\sinh\eta\right]}H\left(\zeta-v_{n,\zeta}(\lambda)\right)+e^{\beta_L\left[2hn-J\pi a_n(\lambda)\sinh\eta\right]}H\left(v_{n,\zeta}(\lambda)-\zeta\right)\,,
\eea
where $H(x)$ is the Heaviside step function. Note that this expression still depends on the exact velocities $v_{n,\zeta}(\lambda)$. In order to evaluate the profiles, we need to compute their leading behaviour, together with the one of the rapidity distribution functions $\rho_{n,\zeta}(\lambda)$. Since $\eta_{n,\zeta}(\lambda)$ diverges exponentially as $T\to0$ for $n\geq 1$, from \eqref{eq:tba} we have
\bea
\rho^{t\,0}_{n,\zeta}(\lambda)\equiv \lim_{T\to 0}\rho^t_{n,\zeta}(\lambda)&=&a_n(\lambda)\,,
\label{eq:rhotot_ferro}
\eea
and in the same way
\bea
v^0_{n,\zeta}(\lambda) \equiv \lim_{T\to 0}v_{n,\zeta}(\lambda)&=&-\frac{\sinh(\eta)}{2} \frac{a^{\prime}_n(\lambda)}{a_n(\lambda)}\,.
\label{eq:final_vn_ferro}
\eea
Finally, using these equations we arrive at the following leading behaviour for the rapidity distribution functions
\bea
\rho_{n,\zeta}(\lambda) \simeq \frac{\rho^{t\,0}_{n,\zeta}(\lambda)}{\eta_{n,\zeta}(\lambda)}&=& a_n(\lambda)\left(e^{-\beta_R\left[2h n-J\pi a_n(\lambda)\sinh\eta\right]}H\left(\zeta+\frac{\sinh(\eta)}{2} \frac{a^{\prime}_n(\lambda)}{a_n(\lambda)}\right)\right.\nonumber\\
&&+\left.e^{-\beta_L\left[2hn-J\pi a_n(\lambda)\sinh\eta\right]}H\left(-\frac{\sinh(\eta)}{2} \frac{a^{\prime}_n(\lambda)}{a_n(\lambda)}-\zeta\right)\right)\,.
\label{eq:final_rhon_ferro}
\eea

\begin{figure}
	\includegraphics[width=0.85\textwidth]{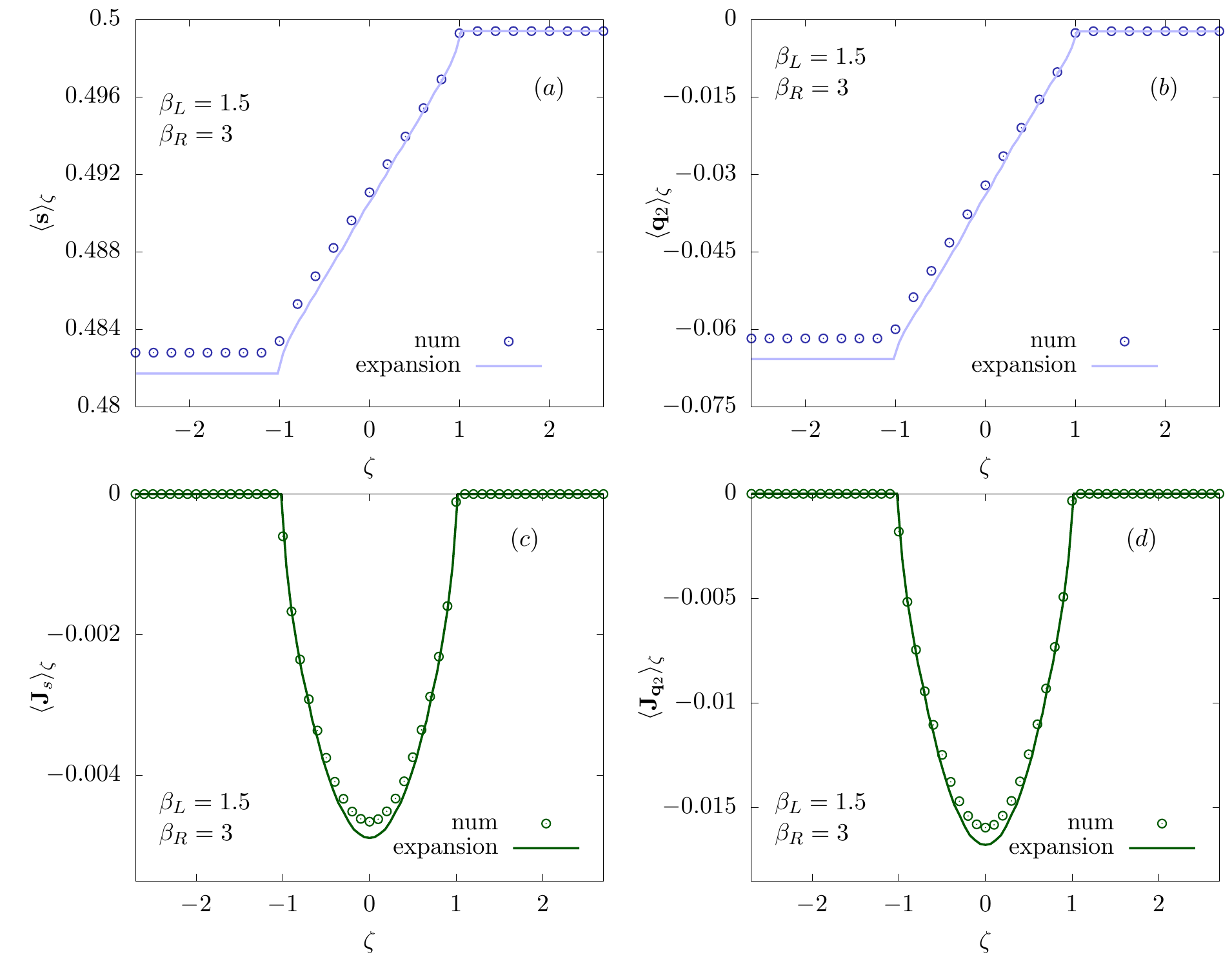}
	\caption{Low-temperature profiles for local densities and currents in the ferromagnetic phase. Plots $(a)$ and $(c)$ correspond to the spin, while plots $(b)$ and $(d)$ to the second local conserved charge ${\boldsymbol{q}_2}$. The parameters of the quench are chosen to be $\Delta=3$, $h=3$, while $\beta_L=1.5$, $\beta_R=3$. The figure shows the comparison between the explicit numerical solution of the continuity equations and the analytic results \eqref{eq:final_charge_den_ferro} and \eqref{eq:final_charge_curr_ferro}.}
	\label{fig:profiles_ferro}
\end{figure}

We have now all the ingredients to compute the leading behaviour of the profiles in the ferromagnetic regime. We start by considering Eqs.~\eqref{eq:chargeprofile} and \eqref{eq:currentprofile} and noting that, at the leading order, we can keep only the first term in each infinite sum. This is because $\rho_{n,\zeta}(\lambda)$ for $n\geq 2$ is exponentially smaller than $\rho_{1,\zeta}(\lambda)$. After some simple algebra, one obtains
\bea
\langle \boldsymbol{q}\rangle_\zeta&=&\int_{-\pi/2}^{\pi/2}d\lambda\ q_1(\lambda)a_1(\lambda)\Psi(\zeta,\lambda)\,,\label{eq:final_charge_den_ferro}\\
\langle\boldsymbol{J}_{\boldsymbol{q}}\rangle_\zeta&=&-\frac{\sinh(\eta)}{2}\int_{-\pi/2}^{\pi/2}d\lambda\ q_1(\lambda)a^{\prime}_1(\lambda)\Psi(\zeta,\lambda)\,,\label{eq:final_charge_curr_ferro}
\eea
where we defined
\bea
\Psi(\zeta,\lambda)&=&e^{-\beta_R\left[2h-J\pi a_1(\lambda)\sinh\eta\right]}H\left(\zeta+\frac{\sinh(\eta) a_1^{\prime}(\lambda)}{2a_1(\lambda)}\right)+e^{-\beta_L\left[2h-J\pi a_1(\lambda)\sinh\eta\right]}H\left(-\frac{\sinh(\eta) a_1^{\prime}(\lambda)}{2a_1(\lambda)}-\zeta\right)\,.
\eea
The final result is expressed in terms of simple integrals which are readily evaluated numerically. 
In Fig.~\ref{fig:profiles_ferro} we compare the analytic prediction for the low-temperature limit with the exact numerical evaluation. We see that the profiles are non-constant functions of $\zeta$ inside the light cone, in contrast with the low-temperature limit of the critical phase. The densities of spin and local charges display smooth monotonic profiles, which are nicely captured by our analytic formulae. The latter do not display non-analytic points inside the light cone: this is because bound states of excitations are not produced after the quench for small enough temperatures. 
It is interesting to note that Eqs.~\eqref{eq:final_charge_den_ferro}--\eqref{eq:final_charge_curr_ferro} have a formal similarity with analytic formulae derived in the corresponding quench setting in the free Ising model \cite{Korm17,PeGa17}. One can interpret this result as follows. For small enough temperatures, the density of excitations is small and their behaviour is not influenced qualitatively by the presence of interaction. The latter is mainly responsible for a dressing of the bare quantities carried by the quasi-particles.
The asymptotic small-temperature regime is reached for fairly large values of $T$. Indeed, using the quench parameters corresponding to Fig.~\ref{fig:profiles_ferro} the analytic formulae above provide excellent approximations already for $T\sim 1$.

\subsection{Antiferromagnetic order}
\label{sec:lowtantiferro}
Let us now consider the antiferromagnetic regime. Once again, using the properties of the ground-state dressed energies, which are summarised in Sec.~\ref{sec:ground}, one can immediately write down the leading behaviour of the functions $\eta_{n}(\lambda)$ outside the light cone. It follows from \eqref{eq:leading_eps1_anti} that 
\bea
\eta^{L/R}_1(\lambda)&\simeq &e^{-\beta_{L/R} J\pi\sinh(\eta)s(\lambda)+\beta_{L/R} h}\,,\label{eq:leading_1}
\eea
while using \eqref{eq:limitepsilonn} and the identity
\be
\left[\left(a_{n+1}+a_{n-1}\right)\ast s\right](\lambda)=a_n(\lambda)\,,
\ee
one can compute
\bea
\eta^{L/R}_n(\lambda)&\simeq&e^{2\beta_{L/R}(n-1)h}\,.\label{eq:leading_2}
\eea
In the following we will only be interested in the leading contributions to the profiles of local observables, so that \eqref{eq:leading_1} and \eqref{eq:leading_2} are sufficient for our purposes. However, one could in principle wish to compute higher terms in the expansions. In order to facilitate this problem, it is convenient to replace the expressions \eqref{eq:leading_1} and \eqref{eq:leading_2} with different ones, which share the same leading behaviour, namely
\bea
\eta_1^{L/R}(\lambda)&\simeq &\bar{\eta}_1^{L/R}(\lambda)\equiv \frac{\sinh{(2\beta_{L/R} h)}}{\sinh(\beta_{L/R} h)}e^{-J\pi\beta_{L/R} \sinh(\eta)s(\lambda)}\,,\label{eq:new_leading_1}\\
\eta_n^{L/R}(\lambda)&\simeq &\bar{\eta}_n^{L/R}(\lambda)\equiv \frac{\sinh^2(\beta_{L/R} hn)}{\sinh^2(\beta_{L/R} h)}-1\,.\label{eq:new_leading_2}
\eea
One could now proceed to derive higher order corrections setting 
\be
\eta^{L/R}_n(\lambda)=\bar{\eta}^{L/R}_n(\lambda)\left[1+\delta^{L/R}_n(\lambda)\right]\,,
\label{eq:perturbative_expression}
\ee
and computing $\delta^{L/R}_n(\lambda)$ perturbatively (it is clear that $\delta^{L/R}_n(\lambda)\to 0$ as $T\to 0$). This is explained in more detail in Appendix~\ref{sec:appendix_expansions}, where the leading term of $\delta^{L/R}_n(\lambda)$ is derived. This also provides an estimation of the error introduced by approximating $\eta^{L/R}_n(\lambda)$ with $\bar{\eta}^{L/R}_n(\lambda)$. For clarity of presentation, however, in the following calculations we neglect contributions coming from $\delta^{L/R}_n(\lambda)$. Using the expressions \eqref{eq:new_leading_1} and \eqref{eq:new_leading_2} we arrive at 
\begin{align}
\eta_{1,\zeta}(\lambda)&\simeq \frac{\sinh{(2\beta_R h)}}{\sinh(\beta_R h)}e^{-J\pi \beta_R\sinh\eta s(\lambda)}H(\zeta-v_{1,\zeta}(\lambda))+\frac{\sinh{(2\beta_L h)}}{\sinh(\beta_L h)}e^{-J \pi \beta_L\sinh\eta s(\lambda)}H(v_{1,\zeta}(\lambda)-\zeta)\,,\label{eq:final_leading_eta1}\\
\eta_{n,\zeta}(\lambda)&\simeq \left[\frac{\sinh^2(\beta_R h n)}{\sinh^2(\beta_R h)}-1\right]H(\zeta-v_{n,\zeta}(\lambda))+\left[\frac{\sinh^2(\beta_L h n)}{\sinh^2(\beta_L h)}-1\right]H(v_{n,\zeta}(\lambda)-\zeta)\,.\label{eq:final_leading_eta2}
\end{align}
As in the previous section, this expression needs to be simplified as it displays a dependence on the exact velocities $v_{n,\zeta}(\lambda)$, which are unknown. Moreover, to compute the profiles we also have to determine the leading behaviour of the rapidity distribution functions $\rho_{n,\zeta}(\lambda)$. For these quantities, the dependence on the ray $\zeta$ is highly non-trivial, as it enters in the Bethe equations through integrals involving the functions $\eta_{n,\zeta}(\lambda)$. While this makes it difficult to set up an iterative scheme to systematically compute higher order corrections, it is still possible to obtain the leading term fairly easily.

First note that $\eta_{1,\zeta}(\lambda)\to 0$, while  $\eta_{n,\zeta}(\lambda)\to \infty$ exponentially for $T\to 0$. From \eqref{eq:tba}, this implies
\bea
\rho^{0}_{1,\zeta}(\lambda) \equiv \lim_{T\to 0}\rho_{1,\zeta}(\lambda)&=&\rho^{t\,0}_{1,\zeta}(\lambda)=s(\lambda)\,\label{eq:gs_rho_0},\\
\rho^{0}_{n,\zeta}(\lambda) \equiv \lim_{T\to 0}\rho_{n,\zeta}(\lambda)&=&0\,.
\eea
The next step is to rewrite the TBA equations~\eqref{eq:tba} in the following ``decoupled'' form~\cite{takahashi}
\bea 
\rho^t_{n,\zeta}(\lambda) &=& \delta_{n,1} s(\lambda) +  s \ast  \left( \rho^t_{n-1,\zeta} \frac{\eta_{n-1,\zeta}}{1 + \eta^t_{n-1,\zeta}} + \rho^t_{n+1,\zeta}  \frac{\eta_{n+1,\zeta}}{1 + \eta_{n+1,\zeta}} \right)(\lambda)\,,\label{eq:BT_decoupled}
\eea
with the convention $\rho^t_{0,\zeta}(\lambda)\equiv 0$. For $n\geq 2$, we have that $\eta_{n,\zeta}(\lambda)$ always diverges exponentially. Accordingly, at the leading order we can write 
\bea 
\label{eq:gappedrhot}
\rho^t_{n,\zeta}(\lambda) &=&  s \ast  \left( \rho^t_{n-1,\zeta} + \rho^t_{n+1,\zeta} \right)(\lambda)\,,\qquad n\geq 3\,.
\eea
These equations are conveniently analysed by taking the Fourier transform, which we define as
\bea
\hat{f}(k) &=& \frak{F}[f](k)=\int_{-\pi/2}^{\pi/2}d\lambda e^{2ik\lambda}f(\lambda) \,, \label{eq:convention_FT}\\
f(\lambda) &=&\frac{1}{\pi}\sum_{k\in \mathbb{Z}}e^{-2ik\lambda}\hat{f}(k)\,.\label{eq:convention_IFT}
\eea
The Fourier transform of Eq.~\eqref{eq:gappedrhot} reads as
\be
\left(e^{-|k|\eta}+e^{|k|\eta}\right)\hat{\rho}^t_{n,\zeta}(k) =  \left( \hat{\rho}^t_{n+1,\zeta}(k) +  \hat{\rho}^t_{n-1,\zeta}(k)\right)\,,\qquad n\geq 3\,.
\label{eq:to_solve}
\ee
This is a second order difference equation, whose general solution depends on two parameters. In particular, it is straightforward to verify that the two-parameter sequence
\be
\hat{\rho}^t_{n,\zeta}(k)=A(k)e^{n|k|\eta}+B(k)e^{-n|k|\eta}\,,\qquad n\geq 3\,,
\ee
gives us the general solution to \eqref{eq:to_solve}. Assuming that $\rho_n^t(\lambda)$ does not grow to infinity we can immediately set $A(k)=0$. In order to fix the parameter $B(k)$ we consider Eq.~\eqref{eq:BT_decoupled} for $n=2$, which at the leading order in $\beta$ is rewritten as
\bea 
\rho^t_{2,\zeta}(\lambda)  &=&  s \ast  \left( \rho^t_{1,\zeta} \eta_{1,\zeta} + \rho^t_{3,\zeta} \right)(\lambda)\,.
\eea
Since $\eta_{1,\zeta}$ is already exponentially vanishing, we can substitute $\rho^t_{1,\zeta}(\lambda)$ with its zero-temperature limit in \eqref{eq:gs_rho_0}. Hence, we rewrite
\bea
\rho^t_{2,\zeta}(\lambda)  &=&  s \ast  \left( s\eta_{1,\zeta} \right)(\lambda)+ s \ast\rho^t_{3,\zeta}(\lambda)\,.
\eea
This equation can now be easily solved in Fourier space to determine the parameter $B(k)$. Putting everything together, and transforming back to real space, one obtains
\bea
\rho^t_{n,\zeta}(\lambda)&=&a_{n+1}\ast \left[s \eta_{1,\zeta}\right](\lambda)\,,\qquad n\geq 2\,,\label{eq:leading_rho_n}
\eea
where $a_n(\lambda)$ is defined in \eqref{eq:afunction}.

Equations~\eqref{eq:dressedenderivative} can also be brought to a partially decoupled form as follows 
\begin{equation}
[\rho_j(\lambda)+ \rho^h_j(\lambda)]v_j(\lambda) = -\frac{\sinh\eta}{2} s^{\prime}(\lambda)\delta_{j,1} +s \ast  ( \rho^h_{j-1}v_{j-1} + \rho^h_{j+1}v_{j+1})(\lambda)\,.
\label{eq:gapped_vel_decoupled}
\end{equation}
Using these equations we can proceed in a completely analogous way to compute the leading behaviour of the function $\rho^t_{n,\zeta}(\lambda) v_{n,\zeta}(\lambda)$. Repeating the steps above one finds
\bea
v_{1,\zeta}(\lambda)&=&-\frac{\sinh\eta s^{\prime}(\lambda)}{2s(\lambda)}\,,\label{eq:leading_v_1}\\
v_{n,\zeta}(\lambda)&=&-\frac{\sinh\eta}{2} \frac{a_{n+1}\ast  \left[ s^{\prime}\eta_{1,\zeta}\right](\lambda)}{a_{n+1}\ast [s\eta_{1,\zeta}](\lambda)}\,,\qquad n\geq 2\,.
\label{eq:leading_v_n}
\eea
\begin{figure}
	\includegraphics[width=0.85\textwidth]{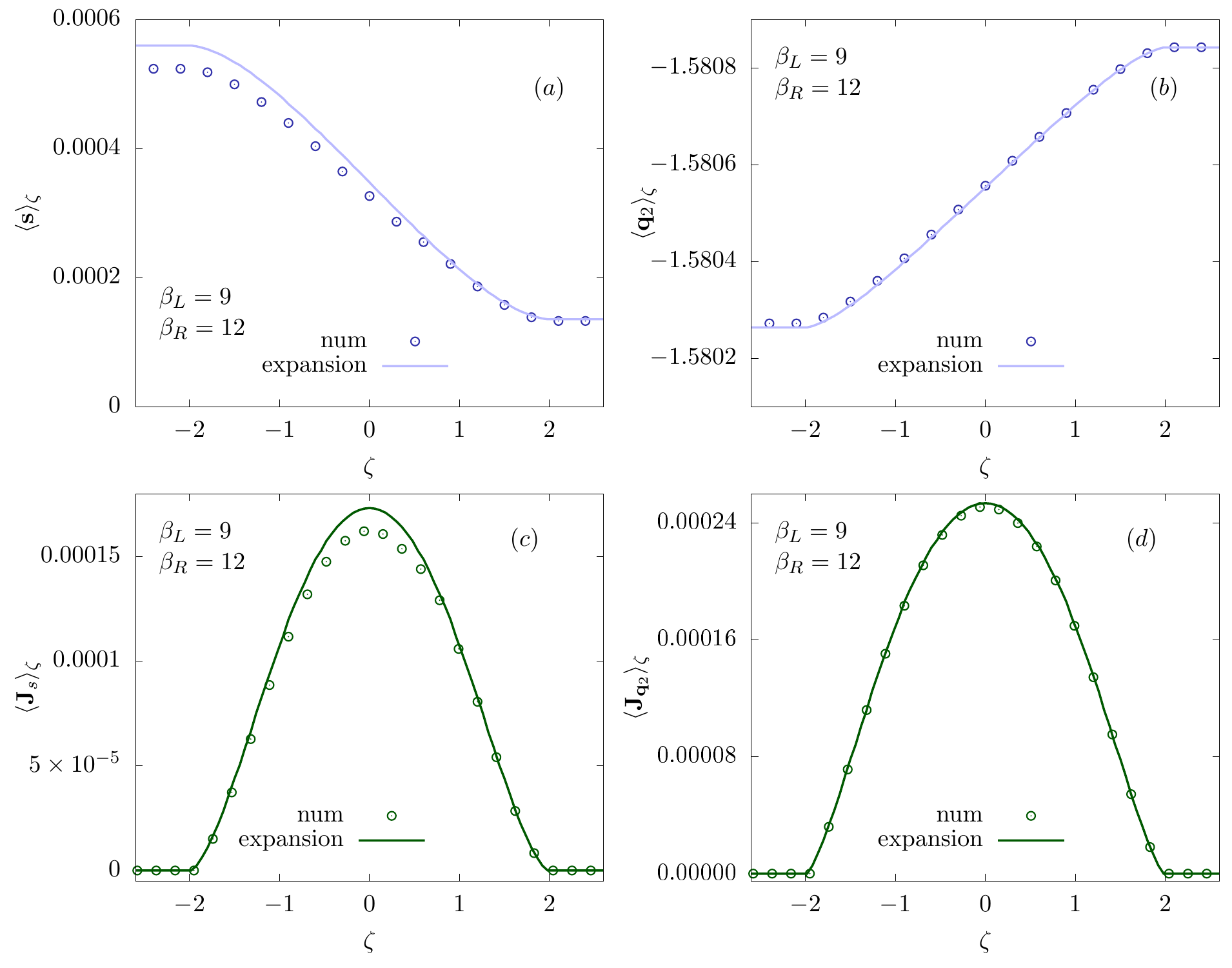}
	\caption{Low-temperature profiles for local densities and currents in the antiferromagnetic phase. Plots $(a)$ and $(c)$ correspond to the spin, while plots $(b)$ and $(d)$ to the second local conserved charge ${\boldsymbol{q}_2}$. The parameters of the quench are chosen to be $\Delta=3$, $h=0.2$. The figure shows the comparison between the explicit numerical solution of the continuity equations and the analytic results \eqref{eq:final_charge_den}--\eqref{eq:final_spin_curr}.}
	\label{fig:densities_antiferro}
\end{figure} 
We have now all the elements to explicitly write down a formula for the profiles of local observables. Indeed, at the leading order we can substitute in Eq.~\eqref{eq:final_leading_eta1} and \eqref{eq:final_leading_eta2} the leading behaviour of the velocities found above and use Eqs.~\eqref{eq:chargeprofile} and \eqref{eq:currentprofile} to find the profiles of charge densities and currents. However, since we are interested in expectation values of the density and current of the local charges $\{\boldsymbol{Q}_m\}_{m=1}^\infty$ (\emph{cf}. Sec.~\ref{sec:charges}) and of the magnetization $\boldsymbol{S}^z$, instead of Eqs.~\eqref{eq:chargeprofile} and \eqref{eq:currentprofile} it is convenient to use the following simplified formulae derived in Appendix~\ref{app:simpformulae}
\bea
\braket{\boldsymbol{q}_{m+1}}_\zeta&=&\frac{\pi {\sinh^m\left(\eta\right)}}{(-2)^{m-1}}\int_{-\pi/2}^{\pi/2}d\lambda s^{(m-1)}(\lambda)\left[\rho^{h}_{1,\zeta}(\lambda)-a_1(\lambda)\right]\,,\label{eq:partial_1}\\
\braket{\boldsymbol{J}_{m+1}}_\zeta&=&\frac{\pi {\sinh^m\left(\eta\right)}}{(-2)^{m-1}}\int_{-\pi/2}^{\pi/2}d\lambda s^{(m-1)}(\lambda)\left[\rho^{h}_{1,\zeta}(\lambda)v_1(\lambda)+\frac{\sinh\eta}{2}a^\prime_1(\lambda)\right]\,,\label{eq:partial_2}\\
\braket{\boldsymbol{s}}_\zeta&=&\frac{1}{2}\lim_{n\to\infty}\int_{-\pi/2}^{\pi/2}\rho^{t}_{n,\zeta}(\lambda)\,,\\
\braket{\boldsymbol{J}_s}_\zeta&=&\frac{1}{2}\lim_{n\to\infty}\int_{-\pi/2}^{\pi/2}\rho^{t}_{n,\zeta}(\lambda)v_{n,\zeta}(\lambda)\,,\label{eq:partial_4}
\eea
where we introduced
\bea
s^{(n)}(\lambda)&=&\frac{d^n}{d\lambda^n}s(\lambda)\,,\\
s^{(0)}(\lambda) &=& s(\lambda)\,.
\eea
Using 
\be
\rho^h_{1,\zeta}(\lambda)\simeq \rho^{t\,0}_{1,\zeta}(\lambda)\eta_{1,\zeta}(\lambda)\,,
\ee
and substituting the results \eqref{eq:gs_rho_0}, \eqref{eq:leading_rho_n}, \eqref{eq:leading_v_1}, and \eqref{eq:leading_v_n} in \eqref{eq:partial_1}--\eqref{eq:partial_4} we find  
\bea
\langle \boldsymbol{q}_{m}\rangle_{\zeta}&=& \frak d_{0}[q^{(m)}_1]+ \frac{\sinh^{m-1}\left(\eta\right)\pi}{(-2)^{m-2}}\int_{-\pi/2}^{\pi/2}d\lambda s^{(m-2)}(\lambda)s(\lambda)\Phi(\zeta,\lambda)\,,\label{eq:final_charge_den}\\
\langle \boldsymbol{J}_{\boldsymbol{q}_{m}}\rangle_\zeta&=&\frak
j_{0}[q^{(m)}_1]-\frac{\sinh^{m}(\eta)\pi}{(-2)^{m-1}}\int_{-\pi/2}^{\pi/2}d\lambda s^{(m-2)}(\lambda)s^{(1)}(\lambda)\Phi(\zeta,\lambda)\,,\label{eq:final_charge_curr}\\
\langle \boldsymbol{s}\rangle_\zeta&=& \frac{1}{2}\int_{-\pi/2}^{\pi/2}d\lambda s(\lambda)\Phi(\zeta,\lambda)\,,\label{eq:final_spin_den}\\
\langle \boldsymbol{J}_s\rangle_\zeta&=&-\frac{\sinh(\eta)}{4}\int_{-\pi/2}^{\pi/2}d\lambda s^{(1)}(\lambda)\Phi(\zeta,\lambda)\,.\label{eq:final_spin_curr}
\eea
Here we have introduced the symbols $\frak d_{0}[q]$ and $\frak j_{0}[q]$ (\emph{cf}. Eqs.~\eqref{eq:gsvalues}) indicating the expectation value of the charge density $\boldsymbol{q}$ and the corresponding current on the ground state; moreover we defined 
\bea
\Phi(\zeta,\lambda)&=&\frac{\sinh{(2\beta_R h)}}{\sinh(\beta_R h)}e^{-J \pi \beta_R\sinh\eta s(\lambda)}H\left[\zeta+\frac{\sinh(\eta) s^{\prime}(\lambda)}{2s(\lambda)}\right]\nonumber\\
&+&\frac{\sinh{(2\beta_L h)}}{\sinh(\beta_L h)}e^{-J \pi \beta_L\sinh\eta s(\lambda)}H\left[-\frac{\sinh(\eta) s^{\prime}(\lambda)}{2s(\lambda)}-\zeta\right]\,.
\eea
and used
\bea
\lim_{n\to\infty}\rho_{n,\zeta}^t(\lambda)= \frac{1}{\pi}\int_{-\pi/2}^{\pi/2}{\rm d}\mu\, s(\mu)\eta_{1,\zeta}(\mu)\,,\qquad\qquad
\lim_{n\to\infty}v_{n,\zeta}(\lambda)= \frac{-\sinh\eta}{2}\frac{\int_{-\pi/2}^{\pi/2}{\rm d}\mu\, s^{\prime}(\mu)\eta_{1,\zeta}(\mu)}{\int_{-\pi/2}^{\pi/2}{\rm d}\mu\, s(\mu)\eta_{1,\zeta}(\mu)}\,.
\eea
We see that, once again, the final result is expressed in terms of simple integrals which are readily evaluated numerically. 
In Fig.~\ref{fig:densities_antiferro} we compare the analytic prediction for the low-temperature limit with the exact numerical evaluation. As in the ferromagnetic phase, the profiles are non-constant functions of $\zeta$, which do not display non-analytic points inside the light cone. In particular, Eqs.~\eqref{eq:final_charge_den}--\eqref{eq:final_spin_curr} have the same structure of those encountered in the ferromagnetic phase. As we have already stressed, these equations are in turn analogous to the formulae derived in the case of the free Ising model \cite{Korm17,PeGa17}, \emph{cf.} the discussion at the end of the previous section.
As expected, for the same values of initial temperatures the sign of the spin currents is opposite with respect to the ferromagnetic case. This is due to the different structure of the ground state. 
Finally, we observe that higher order corrections are more important for the profiles of spin density and currents, while they are seen to be smaller for other local charges. In any case, we see from Fig.~\ref{fig:densities_antiferro} that \eqref{eq:final_spin_den} and \eqref{eq:final_spin_curr} are an excellent approximation to the exact profiles already for $T\sim 0.1$.

\section{Conclusions}
\label{sec:conclusions}

We have studied low-temperature transport properties of the XXZ spin-$1/2$ Heisenberg chain. We have focused on the protocol where two semi-infinite chains are prepared in thermal states at small but different temperatures and suddenly joined together. In this setting, expectation values of local observables determine, at large times and distances from the junction, well-defined profile functions of the ratio $\zeta=x/t$. These can be fully investigated by means of the recently introduced generalised hydrodynamic approach \cite{BCDF16,CaDY16}. We have performed a systematic analysis of the low-temperature properties arising in different regions of the phase diagram by developing analytical low-temperature expansions. 

In the critical region, we recover explicitly the CFT predictions for the transport of energy \cite{BeDo12}. The profiles of energy density and current display a three-step form, where the explicit value of the plateau inside the light cone depends on universal parameters. Conversely, the behaviour of generic observables deviates from this picture: a region of width $\sim T$ at the edges of the light cone emerges, where the dependence on $\zeta$ is non-constant but continuous. This recovers the non-linear Luttinger liquid prediction of Ref.~\cite{BPC:short}. As the parameters of the Hamiltonian are tuned to be outside the critical region, leading corrections to the zero-temperature limit become exponentially small in the temperature. Both in the ferromagnetic and antiferromagnetic regions, we have derived analytic formulae for the leading order in $T$ of the profiles, which correspond to non-trivial functions. These are in good agreement with the exact predictions also for small but finite values of $T$.

Our calculations explicitly show how the general hydrodynamic equations can give rise to different qualitative behaviour depending on model-dependent parameters. While we considered the prototypical case of the interacting XXZ chains, similar low-temperature expansions could be naturally generalised to more complicated systems, including, for instance, ``nested" spin chains and multi-component quantum gases, where interesting transport features are expected to appear \cite{IlDe17,IlDe2-17}. These questions are currently under investigation~\cite{MPBC:lowt}.

\begin{acknowledgments} %
We are grateful to Maurizio Fagotti for collaboration at an early stage of this project and for many valuable comments. We also thank Jacopo De Nardis, Paquale Calabrese, Fabian Essler and Marton Mesty\'an for inspiring discussions. BB acknowledges the financial support by the ERC under Starting Grant 279391 EDEQS and under the Advanced Grant 694544 OMNES.
\end{acknowledgments}%

\appendix

\section{Details on the low-temperature expansion in the gapless phase}
\label{app:detailslowtemperature}

\subsection{Sommerfeld expansion of the integral \eqref{eq:integralbeta}}
\label{app:sommerfeldoneT}

Let us consider the integral 
\be
G(\beta)=\int_{0}^{\pi/2}\!\!{\rm d}\mu\,\vartheta_1(\mu) f(\mu),
\ee
where 
\be
\vartheta_1(\lambda)=\frac{1}{1+e^{\beta {\varepsilon_1(\lambda)}}}\,.
\ee
Here $\beta=1/T$, while $\varepsilon_1(\lambda)$ is the thermal dressed energy and we have (\emph{cf}.~\eqref{eq:DEesp})
\be
\varepsilon_1(\lambda)=\varepsilon_1^{0}(\lambda)+\frac{\pi^2 T^2}{6 \varepsilon_1^{(0)\prime}(B)} U(\lambda) + O(T^4)\,.
\ee
Integrating by parts we have 
\be
G(\beta)=\vartheta_1(\mu) F(\mu)\Bigr|_0^{\pi/2}-\int_{0}^{\pi/2}\!\!{\rm d}\mu\,\left(\frac{\rm d\,\,}{{\rm d}\mu}\vartheta_1(\mu)\right)F(\mu)\,,
\ee
where we defined
\be
F(\lambda)=\int^\lambda_0\!\!{\rm d}\nu\, f(\nu)\,.
\ee
The first term is exponentially suppressed in $T$ and can safely be neglected. Focussing on the second term we observe that $(d/d\mu)\vartheta_1(\mu)$ is strongly peaked around $B'$, the zero of the dressed energy -- it approaches a Dirac delta function in the limit $T\rightarrow0$. We then Taylor expand the function $F(\mu)$ around $B'$. The result is written as 
\be
G(\beta)=\sum_{n=0}^{\infty} M_n \frac{{\rm d}^nF}{{\rm d}\mu^n}(\mu)\Bigr|_{\mu=B'}
\ee
where we defined
\bea
 M_n=-\frac{1}{n!}\int_{0}^{\pi/2}\!\!{\rm d}\mu\,\left(\frac{\rm d\,\,}{{\rm d}\mu}\vartheta_1(\mu)\right){(\mu-B')^n}=\frac{1}{n!}\int_{0}^{\pi/2}\!\!{\rm d}\mu\,\frac{\beta \varepsilon_1'(\mu) e^{\beta \varepsilon_1(\mu)}}{(1+e^{\beta \varepsilon_1(\mu)})^2}{(\mu-B')^n}\,.
\eea
Computing the first non trivial order contributions we find
\begin{align}
M_0&=\int_{0}^{\pi/2}\!\!{\rm d}\mu\,\frac{\beta \varepsilon_1'(\mu) e^{\beta \varepsilon_1(\mu)}}{(1+e^{\beta \varepsilon_1(\mu)})^2}=1+O(e^{-\beta})\,,\\
M_1&=\int_{0}^{\pi/2}\!\!{\rm d}\mu\,\frac{\beta \varepsilon_1'(\mu) e^{\beta \varepsilon_1(\mu)}}{(1+e^{\beta \varepsilon_1(\mu)})^2}{(\mu-B')}=\frac{\varepsilon_1^{0\,\prime\prime}(B') T^2}{ (\varepsilon_1^{0\,\prime}(B))^3}\left(\int_{-\infty}^\infty\!\!{\rm d}\mu\,\left[\frac{\mu^2 e^{\mu}}{(1+e^{\mu})^2}-\frac{\mu^3 e^\mu(e^\mu-1)}{2(e^\mu+1)^3}\right]\right)+O(T^4)\,,\nn
 &=-\frac{\pi^2 T^2 \varepsilon_1^{0\,\prime\prime}(B)}{6 (\varepsilon_1^{0\,\prime}(B))^3}+O(T^4)\,,\\
M_2&=\frac{1}{2}\int_{0}^{\pi/2}\!\!{\rm d}\mu\,\frac{\beta \varepsilon_1'(\mu) e^{\beta \varepsilon_1(\mu)}}{(1+e^{\beta \varepsilon_1(\mu)})^2}{(\mu-B')^2}=\frac{T^2}{2 (\varepsilon_1^{0\,\prime}(B))^2}\int_{-\infty}^\infty\!\!{\rm d}\mu\,\frac{\mu^2e^{\mu}}{(1+e^{\mu})^2}+O(T^4)=\frac{\pi^2 T^2}{6 (\varepsilon_1^{0\,\prime}(B))^2}+O(T^4)\,,\\
M_n&=O(T^4)\qquad n>2\,.
\end{align}
Here we introduced $B$ which is the zero of the ground state dressed energy $\varepsilon_1^{0}(\lambda)$. From Eq.~\eqref{eq:DEesp} we find 
\be
B'-B=-\frac{\pi^2 T^2}{6 (\varepsilon_1^{0\,\prime}(B))^2} U(B)+O(T^4)\,.
\ee
Putting everything together and using 
\be
\int^{B'}_0\!\!\!\!{\rm d}\mu\, f(\mu)=\int^{B}_0\!\!\!\!{\rm d}\mu\, f(\mu)-\frac{\pi^2 T^2}{6 (\varepsilon_1^{0\,\prime}(B))^2} U(B) f(B) + O(T^4)\,,
\ee
we find 
\bea
G(\beta)=\int^{B}_0\!\!\!\!{\rm d}\mu\, f(\mu)+\frac{\pi^2 T^2}{6 (\varepsilon_1^{0\,\prime}(B))^2}\left(f'(B)-\left(\frac{\varepsilon_1^{0\,\prime\prime}(B)}{\varepsilon_1^{0\,\prime}(B)}+U(B)\right)f(B)\right)+O(T^4)\,.
\eea
Using this result we immediately have  
\bea
I(\beta)&=&\int_{-\pi/2}^{\pi/2}\!\!{\rm d}\mu\,\vartheta_1(\mu) f(\mu)\nn
&=&\int^{B}_{-B}\!\!\!\!{\rm d}\mu\, f(\mu)+\frac{\pi^2 T^2}{6 (\varepsilon_1^{0\,\prime}(B))^2}\left[f'(B)-f'(-B)-\left(\frac{\varepsilon_1^{0\,\prime\prime}(B)}{\varepsilon_1^{0\,\prime}(B)}+U(B)\right)(f(B)+f(-B))\right]+O(T^4)\,.
\eea

\subsection{Sommerfeld expansion of the integral \eqref{eq:sommerfeldtwot}}
\label{app:integraltwoT}
Let us consider 
\begin{align}
I(\beta,r,\xt)=&\int\limits_{-\pi/2}^{\pi/2}\!\!\!{\rm d}\lambda\,\,\vartheta_{1,\xt}(\lambda)f(\lambda)=\int\limits_{-\pi/2}^{\pi/2}{\rm d}\lambda\,\,\vartheta_{1}^{L}(\lambda)  f(\lambda)H(v_{1,\xt}(\lambda)-\xt)+\vartheta_{1}^{R}(\lambda)f(\lambda) H(\xt-v_{1,\xt}(\lambda))\nn
=&\int\limits_{0}^{\pi/2}{\rm d}\lambda\,\,\vartheta_{1}^L(\lambda)  f(\lambda)H(v_{\xt,1}(\lambda)-\xt)+\vartheta_{1}^R(\lambda)f(\lambda) H(\xt-v_{1,\xt}(\lambda))\notag\\
&+\int\limits_{0}^{\pi/2}{\rm d}\lambda\,\,\vartheta_{1}^L(\lambda)  f(-\lambda)H(v_{\xt,1}(-\lambda)-\xt)+\vartheta_{1}^R(\lambda)f(-\lambda) H(\xt-v_{1,\xt}(-\lambda))\,,
\label{eq:inttwoTapp}
\end{align}
If the ray $\xt$ is $O(T^0)$ away from $\pm v^{0}_{1}(B)$, \emph{i.e.}
\be
\lim_{T\rightarrow0}|\xt\pm v^{0}_{1}(B)|\neq0\,,
\ee
we can apply the result of the above section to each of the four contributions on the r.h.s. of \eqref{eq:inttwoTapp}, the result reads as   
\bea
 I(\beta,r,\xt)&=&\int^{B}_{-B}\!\!{\rm d}\lambda\, f(\lambda)+\frac{\pi^2 T^2}{6 (\varepsilon_1^{0\,\prime}(B))^2}\left(f^{0\,\prime}(B)-\frac{\varepsilon_1^{0\,\prime\prime}(B)}{\varepsilon_1^{0\,\prime}(B)}f^{0}(B)-U(B)f^{0}(B)\right)H(v^{0}_{1}(B)-\xt)\nn
&&+\frac{\pi^2 T^2}{6 (\varepsilon_1^{0\,\prime}(B))^2}\left(-f^{0\,\prime}(-B)-\frac{\varepsilon_1^{0\,\prime\prime}(B)}{\varepsilon_1^{0\prime}(B)}f^{0}(-B)-U(B)f^{0}(-B)\right)H(-v^{0}_{1}(B)-\xt)\nn
&&+\frac{\pi^2 r^2 T^2}{6 (\varepsilon_1^{0\,\prime}(B))^2}\left(f^{0\,\prime}(B)-\frac{\varepsilon_1^{0\,\prime\prime}(B)}{\varepsilon_1^{0\,\prime}(B)}f^{0}(B)-U(B)f^{0}(B)\right)H(\xt-v^{0}_{1}(B))\nn
&&+\frac{\pi^2 r^2 T^2}{6 (\varepsilon_1^{0\,\prime}(B))^2}\left(-f^{0\,\prime}(-B)-\frac{\varepsilon_1^{0\,\prime\prime}(B)}{\varepsilon_1^{0\,\prime}(B)}f^{0}(-B)-U(B)f^{0}(-B)\right)H(\xt+v^{0}_{1}(B))\,.
\label{eq:naiveexpansion}
\eea
Here we want to consider also the region 
\be
|\xt\pm v^{0}_{1}(B)|=O(T)\,.
\label{eq:regionofinterest}
\ee
Using the above approach also to this region would give an additional unphysical contribution of the form
\be
\frac{\pi^2  v^{0\,\prime}_{1}(B) T^2}{6 (\varepsilon_1^{0\,\prime}(B))^2}\left(1-r^2\right)\left(f^0(B) \delta(\xt-v^{0}_1(B))-f^0(-B)\delta(\xt+v^{0}_1(B))\right)\,,
\label{eq:illdefined}
\ee
coming from the derivative of the $\theta_H(x)$. This contribution is spurious; it is the result of having taken separately the zero temperature limit in a product of the form 
\be
T^2 \delta (T, x)
\ee
where $\delta (T, x)$ is such that 
\be
\lim_{T\rightarrow 0}\delta(T,x)=\delta(x)\,.
\ee
To find the correct expression for $\delta (T, x)$ we need a more accurate treatment of the expansion. Let us now consider again \eqref{eq:inttwoTapp}. Considering $\xt>0$ and solving the Heaviside functions we have 
\begin{align}
I(\beta_L,\beta_R,\xt)=&\int\limits_{\lambda_\xt}^{\lambda^{\textsc{ii}}_\xt}{\rm d}\lambda\,\,\,\vartheta^{\rm L}_1(\lambda) f(\lambda)+\int\limits_{-\pi/2}^{\lambda_\xt}{\rm d}\lambda\,\,\vartheta^{\rm R}_1(\lambda) f(\lambda)+\int\limits_{\lambda^{\textsc{ii}}_\xt}^{\pi/2}{\rm d}\lambda\,\,\vartheta^{\rm R}_1(\lambda) f(\lambda)\,.
\label{eq:integraltwoTapp}
\end{align}
Here we assumed that  
\be
v_{\xt,1}(\lambda)-\xt=0\,,
\label{eq:solutiontheta}
\ee
has only two solutions and we denoted them by $\lambda_\xt$ and $\lambda^{\textsc{ii}}_\xt$. The solutions $\lambda_\xt$ and $\lambda^{\textsc{ii}}_\xt$ are labeled such that 
\be
\lambda^{\textsc{ii}}_\xt\neq \pm B\,,\qquad\qquad \forall\,\xt\,,
\ee
and for definiteness we have taken  $\lambda^{\textsc{ii}}_\xt>\lambda_\xt$. Let us develop an asymptotic expansion of the integrals on the r.h.s. of \eqref{eq:integraltwoTapp}. Focussing on the first term on the first line and integrating by parts we have 
\be
\int\limits_{\lambda_\xt}^{\lambda^{\textsc{ii}}_\xt}{\rm d}\lambda\,\,\vartheta^{\rm L}_1(\lambda) f(\lambda) =  \vartheta^{\rm L}_1(\lambda^{\textsc{ii}}_\xt)\int\limits_{\lambda_\xt}^{\lambda^{\textsc{ii}}_\xt}{\rm d}\lambda\,\, f(\lambda) - \int\limits_{\lambda_\xt}^{\lambda^{\textsc{ii}}_\xt}{\rm d}\lambda\,\,\vartheta^{\rm L\,\prime}_1(\lambda) \int\limits_{\lambda_\xt}^{\lambda}{\rm d}\mu\,\,f(\mu) \,.
\label{eq:integralbyparts}
\ee
The first term in the r.h.s. is exponentially suppressed in $1/T$ for low temperatures and can be safely neglected. Considering the second term we proceed as in the previous section and expand the function
\be
F(\lambda)=\int\limits_{\lambda_\xt}^{\lambda} {\rm d}\mu\,\,f(\mu)\,, 
 \ee
in power series around $B'_L$, the value around which $\vartheta^{\rm L\,\prime}_1(\lambda)$ is peaked. Plugging this series expansion in \eqref{eq:integralbyparts} we find 
\be
\int\limits_{\lambda_\xt}^{\lambda^{\textsc{ii}}_\xt}{\rm d}\lambda\,\,\vartheta^{\rm L}_1(\lambda) f(\lambda) =\sum_{n=0}^{\infty} M_n(\lambda_\xt,\lambda^{\textsc{ii}}_\xt) \frac{{\rm d}^nF}{{\rm d}\mu^n}(\mu)\Bigr|_{\mu=B'_L}\,,
\label{eq:appltsumL}
\ee  
where 
\be
M_n(\lambda_\xt,\lambda^{\textsc{ii}}_\xt)=-\frac{1}{n!}\int\limits_{\lambda_\xt}^{\lambda^{\textsc{ii}}_\xt}{\rm d}\mu\,\,\vartheta^{\rm L\,\prime}_1(\lambda)(\lambda-B_L')^n\,. 
 \ee
Computing the first non trivial order in $T$ we have  
\begin{align}
M_0(\lambda_\xt,\lambda^{\textsc{ii}}_\xt)&=\int\limits_{\lambda_\xt}^{\lambda^{\textsc{ii}}_\xt}{\rm d}\mu\,\frac{\beta \varepsilon_1'(\mu) e^{\beta \varepsilon_1(\mu)}}{(1+e^{\beta \varepsilon_1(\mu)})^2}=\vartheta^{\rm L}_1(\lambda_\xt)-\vartheta^{\rm L}_1(\lambda^{\textsc{ii}}_\xt)\,,\\
M_1(\lambda_\xt,\lambda^{\textsc{ii}}_\xt)&=\int\limits_{\lambda_\xt}^{\lambda^{\textsc{ii}}_\xt}{\rm d}\mu\,\frac{\beta \varepsilon_1'(\mu) e^{\beta \varepsilon_1(\mu)}}{(1+e^{\beta \varepsilon_1(\mu)})^2}{(\mu-B'_L)}\nn
&=\frac{T}{\varepsilon_1^{0\,\prime}(B)}\int\limits_{\beta\varepsilon^{0\,\prime}_1(B)  \lambda_\xt^-}^\infty\!\!{\rm d}\mu\,\frac{ \mu e^{\mu}}{(1+e^{\mu})^2}+\frac{\varepsilon_1^{0\,\prime\prime}(B) T^2}{ (\varepsilon_1^{0\,\prime}(B))^3}\int\limits_{\beta \varepsilon^{0\,\prime}_1(B)  \lambda_\xt^-}^\infty\!\!{\rm d}\mu\,\left[\frac{\mu^2 e^{\mu}}{(1+e^{\mu})^2}-\frac{\mu^3 e^\mu(e^\mu-1)}{2(e^\mu+1)^3}\right]+O(T^3)\nn
 &=\frac{\pi^2 T}{3\varepsilon_1^{0\,\prime}(B)} \mathcal F\left(\beta \varepsilon^{0\,\prime}_1(B)  \lambda_\xt^-\right)-\frac{\pi^2 T^2 \varepsilon_1^{0\,\prime\prime}(B)}{6 (\varepsilon_1^{0\,\prime}(B))^3}  \mathcal G (\beta \varepsilon^{0\,\prime}_1(B)  \lambda_\xt^-)-\frac{\pi^2 T^2 \varepsilon_1^{0\,\prime\prime}(B)}{2 (\varepsilon_1^{0\,\prime}(B))^3}  \mathcal H (\beta \varepsilon^{0\,\prime}_1(B)  \lambda_\xt^-)+O(T^3)\,,\\
M_2(\lambda_\xt,\lambda^{\textsc{ii}}_\xt)&=\frac{1}{2}\int\limits_{\lambda_\xt}^{\lambda^{\textsc{ii}}_\xt}{\rm d}\mu\,\frac{\beta \varepsilon_1'(\mu) e^{\beta \varepsilon_1(\mu)}}{(1+e^{\beta \varepsilon_1(\mu)})^2}{(\mu-B'_L)^2}=\frac{T^2}{2 (\varepsilon_1^{0\,\prime}(B))^2}\int\limits_{\beta \varepsilon^{0\,\prime}_1(B)  \lambda_\xt^-}^\infty\!\!\!\!{\rm d}\mu\,\frac{\mu^2e^{\mu}}{(1+e^{\mu})^2}+O(T^3)\notag\\
&=\frac{\pi^2 T^2}{6 (\varepsilon_1^{0\,\prime}(B))^2} \mathcal G (\beta \varepsilon^{0\,\prime}_1(B)  \lambda_\xt^-)+O(T^3)\,,\\
M_n(\lambda_\xt,\lambda^{\textsc{ii}}_\xt)&=O(T^3)\,,\qquad\qquad n>2\,,
\end{align}
where we introduced 
\be
 \lambda_\xt^-\equiv\lambda_\xt-B\,.
\ee
Moreover, we defined the following functions 
\begin{align}
 {\cal F} (z) &\equiv\frac{3}{\pi^2}\log[1 + e^z]- \frac{3ze^{z}}{\pi^2(1 + e^z)}\,, \\
{\cal G} (z) &\equiv\frac{6}{\pi^2} \text{Li}_2\left(-e^z\right)-\frac{3}{\pi^2}\frac{e^z z^2}{e^z+1}+\frac{6z}{\pi^2} \log \left(e^z+1\right)+1\,, \label{eq:defGfun}\\
  {\cal H} (z) &\equiv\frac{e^z  z^3}{{\pi^2} \left(e^z+1\right)^2}\,.
\end{align}
Substituting in \eqref{eq:appltsumL} we find 
\begin{align}
\int\limits_{\lambda_\xt}^{\lambda^{\textsc{ii}}_\xt}{\rm d}\lambda\,\,\vartheta^{\rm L}_1(\lambda) f(\lambda)=&\int\limits_{\lambda_\xt}^{B'_L}{\rm d}\lambda\,\, f(\lambda)\vartheta^{\rm L}_1(\lambda_\xt)-\frac{\pi^2 T^2 \varepsilon_1^{0\,\prime\prime}(B)}{2 (\varepsilon_1^{0\,\prime}(B))^3}   {\cal H} (\beta \varepsilon^{0\,\prime}_1(B)  \lambda_\xt^-) f(B) + \frac{\pi^2 T f(B)}{3\varepsilon_1^{0\,\prime}(B)}  {\cal F}\left(\beta \varepsilon^{0\,\prime}_1(B)  \lambda_\xt^-\right)\nn
&+\frac{\pi^2 T^2}{6 (\varepsilon_1^{0\,\prime}(B))^2}\left(f'(B)-\frac{\varepsilon_1^{0\,\prime\prime}(B)}{\varepsilon_1^{0\,\prime}(B)} f(B)\right)  {\cal G} (\beta \varepsilon^{0\,\prime}_1(B)  \lambda_\xt^-)+O(T^3)\,.
\label{eq:result1}
\end{align}
Considering the second contribution on the first line of the r.h.s. of \eqref{eq:integraltwoTapp} and proceeding as above we find 
\begin{align}
\int\limits_{-\pi/2}^{\lambda_\xt}{\rm d}\lambda\,\,\vartheta^{\rm R}_1(\lambda) f(\lambda)=&\int\limits_{-B'_R}^{B_R'}{\rm d}\lambda\,\, f(\lambda)+\int\limits_{B_R'}^{\lambda_\xt}{\rm d}\lambda\,\, f(\lambda)\vartheta^{\rm R}_1(\lambda_\xt)+\frac{\pi^2 r^2 T^2 \varepsilon_1^{0\,\prime\prime}(B)}{2 (\varepsilon_1^{0\,\prime}(B))^3}   {\cal H} \left(\frac{\beta}{r}\varepsilon^{0\,\prime}_1(B)  \lambda_\xt^-\right)f(B)\nn
&-\frac{\pi^2 r T f(B)}{3\varepsilon_1^{0\,\prime}(B)}  {\cal F}\left(\frac{\beta}{r}\varepsilon^{0\,\prime}_1(B)  \lambda_\xt^-\right)-\frac{\pi^2 r^2 T^2}{6 (\varepsilon_1^{0\,\prime}(B))^2}\left(f'(B)-\frac{\varepsilon_1^{0\,\prime\prime}(B)}{\varepsilon_1^{0\,\prime}(B)} f(B)\right){\cal G} \left(\frac{\beta}{r}\varepsilon^{0\,\prime}_1(B)  \lambda_\xt^-\right)\nn
&+\frac{\pi^2 r^2 T^2}{6 (\varepsilon_1^{0\,\prime}(B))^2}\left(f'(B)-\frac{\varepsilon_1^{0\,\prime\prime}(B)}{\varepsilon_1^{0\,\prime}(B)} f(B)-f'(-B)-\frac{\varepsilon_1^{0\,\prime\prime}(B)}{\varepsilon_1^{0\,\prime}(B)} f(-B)\right)+O(T^3)\,.
\label{eq:result3}
\end{align}
Note that the third term on the first line of \eqref{eq:integraltwoTapp} is exponentially suppressed because $\lambda^{\textsc{ii}}_\xt>B$. Repeating the same for $\xt<0$ and collecting all together we obtain  
\begin{align}
I(\beta,r,\xt)&=\int_{-B}^{B}{\rm d}\lambda\,\, f(\lambda)+\frac{\pi^2 T^2}{6 (\varepsilon_1^{0\,\prime}(B))^2}\left[\left(f'(B)-\frac{\varepsilon_1^{0\,\prime\prime}(B)}{\varepsilon_1^{0\,\prime}(B)} f(B)\right)  {\cal G} ({\beta}\varepsilon^{0\,\prime}_1(B)  \lambda_\xt^-)-U(B)f(B)  {\cal I} ({\beta}\varepsilon^{0\,\prime}_1(B)  \lambda_\xt^-)\right]\nn
&+\frac{\pi^2 r^2 T^2}{6 (\varepsilon_1^{0\,\prime}(B))^2}\left[\left(f'(B)-\frac{\varepsilon_1^{0\,\prime\prime}(B)}{\varepsilon_1^{0\,\prime}(B)} f(B)\right)\left(1-  {\cal G} \left(\frac{\beta}{r}\varepsilon^{0\,\prime}_1(B)  \lambda_\xt^-\right)\right)-U(B)f(B)\left(1-  {\cal I} \left(\frac{\beta}{r}\varepsilon^{0\,\prime}_1(B)  \lambda_\xt^-\right)\right)\right]\nn
&+\frac{\pi^2 T^2}{6 (\varepsilon_1^{0\,\prime}(B))^2}\left[\left(-f'(-B)-\frac{\varepsilon_1^{0\,\prime\prime}(B)}{\varepsilon_1^{0\,\prime}(B)} f(-B)\right)  {\cal G} ({\beta}\varepsilon^{0\,\prime}_1(B)  \lambda_\xt^+)-U(B)f(-B)  {\cal I}({\beta}\varepsilon^{0\,\prime}_1(B) \lambda_\xt^+)\right]\nn
&+\frac{\pi^2 r^2 T^2}{6 (\varepsilon_1^{0\,\prime}(B))^2}\left[\left(-f'(-B)-\frac{\varepsilon_1^{0\,\prime\prime}(B)}{\varepsilon_1^{0\,\prime}(B)} f(-B)\right)\left(1- {\cal G} \left(\frac{\beta}{r}\varepsilon^{0\,\prime}_1(B) \lambda_\xt^+\right)\right)-U(B)f(-B)\left(1-  {\cal I} \left(\frac{\beta}{r}\varepsilon^{0\,\prime}_1(B)  \lambda_\xt^+\right)\right)\right]\nn
&-\frac{\pi^2 T(r^2-1)}{6\varepsilon^{0\,\prime}_1(B)}\left(f(B)  {\cal D}_r \left({\beta}\varepsilon^{0\,\prime}_1(B)  \lambda_\xt^-\right)-f(-B)  {\cal D}_r \left({\beta}\varepsilon^{0\,\prime}_1(B)  \lambda_\xt^+ \right)\right)\nn
&-\frac{7\pi^4 T^2 (r^4-1) }{120  (\varepsilon^{0\,\prime}_1(B))^2}(f'(B)   {\cal E}_r({\beta}\varepsilon^{0\,\prime}_1(B)  \lambda_\xt^-)-f'(-B)  {\cal E}_r({\beta}\varepsilon^{0\,\prime}_1(B)  \lambda_\xt^+))+O(T^3)\,.
\label{eq:nonnaiveexpansion}
\end{align}
Here we considered the case $|\lambda_\xt|<|\lambda^{\textsc{ii}}_\xt|$ and defined $\lambda_\xt^+\equiv\lambda_\xt+B$. We also introduced the functions 
\begin{align}
{\cal I}(z)&\equiv\frac{1}{1+e^{z}}- \frac{z e^z}{(1+e^z)^2}\,, \label{eq:Ifundef}\\
{\cal D}_r (z)&\equiv\frac{6}{\pi^2(1-r^2)} \log(1+e^{z})-\frac{6r}{\pi^2(1-r^2)}\log(1+e^{z/r})\,,\label{eq:Dfundef}\\
 {\cal E}_r(z)&\equiv \frac{60}{7 \pi^4(r^4-1)}\left(\frac{z^2}{1+e^{z}}-\frac{z^2}{1+e^{z/r}}\right)\,.\label{eq:Efundef}
\end{align}
Equation \eqref{eq:nonnaiveexpansion} gives a low temperature expansion of $I(\beta,r,\xt)$ including up to $O(T^2)$ terms. It generically requires, however, the full solution of the transport problem: to find $\lambda_\xt$ one has to find and invert the velocity $v_{\xt,1}(\lambda)$. To avoid this we proceed as follows. Let us focus on $ \lambda_\xt^-$ and note that the non-trivial variation of the functions in  \eqref{eq:nonnaiveexpansion} happens for $\lambda^-_\xt\sim T $: this can happen only if $\xt-v^0_1(B)=O(T)$. We can then expand $\lambda_\xt$ around $v^0_1(B)$; using the known formulae for the derivatives of the inverse function we have 
\be
\lambda_\xt=\lambda_{v^0_1(B)}+\left(\frac{\xt-v^{0}_1(B)}{v^{0\,\prime}_1(B)}\right)-\frac{v^{0\,\prime\prime}_1(B)}{2v^{0\,\prime}_1(B)}\left(\frac{\xt-v^{0}_1(B)}{v^{0\,\prime}_1(B)}\right)^2+O(T^3)\,.
\label{eq:taylorlambdaxt}
\ee
From the expansion \eqref{eq:nonnaiveexpansion} one can see that the first finite temperature correction to the velocity $v_{1,\xt}(\lambda)$ is $O(T^2)$. To see that we can focus on the $O(T)$ corrections to $\rho_{1,\xt}^t(\lambda)$ and $v_{1,\xt}(\lambda)\rho_{1,\xt}^t(\lambda)$, using \eqref{eq:nonnaiveexpansion} we find 
\be
\delta(v_{1,\xt}\rho_{1,\xt}^t)(\lambda)= v_{1}^0(\lambda)\delta \rho_{1,\xt}^t(\lambda)+O(T^2)\qquad\Rightarrow\qquad\delta v_{1,\xt}(\lambda)= O(T^2)\,.
\ee
This implies that 
\be
\lambda_{v^0_1(B)}= B + T^2 \delta_-\,,
\label{eq:lambdapxt}
\ee 
where $\delta_-$ will be determined later. Substituting \eqref{eq:taylorlambdaxt} and  \eqref{eq:lambdapxt} in \eqref{eq:nonnaiveexpansion} and proceeding analogously for $\lambda_\xt^+$ we find 
\begin{align}
I(\beta,r,\xt)&=\int_{-B}^{B}{\rm d}\lambda\,\, f(\lambda)+\frac{\pi^2 T^2r^2}{6 (\varepsilon_1^{0\,\prime}(B))^2}\left[f'(B)-f'(-B)-\left(\frac{\varepsilon_1^{0\,\prime\prime}(B)}{\varepsilon_1^{0\,\prime}(B)} +U(B)\right)\left(f(B)+f(-B)\right)\right]\nn
&+\frac{\pi^2 T^2}{6 (\varepsilon_1^{0\,\prime}(B))^2}\left[\left(f'(B)-\frac{\varepsilon_1^{0\,\prime\prime}(B)}{\varepsilon_1^{0\,\prime}(B)} f(B)\right)  {\cal G} \left(\varepsilon^{0\,\prime}_1(B) \frac{\xt-v^{0}_1(B)}{T |v^{0\,\prime}_1(B)|}\right)-U(B)f(B)  {\cal I} \left(\varepsilon^{0\,\prime}_1(B) \frac{\xt-v^{0}_1(B)}{T |v^{0\,\prime}_1(B)|}\right)\right]\nn
&-\frac{\pi^2 r^2 T^2}{6 (\varepsilon_1^{0\,\prime}(B))^2}\left[\left(f'(B)-\frac{\varepsilon_1^{0\,\prime\prime}(B)}{\varepsilon_1^{0\,\prime}(B)} f(B)\right){\cal G}\left(\varepsilon^{0\,\prime}_1(B) \frac{\xt-v^{0}_1(B)}{T r|v^{0\,\prime}_1(B)|}\right) -U(B)f(B){\cal I} \left(\varepsilon^{0\,\prime}_1(B) \frac{\xt-v^{0}_1(B)}{T r|v^{0\,\prime}_1(B)|}\right) \right]\nn
&-\frac{\pi^2 T^2}{6 (\varepsilon_1^{0\,\prime}(B))^2}\left[\left(f'(-B)+\frac{\varepsilon_1^{0\,\prime\prime}(B)}{\varepsilon_1^{0\,\prime}(B)} f(-B)\right)  {\cal G} \left(\varepsilon^{0\,\prime}_1(B) \frac{\xt+v^{0}_1(B)}{T |v^{0\,\prime}_1(B)|}\right)+U(B)f(-B)  {\cal I} \left(\varepsilon^{0\,\prime}_1(B) \frac{\xt+v^{0}_1(B)}{T |v^{0\,\prime}_1(B)|}\right)\right]\nn
&+\frac{\pi^2 r^2 T^2}{6 (\varepsilon_1^{0\,\prime}(B))^2}\left[\left(f'(-B)+\frac{\varepsilon_1^{0\,\prime\prime}(B)}{\varepsilon_1^{0\,\prime}(B)} f(-B)\right){\cal G} \left(\varepsilon^{0\,\prime}_1(B) \frac{\xt+v^{0}_1(B)}{T r| v^{0\,\prime}_1(B)|}\right)+U(B)f(-B){\cal I} \left(\varepsilon^{0\,\prime}_1(B) \frac{\xt+v^{0}_1(B)}{T r|v^{0\,\prime}_1(B)|}\right)\right]\nn
&-\text{sgn}(v^{0\,\prime}_1(B))\frac{\pi^2 T(r^2-1)}{6\varepsilon^{0\,\prime}_1(B)}\left(f(B)  {\cal D}_r \left(\varepsilon^{0\,\prime}_1(B) \frac{\xt-v^{0}_1(B)}{T v^{0\,\prime}_1(B)}\right)-f(-B)  {\cal D}_r \left(\varepsilon^{0\,\prime}_1(B) \frac{\xt+v^{0}_1(B)}{T v^{0\,\prime}_1(B)} \right)\right)\nn
&-\frac{7\pi^4 T^2 (r^4-1) }{120  (\varepsilon^{0\,\prime}_1(B))^2}\left[\left(f'(B)-\frac{v^{0\,\prime\prime}_1(B)}{v^{0\,\prime}_1(B)}f(B)\right)   {\cal E}_r\left(\varepsilon^{0\,\prime}_1(B) \frac{\xt-v^{0}_1(B)}{T |v^{0\,\prime}_1(B)|}\right)\right]\nn
&+\frac{7\pi^4 T^2 (r^4-1) }{120  (\varepsilon^{0\,\prime}_1(B))^2}\left[\left(f'(-B)+\frac{v^{0\,\prime\prime}_1(B)}{v^{0\,\prime}_1(B)}f(-B)\right)   {\cal E}_r\left(\varepsilon^{0\,\prime}_1(B) \frac{\xt+v^{0}_1(B)}{T |v^{0\,\prime}_1(B)|}\right)\right]\nn
&-\text{sgn}(v^{0\,\prime}_1(B))\frac{\pi^2 T^2(r^2-1)}{6}\left(f(B) \delta_-  {\cal P}_r \left(\varepsilon^{0\,\prime}_1(B) \frac{\xt-v^{0}_1(B)}{T |v^{0\,\prime}_1(B)|}\right)-f(-B) \delta_+  {\cal P}_r \left(\varepsilon^{0\,\prime}_1(B) \frac{\xt+v^{0}_1(B)}{T |v^{0\,\prime}_1(B)|} \right)\right)\,.
\label{eq:nonnaiveexpansionfinal}
\end{align}
Here we extended our results also to the case $|\lambda_\xt|>|\lambda^{\textsc{ii}}_\xt|$, we also introduced the functions 
\begin{align}
{\cal P}_r(z)&\equiv{\cal D}'_r(z)=\frac{6}{\pi^2(r^2-1)}\left(\frac{e^{z/r}}{1+e^{z/r}}-\frac{e^{z}}{1+e^{z}}\right)\,,
\label{eq:Pfundef}
\end{align}
and defined 
\be
\delta_+=\frac{\lambda_{-v^0_1(B)}+B}{T^2}\text{sgn}(v^{0\,\prime}_1(B))\,.
\label{eq:lambdamxt}
\ee
First we note that for 
\be
\lim_{T\rightarrow0}|\xt\pm v^{0}_{1}(B)|\neq0\,,
\ee
we recover \eqref{eq:naiveexpansion}. The ill-defined Dirac delta term \eqref{eq:illdefined} is regularised by 
\be
-\text{sgn}(v^{0\,\prime}_1(B))\frac{\pi^2 T(r^2-1)}{6\varepsilon_1'(B)}\left(f(B)  {\cal D}_r \left(\varepsilon^{0\,\prime}_1(B) \frac{\xt-v^{0}_1(B)}{T v^{0\,\prime}_1(B)}\right)-f(-B)  {\cal D}_r \left(\varepsilon^{0\,\prime}_1(B) \frac{\xt+v^{0}_1(B)}{T v^{0\,\prime}_1(B)} \right)\right)\,.
\ee
The expression \eqref{eq:nonnaiveexpansionfinal}, however, can also interpreted in the distributional sense, \emph{i.e.}, one sees $I(\beta_L,\beta_R,\xt)$ as a distribution in the variable $\xt$ and considers its integrals over $\xt$ multiplied by some smooth function $f(\xt)$. In this case, up to $O(T^3)$ terms, the expansion \eqref{eq:nonnaiveexpansion} agrees with  \eqref{eq:naiveexpansion} supplemented with the Dirac delta term \eqref{eq:illdefined}. Indeed we have 
\begin{align}
&\lim_{T \rightarrow 0}   {\cal G} \left(\frac{z}{T}\right)  =\lim_{T \rightarrow 0}   {\cal I} \left(\frac{z}{T}\right) = \theta_H(z)\,,\\
&\lim_{T \rightarrow 0}\frac{1}{T}   {\cal D}_r \left(\frac{z}{T}\right) = \delta(z)\,,\\
&\lim_{T \rightarrow 0}\frac{1}{T^2}   {\cal P}_r \left(\frac{z}{T}\right) =\lim_{T \rightarrow 0}\frac{1}{T^2}   {\cal E}_r \left(\frac{z}{T}\right) = \delta'(z)\,.
\end{align}
Let us now find the terms $\delta_-$ and $\delta_+$ and, in turn, fully determine \eqref{eq:nonnaiveexpansionfinal}. From their definitions \eqref{eq:lambdapxt} and \eqref{eq:lambdamxt} we immediately find 
\be
\delta_-=  \frac{v_1^0(B)-v_{v_1^{0}(B),1}(B)}{T^2 |v_1^{0\,\prime}(B)|}\qquad\qquad\delta_+=  \frac{v_1^0(-B)-v_{- v_1^{0}(B),1}(-B)}{T^2 |v_1^{0\,\prime}(B)|}\,.
\ee
Using \eqref{eq:nonnaiveexpansionfinal} to find the first corrections to the velocity we finally arrive at  
\be
\delta_\pm=-\frac{\pi^2 v_1^0(B) C_{r,\pm}(\mp B)}{12 (\varepsilon^{0\,\prime}_1(B) )^2 |v_1^{0\,\prime}(B)|}\,,
\ee
where 
\begin{align}
C_{r,-}(\lambda)=&-\int_{-B}^{B}{\rm d}\mu\, a_2(\lambda-\mu)\, C_{r,-}(\mu)+4r^2 a'_2(\lambda + B)- 4 r^2 a_2(\lambda + B)U(B)\\
&-\frac{v_1^{0\,\prime}(B)}{v_1^0(B)}\left(2 r^2 a_2(\lambda+B)+(1+r^2)a_2(\lambda-B)\right)\,,\notag\\
C_{r,+}(\lambda)=&-\int_{-B}^{B}{\rm d}\mu\, a_2(\lambda-\mu)\, C_{r,+}(\mu)+ 4 a_2'(\lambda - B)+4 U(B)a_2(\lambda - B)\notag\\
&+\frac{v_1^{0\,\prime}(B)}{v_1^0(B)}\left((1+ r^2) a_2(\lambda+B)+2 a_2(\lambda-B) \right)\,.
\end{align}

\subsection{Profile expansion in the gapless phase}
\label{app:tbaidentitieslowt}

In this appendix we explicitly determine the profiles of charges and currents in the gapless phase including up to $O(T^2)$. Let us consider the charge and current density profiles \eqref{eq:chargedensityprofilelt} and \eqref{eq:currentprofilelt}, at low temperatures we have 
 \begin{align}
\braket{{\boldsymbol q}}_\xt =& \int_{-\pi/2}^{\pi/2} \!\!\!{\rm d}\lambda \,\, q(\lambda) \vartheta_{1,\xt}(\lambda) \rho^{t\,0}_{1,\xt}(\lambda)+\int_{-B}^{B} \!\!\!{\rm d}\lambda \,\, q(\lambda) \delta \rho^t_{1,\xt}(\lambda)\notag\\
&+\frac{\pi^2 T (1-r^2)v^{0\,\prime}_1(B)}{6\varepsilon_1'(B)|v^{0\,\prime}_1(B)|}\int_{-B}^{B} \!\!\!{\rm d}\lambda \frac{\rm d}{{\rm d}\lambda}\left(q(\lambda)\delta\rho^{t}_{1,\xt}(\lambda) {\cal D}_r \left({\varepsilon^{0\,\prime}_1(\lambda)}\frac{\xt-v^0_1(\lambda)}{T v_1^{0\,\prime}(\lambda)}\right)\right)+O(T^3)\,,\\
\braket{\boldsymbol J_{\boldsymbol q}}_\xt =&\int_{-\pi/2}^{\pi/2} \!\!\!{\rm d}\lambda \,\, q(\lambda) \vartheta_{1,\xt}(\lambda) v^0_{1,\xt}(\lambda) \rho^{t\,0}_{1,\xt}(\lambda)+\int_{-B}^{B} \!\!\!{\rm d}\lambda \,\, q(\lambda) \delta\left(v_{1,\xt} \rho^{t}_{1,\xt}\right)(\lambda)\notag\\
&+\frac{\pi^2 T (1-r^2)v^{0\,\prime}_1(B)}{6\varepsilon_1'(B)|v^{0\,\prime}_1(B)|}\int_{-B}^{B} \!\!\!{\rm d}\lambda \frac{\rm d}{{\rm d}\lambda}\left(q(\lambda)\delta(v_{1,\xt}\rho^{t}_{1,\xt})(\lambda) {\cal D}_r \left({\varepsilon^{0\,\prime}_1(\lambda)}\frac{\xt-v^0_1(\lambda)}{T v_1^{0\,\prime}(\lambda)}\right)\right)+O(T^3)\,,
\end{align}
The first term on the r.h.s. of these equations is treated using the expansion \eqref{eq:nonnaiveexpansionfinal}, while the second by plugging in the explicit form of the corrections to $\rho^t_{1,\xt}(\lambda)$ and $v_{1,\xt}(\lambda)$. Using the results of the previous appendix to expand the integrals in \eqref{eq:tbalt} and  \eqref{eq:velocitieslt}, the corrections to  $\rho^t_{1,\xt}(\lambda)$ and $v_{1,\xt}(\lambda)$ are written as 
\begin{align}
\delta \rho^t_{1,\xt}(\lambda) &=-\text{sgn}(v^{0\,\prime}_1(B))\frac{\pi T(r^2-1)}{12 v^{0}_1(B)}\left(K_-(\lambda)  {\cal D}_r \left(\varepsilon^{0\,\prime}_1(B) \frac{\xt-v^{0}_1(B)}{T v^{0\,\prime}_1(B)}\right)-K_+(\lambda)   {\cal D}_r \left(\varepsilon^{0\,\prime}_1(B) \frac{\xt+v^{0}_1(B)}{T v^{0\,\prime}_1(B)} \right)\right)\nn
&+\frac{\pi T^2r^2}{12 \varepsilon_1^{0\,\prime}(B) v^{0}_1(B)}\left[Z_{-}(\lambda)-Z_{+}(\lambda)-\left(\frac{v^{0\,\prime}_1(B)}{v^{0}_1(B)} +U(B)\right)\left(K_-(\lambda)+K_+(\lambda)\right)\right]\nn
&+\frac{\pi T^2}{12 \varepsilon_1^{0\,\prime}(B) v^{0}_1(B)}\left[\left(Z_{-}(\lambda)-\frac{v^{0\,\prime}_1(B)}{v^{0}_1(B)} K_-(\lambda)\right)  {\cal G} \left(\varepsilon^{0\,\prime}_1(B) \frac{\xt-v^{0}_1(B)}{T |v^{0\,\prime}_1(B)|}\right)-K_-(\lambda)  U(B) {\cal I} \left(\varepsilon^{0\,\prime}_1(B) \frac{\xt-v^{0}_1(B)}{T |v^{0\,\prime}_1(B)|}\right)\right]\nn
&-\frac{\pi r^2 T^2}{12 \varepsilon_1^{0\,\prime}(B) v^{0}_1(B)}\left[\left(Z_{-}(\lambda)-\frac{v^{0\,\prime}_1(B)}{v^{0}_1(B)}K_-(\lambda)\right){\cal G}\left(\varepsilon^{0\,\prime}_1(B) \frac{\xt-v^{0}_1(B)}{T r|v^{0\,\prime}_1(B)|}\right) -K_-(\lambda) U(B) {\cal I} \left(\varepsilon^{0\,\prime}_1(B) \frac{\xt-v^{0}_1(B)}{T r|v^{0\,\prime}_1(B)|}\right) \right]\nn
&-\frac{\pi T^2}{12 \varepsilon_1^{0\,\prime}(B) v^{0}_1(B)} \left[\left(Z_{+}(\lambda)+\frac{v^{0\,\prime}_1(B)}{v^{0}_1(B)} K_+(\lambda)\right)  {\cal G} \left(\varepsilon^{0\,\prime}_1(B) \frac{\xt+v^{0}_1(B)}{T |v^{0\,\prime}_1(B)|}\right)+K_+(\lambda)  U(B) {\cal I} \left(\varepsilon^{0\,\prime}_1(B) \frac{\xt+v^{0}_1(B)}{T |v^{0\,\prime}_1(B)|}\right)\right]\nn
&+\frac{\pi r^2 T^2}{12 \varepsilon_1^{0\,\prime}(B) v^{0}_1(B)}\left[\left(Z_{+}(\lambda)+\frac{v^{0\,\prime}_1(B)}{v^{0}_1(B)}K_+(\lambda)\right){\cal G} \left(\varepsilon^{0\,\prime}_1(B) \frac{\xt+v^{0}_1(B)}{T r| v^{0\,\prime}_1(B)|}\right)+K_+(\lambda) U(B){\cal I} \left(\varepsilon^{0\,\prime}_1(B) \frac{\xt+v^{0}_1(B)}{T r|v^{0\,\prime}_1(B)|}\right)\right]\nn
&+\frac{\pi^3 T^2 (r^2-1)}{144 \varepsilon^{0\,\prime}_1(B) v_1^{0\,\prime}(B)}\left[K_-(\lambda) C_{r,+}(- B)  {\cal P}_r \left(\varepsilon^{0\,\prime}_1(B) \frac{\xt-v^{0}_1(B)}{T |v^{0\,\prime}_1(B)|}\right)-K_+(\lambda) C_{r,-}(B)  {\cal P}_r \left(\varepsilon^{0\,\prime}_1(B) \frac{\xt+v^{0}_1(B)}{T |v^{0\,\prime}_1(B)|} \right)\right]\nn
&-\frac{7\pi^3 T^2 (r^4-1) }{240  \varepsilon^{0\,\prime}_1(B)  v^{0}_1(B)}\left[\left(Z_{-}(\lambda)-\left(\frac{v^{0\,\prime}_1(B)}{v^{0}_1(B)}+\frac{v^{0\,\prime\prime}_1(B)}{v^{0\,\prime}_1(B)}-\frac{\varepsilon^{0\,\prime\prime}_1(B)}{\varepsilon^{0\,\prime}_1(B)}\right)K_-(\lambda)\right)   {\cal E}_r\left(\varepsilon^{0\,\prime}_1(B) \frac{\xt-v^{0}_1(B)}{T |v^{0\,\prime}_1(B)|}\right)\right]\nn
&+\frac{7\pi^3 T^2 (r^4-1) }{240 \varepsilon^{0\,\prime}_1(B)  v^{0}_1(B)}\left[\left(Z_{+}(\lambda)+\left(\frac{v^{0\,\prime}_1(B)}{v^{0}_1(B)}+\frac{v^{0\,\prime\prime}_1(B)}{v^{0\,\prime}_1(B)}-\frac{\varepsilon^{0\,\prime\prime}_1(B)}{\varepsilon^{0\,\prime}_1(B)}\right)K_+(\lambda)\right)   {\cal E}_r\left(\varepsilon^{0\,\prime}_1(B) \frac{\xt+v^{0}_1(B)}{T |v^{0\,\prime}_1(B)|}\right)\right],
\label{eq:fullexpansionrhot}
\end{align}
\begin{align}
\delta \left(v_{1,\xt}\rho^t_{1,\xt}\right)(\lambda) &=-\text{sgn}(v^{0\,\prime}_1(B))\frac{\pi T(r^2-1)}{12}\left(K_-(\lambda)  {\cal D}_r \left(\varepsilon^{0\,\prime}_1(B) \frac{\xt-v^{0}_1(B)}{T v^{0\,\prime}_1(B)}\right)+K_+(\lambda)   {\cal D}_r \left(\varepsilon^{0\,\prime}_1(B) \frac{\xt+v^{0}_1(B)}{T v^{0\,\prime}_1(B)} \right)\right)\nn
&+\frac{\pi T^2 r^2}{12 \varepsilon_1^{0\,\prime}(B)}\left[Z_{-}(\lambda)+Z_{+}(\lambda)-U(B)\left(K_-(\lambda)-K_+(\lambda)\right)\right]\nn
&+\frac{\pi T^2}{12 \varepsilon_1^{0\,\prime}(B)}\left[Z_{-}(\lambda) {\cal G} \left(\varepsilon^{0\,\prime}_1(B) \frac{\xt-v^{0}_1(B)}{T |v^{0\,\prime}_1(B)|}\right)-K_-(\lambda)  U(B) {\cal I} \left(\varepsilon^{0\,\prime}_1(B) \frac{\xt-v^{0}_1(B)}{T |v^{0\,\prime}_1(B)|}\right)\right]\nn
&-\frac{\pi r^2 T^2}{12 \varepsilon_1^{0\,\prime}(B)}\left[Z_{-}(\lambda){\cal G}\left(\varepsilon^{0\,\prime}_1(B) \frac{\xt-v^{0}_1(B)}{T r|v^{0\,\prime}_1(B)|}\right) -K_-(\lambda) U(B) {\cal I} \left(\varepsilon^{0\,\prime}_1(B) \frac{\xt-v^{0}_1(B)}{T r|v^{0\,\prime}_1(B)|}\right) \right]\nn
&+\frac{\pi T^2}{12 \varepsilon_1^{0\,\prime}(B)} \left[Z_{+}(\lambda) {\cal G} \left(\varepsilon^{0\,\prime}_1(B) \frac{\xt+v^{0}_1(B)}{T |v^{0\,\prime}_1(B)|}\right)+K_+(\lambda)  U(B) {\cal I} \left(\varepsilon^{0\,\prime}_1(B) \frac{\xt+v^{0}_1(B)}{T |v^{0\,\prime}_1(B)|}\right)\right]\nn
&-\frac{\pi r^2 T^2}{12 \varepsilon_1^{0\,\prime}(B) }\left[Z_{+}(\lambda){\cal G} \left(\varepsilon^{0\,\prime}_1(B) \frac{\xt+v^{0}_1(B)}{T r| v^{0\,\prime}_1(B)|}\right)+K_+(\lambda) U(B){\cal I} \left(\varepsilon^{0\,\prime}_1(B) \frac{\xt+v^{0}_1(B)}{T r|v^{0\,\prime}_1(B)|}\right)\right]\nn
&+\frac{\pi^3 v_1^0(B) T^2 (r^2-1)}{144 \varepsilon^{0\,\prime}_1(B) v_1^{0\,\prime}(B)}\left[K_-(\lambda) C_{r,+}(- B)  {\cal P}_r \left(\varepsilon^{0\,\prime}_1(B) \frac{\xt-v^{0}_1(B)}{T |v^{0\,\prime}_1(B)|}\right)-K_+(\lambda) C_{r,-}(B)  {\cal P}_r \left(\varepsilon^{0\,\prime}_1(B) \frac{\xt+v^{0}_1(B)}{T |v^{0\,\prime}_1(B)|} \right)\right]\nn
&-\frac{7\pi^3 T^2 (r^4-1) }{240  \varepsilon^{0\,\prime}_1(B)}\left[\left(Z_{-}(\lambda)-\left(\frac{v^{0\,\prime\prime}_1(B)}{v^{0\,\prime}_1(B)}-\frac{\varepsilon^{0\,\prime\prime}_1(B)}{\varepsilon^{0\,\prime}_1(B)}\right)K_-(\lambda)\right)   {\cal E}_r\left(\varepsilon^{0\,\prime}_1(B) \frac{\xt-v^{0}_1(B)}{T |v^{0\,\prime}_1(B)|}\right)\right]\nn
&+\frac{7\pi^3 T^2 (r^4-1) }{240 \varepsilon^{0\,\prime}_1(B)}\left[\left(Z_{+}(\lambda)+\left(\frac{v^{0\,\prime\prime}_1(B)}{v^{0\,\prime}_1(B)}-\frac{\varepsilon^{0\,\prime\prime}_1(B)}{\varepsilon^{0\,\prime}_1(B)}\right)K_+(\lambda)\right)   {\cal E}_r\left(\varepsilon^{0\,\prime}_1(B) \frac{\xt+v^{0}_1(B)}{T |v^{0\,\prime}_1(B)|}\right)\right].
\label{eq:fullexpansionvelrhot}
\end{align}
Here the functions $\mathcal G(z), \mathcal I(z), \mathcal D_r(z), \mathcal E_r(z)$ and $\mathcal P_r(z)$ are respectively defined in Eqs.~\eqref{eq:defGfun}, \eqref{eq:Ifundef}, \eqref{eq:Dfundef}, \eqref{eq:Efundef} and \eqref{eq:Pfundef}; the functions $K_\pm(\lambda)$ and $Z_\pm(\lambda)$ are defined as the solutions of the following integral equations 
\bea
K_{\pm}(\lambda)&=& -\int_{-B}^B\!{\rm d}\mu\, \ak_2(\lambda-\mu)K_{\pm}(\mu)-\ak_2(\lambda\pm B)\,,\label{eq:defK}\\
Z_{\pm}(\lambda)&=& -\int_{-B}^B\!{\rm d}\mu\, \ak_2(\lambda-\mu)Z_{\pm}(\mu)+\ak'_2(\lambda\pm B)\,.\label{eq:defZ}
\eea
The result for the profiles is then written as 
\begin{align}
\braket{{\boldsymbol q}}_\xt &=\frak d_q^0-\text{sgn}(v^{0\,\prime}_1(B))\frac{\pi T(r^2-1)}{12 v^{0}_1(B)}\left(\frak a^-_q  {\cal D}_r \left(\varepsilon^{0\,\prime}_1(B) \frac{\xt-v^{0}_1(B)}{T v^{0\,\prime}_1(B)}\right)-\frak a^+_q   {\cal D}_r \left(\varepsilon^{0\,\prime}_1(B) \frac{\xt+v^{0}_1(B)}{T v^{0\,\prime}_1(B)} \right)\right)\nn
&+\frac{\pi T^2 r^2}{12 \varepsilon_1^{0\,\prime}(B) v^{0}_1(B)}\left[\frak b^-_q-\frak b^+_q-\left(\frac{v^{0\,\prime}_1(B)}{v^{0}_1(B)} +U(B)\right)\left(\frak a^+_q +\frak a^-_q \right)\right]\nn
&+\frac{\pi T^2}{12 \varepsilon_1^{0\,\prime}(B) v^{0}_1(B)}\left[\left(\frak b^-_q-\frac{v^{0\,\prime}_1(B)}{v^{0}_1(B)} \frak a^-_q\right)  {\cal G} \left(\varepsilon^{0\,\prime}_1(B) \frac{\xt-v^{0}_1(B)}{T |v^{0\,\prime}_1(B)|}\right)-\frak a^-_q  U(B) {\cal I} \left(\varepsilon^{0\,\prime}_1(B) \frac{\xt-v^{0}_1(B)}{T |v^{0\,\prime}_1(B)|}\right)\right]\nn
&-\frac{\pi r^2 T^2}{12 \varepsilon_1^{0\,\prime}(B) v^{0}_1(B)}\left[\left(\frak b^-_q-\frac{v^{0\,\prime}_1(B)}{v^{0}_1(B)} \frak a^-_q\right){\cal G}\left(\varepsilon^{0\,\prime}_1(B) \frac{\xt-v^{0}_1(B)}{T r|v^{0\,\prime}_1(B)|}\right) -\frak a^-_q U(B) {\cal I} \left(\varepsilon^{0\,\prime}_1(B) \frac{\xt-v^{0}_1(B)}{T r|v^{0\,\prime}_1(B)|}\right) \right]\nn
&-\frac{\pi T^2}{12 \varepsilon_1^{0\,\prime}(B) v^{0}_1(B)} \left[\left(\frak b^+_q+\frac{v^{0\,\prime}_1(B)}{v^{0}_1(B)} \frak a^+_q\right)  {\cal G} \left(\varepsilon^{0\,\prime}_1(B) \frac{\xt+v^{0}_1(B)}{T |v^{0\,\prime}_1(B)|}\right)+\frak a^+_q  U(B) {\cal I} \left(\varepsilon^{0\,\prime}_1(B) \frac{\xt+v^{0}_1(B)}{T |v^{0\,\prime}_1(B)|}\right)\right]\nn
&+\frac{\pi r^2 T^2}{12 \varepsilon_1^{0\,\prime}(B) v^{0}_1(B)}\left[\left(\frak b^+_q+\frac{v^{0\,\prime}_1(B)}{v^{0}_1(B)} \frak a^+_q\right){\cal G} \left(\varepsilon^{0\,\prime}_1(B) \frac{\xt+v^{0}_1(B)}{T r| v^{0\,\prime}_1(B)|}\right)+\frak a^+_q U(B){\cal I} \left(\varepsilon^{0\,\prime}_1(B) \frac{\xt+v^{0}_1(B)}{T r|v^{0\,\prime}_1(B)|}\right)\right]\nn
&-\frac{7\pi^3 T^2 (r^4-1) }{240  \varepsilon^{0\,\prime}_1(B)  v^{0}_1(B)}\left[\left(\frak b^-_q-\left(\frac{v^{0\,\prime}_1(B)}{v^{0}_1(B)}+\frac{v^{0\,\prime\prime}_1(B)}{v^{0\,\prime}_1(B)}-\frac{\varepsilon^{0\,\prime\prime}_1(B)}{\varepsilon^{0\,\prime}_1(B)}\right)\frak a^-_q\right)   {\cal E}_r\left(\varepsilon^{0\,\prime}_1(B) \frac{\xt-v^{0}_1(B)}{T |v^{0\,\prime}_1(B)|}\right)\right]\nn
&+\frac{7\pi^3 T^2 (r^4-1) }{240 \varepsilon^{0\,\prime}_1(B)  v^{0}_1(B)}\left[\left(\frak b^+_q+\left(\frac{v^{0\,\prime}_1(B)}{v^{0}_1(B)}+\frac{v^{0\,\prime\prime}_1(B)}{v^{0\,\prime}_1(B)}-\frac{\varepsilon^{0\,\prime\prime}_1(B)}{\varepsilon^{0\,\prime}_1(B)}\right)\frak a^+_q\right)   {\cal E}_r\left(\varepsilon^{0\,\prime}_1(B) \frac{\xt+v^{0}_1(B)}{T |v^{0\,\prime}_1(B)|}\right)\right]\nn
&+\frac{\pi^3  T^2 (r^2-1)}{144 \varepsilon^{0\,\prime}_1(B) v_1^{0\,\prime}(B)}\left[\frak a^-_q C_{r,+}(- B)  {\cal P}_r \left(\varepsilon^{0\,\prime}_1(B) \frac{\xt-v^{0}_1(B)}{T |v^{0\,\prime}_1(B)|}\right)-\frak a^+_q C_{r,-}(B)  {\cal P}_r \left(\varepsilon^{0\,\prime}_1(B) \frac{\xt+v^{0}_1(B)}{T |v^{0\,\prime}_1(B)|} \right)\right]\notag\\
&+\frac{\pi^3 T^2 (1-r^2)^2}{72\varepsilon_1'(B) v^{0}_1(B)}\left[q(B)K_-(B) {\cal D}^2_r \left({\varepsilon^{0\,\prime}_1(B)}\frac{\xt-v^0_1(B)}{T |v_1^{0\,\prime}(B)|}\right)+q(-B)K_+(-B) {\cal D}^2_r \left({\varepsilon^{0\,\prime}_1(B)}\frac{\xt+v^0_1(B)}{T |v_1^{0\,\prime}(B)|} \right)\right]\,.
\label{eq:fullexpansioncharge}
\end{align}
\begin{align}
\braket{{\boldsymbol J_{\boldsymbol q}}}_\xt &=\frak j_q^0-\text{sgn}(v^{0\,\prime}_1(B))\frac{\pi T(r^2-1)}{12}\left(\frak a^-_q  {\cal D}_r \left(\varepsilon^{0\,\prime}_1(B) \frac{\xt-v^{0}_1(B)}{T v^{0\,\prime}_1(B)}\right)+\frak a^+_q   {\cal D}_r \left(\varepsilon^{0\,\prime}_1(B) \frac{\xt+v^{0}_1(B)}{T v^{0\,\prime}_1(B)} \right)\right)\nn
&+\frac{\pi T^2 r^2}{12 \varepsilon_1^{0\,\prime}(B)}\left[\frak b^-_q+\frak b^+_q-U(B)\left(\frak a^+_q -\frak a^-_q \right)\right]\nn
&+\frac{\pi T^2}{12 \varepsilon_1^{0\,\prime}(B) }\left[\frak b^-_q  {\cal G} \left(\varepsilon^{0\,\prime}_1(B) \frac{\xt-v^{0}_1(B)}{T |v^{0\,\prime}_1(B)|}\right)-\frak a^-_q  U(B) {\cal I} \left(\varepsilon^{0\,\prime}_1(B) \frac{\xt-v^{0}_1(B)}{T |v^{0\,\prime}_1(B)|}\right)\right]\nn
&-\frac{\pi r^2 T^2}{12 \varepsilon_1^{0\,\prime}(B)}\left[\frak b^-_q{\cal G}\left(\varepsilon^{0\,\prime}_1(B) \frac{\xt-v^{0}_1(B)}{T r|v^{0\,\prime}_1(B)|}\right) -\frak a^-_q U(B) {\cal I} \left(\varepsilon^{0\,\prime}_1(B) \frac{\xt-v^{0}_1(B)}{T r|v^{0\,\prime}_1(B)|}\right) \right]\nn
&+\frac{\pi T^2}{12 \varepsilon_1^{0\,\prime}(B)} \left[\frak b^+_q  {\cal G} \left(\varepsilon^{0\,\prime}_1(B) \frac{\xt+v^{0}_1(B)}{T |v^{0\,\prime}_1(B)|}\right)+\frak a^+_q  U(B) {\cal I} \left(\varepsilon^{0\,\prime}_1(B) \frac{\xt+v^{0}_1(B)}{T |v^{0\,\prime}_1(B)|}\right)\right]\nn
&-\frac{\pi r^2 T^2}{12 \varepsilon_1^{0\,\prime}(B)}\left[\frak b^+_q {\cal G} \left(\varepsilon^{0\,\prime}_1(B) \frac{\xt+v^{0}_1(B)}{T r| v^{0\,\prime}_1(B)|}\right)+\frak a^+_q U(B){\cal I} \left(\varepsilon^{0\,\prime}_1(B) \frac{\xt+v^{0}_1(B)}{T r|v^{0\,\prime}_1(B)|}\right)\right]\nn
&-\frac{7\pi^3 T^2 (r^4-1) }{240  \varepsilon^{0\,\prime}_1(B) }\left[\left(\frak b^-_q-\left(\frac{v^{0\,\prime\prime}_1(B)}{v^{0\,\prime}_1(B)}-\frac{\varepsilon^{0\,\prime\prime}_1(B)}{\varepsilon^{0\,\prime}_1(B)}\right)\frak a^-_q\right)   {\cal E}_r\left(\varepsilon^{0\,\prime}_1(B) \frac{\xt-v^{0}_1(B)}{T |v^{0\,\prime}_1(B)|}\right)\right]\nn
&-\frac{7\pi^3 T^2 (r^4-1) }{240 \varepsilon^{0\,\prime}_1(B)}\left[\left(\frak b^+_q+\left(\frac{v^{0\,\prime\prime}_1(B)}{v^{0\,\prime}_1(B)}-\frac{\varepsilon^{0\,\prime\prime}_1(B)}{\varepsilon^{0\,\prime}_1(B)}\right)\frak a^+_q\right)   {\cal E}_r\left(\varepsilon^{0\,\prime}_1(B) \frac{\xt+v^{0}_1(B)}{T |v^{0\,\prime}_1(B)|}\right)\right]\nn
&+\frac{\pi^3 v_1^0(B) T^2 (r^2-1)}{144 \varepsilon^{0\,\prime}_1(B) v_1^{0\,\prime}(B) }\left[\frak a^-_q C_{r,+}(- B)  {\cal P}_r \left(\varepsilon^{0\,\prime}_1(B) \frac{\xt-v^{0}_1(B)}{T |v^{0\,\prime}_1(B)|}\right)+\frak a^+_q C_{r,-}(B)  {\cal P}_r \left(\varepsilon^{0\,\prime}_1(B) \frac{\xt+v^{0}_1(B)}{T |v^{0\,\prime}_1(B)|} \right)\right]\notag\\
&+\frac{\pi^3 T^2 (1-r^2)^2}{72\varepsilon_1'(B)}\left[q(B)K_-(B) {\cal D}^2_r \left({\varepsilon^{0\,\prime}_1(B)}\frac{\xt-v^0_1(B)}{T |v_1^{0\,\prime}(B)|}\right)-q(-B)K_+(-B) {\cal D}^2_r \left({\varepsilon^{0\,\prime}_1(B)}\frac{\xt+v^0_1(B)}{T |v_1^{0\,\prime}(B)|} \right)\right]\,.
\label{eq:fullexpansioncurrent}
\end{align}
Here $\frak d_q^0$ and $\frak j_q^0$ are defined in Eq.~\eqref{eq:gsvalues} and we introduced  
\be
\frak a^{\pm}_q=\left[\int^{B}_{-B}\!\!{\rm d}\mu\, q(\mu)K_\pm\left(\mu\right)+q(\mp B)\right]\,,\qquad\qquad \frak b^{\pm}_q=\left[\int^{B}_{-B}\!\!{\rm d}\mu\, q(\mu)Z_\pm\left(\mu\right)+q'(\mp B)\right]\,.
\label{eq:appab}
\ee
The final goal is then to simplify the latter expressions using a chain of TBA identities. In the derivation we make use of the following  shorthand notation.
\begin{itemize}
\item[1)] Functions $w(\mu)$ become vectors $w$ such that 
\be
\label{eq:vecgs}
[  w ](\lambda)= w(\lambda)\,,\qquad\qquad\qquad\lambda\in[-B,B]\,.
\ee
\item[2)] Kernels $A(\lambda,\mu)$ become operators $\hat {A}_g$ such that
\be
[{\hat {A}_g}  w ](\lambda)=\int_{-B}^B{{\rm d}\mu}\,[{\hat {A}_g}](\lambda,\mu) [w ](\mu)=\int_{-B}^B{{\rm d}\mu}\, A(\lambda,\mu) w(\mu)\,.
\ee
The inverse of  $\hat {A}_g$ is the operator $ \hat A_g^{-1}$ satisfying
\be
\int_{-B}^B{{\rm d}\nu}\, [\hat A_g^{-1}](\lambda,\nu)\hat A_g(\nu,\mu)=\delta(\lambda-\mu)\, .
\ee
Thus, the distribution $\delta(\lambda-\mu)$ corresponds to the identity $\hat 1_g$. {Diagonal} operators $\hat w_g$ associated with a function of a single rapidity
\be
[\hat w_g](\lambda,\mu)=\delta(\lambda-\mu)w(\lambda)\, .
\ee
\end{itemize}
Let us consider the first of Eq.~\eqref{eq:appab} and focus on the r.h.s.
\be
\int^{B}_{-B}\!\!{\rm d}\mu\, q(\mu)K_\pm\left(\mu\right)+q(\mp B)=\lim_{x\rightarrow0} \int^{B}_{-B}\!\!{\rm d}\mu\, q^e(x-\mu)K_\pm(\mu)- \int^{B}_{-B}\!\!{\rm d}\mu\, q^o(x-\mu)K_\pm(\mu)+q^e(x\pm B)-q^o(x\pm B)
\label{eq:step1outlc}
\ee
where $q^e$ and $q^o$ are respectively the odd and even parts of $q$. The last two terms can be written as 
\bea
\lim_{x\rightarrow 0}q^e(x\pm B)=\lim_{x\rightarrow0} \hat q^e_g v_{\mp B}(x)\,,\qquad\qquad \lim_{x\rightarrow 0}q^o(x\pm B)=\lim_{x\rightarrow0} \hat q^o_g v_{\mp B}(x)\,,
\label{eq:rewriting}
\eea
where we introduced the vector
\be
[v_A](\lambda)=\delta(\lambda-A)\,,
\ee
together with the operator $\hat{q}^{e}_g$ defined by
\be
[{\hat {q}^e_g}  w ](\lambda)=\int_{-B}^B{{\rm d}\mu}\, q^e(\lambda-\mu) w(\mu)\,.
\ee
Inverting \eqref{eq:defK} we find 
\be
K_\pm = -(\hat 1_g + \widehat{\ak_{2}}_g)^{-1} \widehat{\ak_{2}}_g v_{\mp B}\,.
\label{eq:Kcompact}
\ee 
Plugging \eqref{eq:Kcompact} and  \eqref{eq:rewriting} in \eqref{eq:step1outlc} we find 
\bea
&&\int^{B}_{-B}\!\!{\rm d}\mu\, q(\mu)K_\pm\left(\mu\right)+q(\mp B)=\lim_{x\rightarrow0} \left[\hat{q}^e_g  (\hat 1_g + \widehat{\ak_{2}}_g)^{-1} v_{\mp B}- \hat{q}^o_g  (\hat 1_g + \widehat{\ak_{2}}_g)^{-1} v_{\mp B}\right](x)
\eea
This expression can be simplified further. Considering the first term in the r.h.s. we have  
\begin{align}
&\lim_{x\rightarrow0} \hat{q}^e_g  (\hat 1_g + \widehat{\ak_{2}}_g)^{-1} v_{\mp B}= \int_{-B}^B {\rm d}\mu\,\,  \left[ (\hat 1_g + \widehat{\ak_{2}}_g)^{-1}\right](\mp B,\mu) {q}^e(\mu)=f_{q^e}(\mp B)\,,
\label{eq:step2even}\\
&\lim_{x\rightarrow0} \hat{q}^o_g  (\hat 1_g + \widehat{\ak_{2}}_g)^{-1} v_{\mp B}= -\int_{-B}^B {\rm d}\mu\,\,  \left[ (\hat 1_g + \widehat{\ak_{2}}_g)^{-1}\right](\mp B,\mu) {q}^o(\mu)=-f_{q^o}(\mp B)\,.
\label{eq:step2odd}
\end{align}
Here we used that $(\hat 1_g + \widehat{\ak_{2}}_g)^{-1}$ is symmetric and defined $f_q(\lambda)$ via the following integral equation
\be
f_q(\lambda) = q(\lambda) - \int_{-B}^{B}\mathrm \!\!d\mu \,\,\, \ak_2(\lambda-\mu)f_q(\mu)\,. 
\label{eq:fq}
\ee
So we finally have 
\bea
\frak a^{\pm}_q=\int^{B}_{-B}\!\!{\rm d}\mu\, q(\mu)K_\pm\left(\mu\right)+q(\mp B)=f_{q}(\mp B)\,.
\label{eq:identity1}
\eea 
Proceeding analogously we have 
\bea
&&\int^{B}_{-B}\!\!{\rm d}\mu\, q(\mu)Z_\pm\left(\mu\right)+q'(\mp B)=-\lim_{x\rightarrow0}\lim_{y\rightarrow0} \left[\hat{q}^e_g  (\hat 1_g + \widehat{\ak_{2}}_g)^{-1} w_{\mp B,y}- \hat{q}^o_g  (\hat 1_g + \widehat{\ak_{2}}_g)^{-1} w_{\mp B,y}\right](x)\,,
\eea
where we introduced 
\be
[w_{A,y}](x)=\frac{\delta(x-A+y)-\delta(x-A)}{y}\,.
\ee
In writing \eqref{eq:rewriting} we used 
\be
\lim_{y\rightarrow0} [\hat a_g\,w_{A,y}](x)=\lim_{y\rightarrow0} \frac{1}{y} \int_{-B}^{B} \!\!\!{\rm d}z\,\, a(x-z) (\delta(z-A+y)-\delta(z-A)) = \lim_{y\rightarrow0}\frac{a(x-A+y)-a(x-A)}{y}=a'(x-A)\,.
\ee
Using again the definition \eqref{eq:fq} of $f_q(\lambda)$ we find 
\begin{align}
&\lim_{x\rightarrow0}\lim_{y\rightarrow0} \hat{q}^e_g  (\hat 1_g + \widehat{\ak_{2}}_g)^{-1} w_{\mp B,y}= - \int_{-B}^B {\rm d}\mu\,\,  \partial_\lambda\left[ (\hat 1_g + \widehat{\ak_{2}}_g)^{-1}\right](\lambda,\mu)\Bigl|_{\lambda=\mp B} {q}^e(\mu)=-f'_{q^e}(\mp B)\,,
\label{eq:step2outlc}\\
&\lim_{x\rightarrow0}\lim_{y\rightarrow0} \hat{q}^o_g  (\hat 1_g + \widehat{\ak_{2}}_g)^{-1} w_{\mp B,y}= \int_{-B}^B {\rm d}\mu\,\,  \partial_\lambda\left[ (\hat 1_g + \widehat{\ak_{2}}_g)^{-1}\right](\lambda,\mu)\Bigl|_{\lambda=\mp B} {q}^o(\mu)=f'_{q^o}(\mp B)\,.
\label{eq:step2outlc2}
\end{align}
Putting all together we obtain
\be
\frak b^{\pm}_q=\int^{B}_{-B}\!\!{\rm d}\mu\, q(\mu)Z_\pm\left(\mu\right)+q'(\mp B)=f'_q(\mp B)\,.
\label{eq:identity2}
\ee

\section{Details on the low-temperature expansion in the gapped phase}

\subsection{Small-temperature expansions in the antiferromagnetic regime}\label{sec:appendix_expansions}

In this appendix we provide further details of the small-temperatures expansions presented in Sec.~\ref{sec:lowtantiferro}. As we discussed in the main text, the low temperature analysis of the profiles starts by considering the same problem for the homogeneous case, which is recovered for large absolute values of the ray $\zeta$.  We recall that in this case one can set
\be
\eta_n(\lambda)=\bar{\eta}_n(\lambda)\left[1+\delta_n(\lambda)\right]\,.
\label{eq:appendix_perturbative_expression}
\ee
where $\bar{\eta}_n(\lambda)$ are defined in \eqref{eq:new_leading_1} and \eqref{eq:new_leading_2}. Here, to lighten the notation, we removed the labels ${L/R}$ as the calculation in the left and right thermal states is identical. The functions ${\eta}_n(\lambda)$ and $\bar{\eta}_n(\lambda)$ share the same leading behaviour of \eqref{eq:leading_1} and \eqref{eq:leading_2}, but \eqref{eq:appendix_perturbative_expression} allows one to compute more easily higher order corrections. 
This is illustrated in the following.

It is convenient to introduce the functions
\be
\chi_n(\lambda)=\bar{\eta}_n(\lambda)\delta_n(\lambda)\frac{\sinh(\beta h)}{\sinh(\beta h n)}\,.
\ee 
Assuming $\delta_n(\lambda)\ll1$ we can linearise the TBA equations \eqref{eq:TBA} around $\delta_n(\lambda)\equiv 0$. Dropping terms $O(\delta^2_n(\lambda))$ we obtain
\be
\chi_n(\lambda)=s(\lambda)\ast\left[\frac{\sinh(\beta h (n-1))}{\sinh(\beta h n)}\chi_{n+1}(\lambda)+\frac{\sinh(\beta h (n+1))}{\sinh(\beta h n)}\chi_{n-1}(\lambda)\right]\,.
\ee
Here we used
\be
\log\bar{\eta}_n=\frac{1}{2}\log\left[\left(\bar{\eta}_{n+1}+1\right)\left(\bar{\eta}_{n-1}-1\right)\right]\,.
\ee

This equation can be solved in Fourier space, where it reads
\be
\hat{\chi}_n(k)=\hat{s}(k)\left[\frac{\sinh(\beta h (n-1))}{\sinh(\beta h n)}\hat{\chi}_{n+1}(k)+\frac{\sinh(\beta h (n+1))}{\sinh(\beta h n)}\hat{\chi}_{n-1}(k)\right]\,.
\label{eq:intermediate}
\ee
For each $k$ this is a second order discrete difference equation. We rewrite \eqref{eq:intermediate} by substituting $\hat{s}$ with its analytical expression. We get
\be
\left(e^{-|k|\eta}+e^{|k|\eta}\right)\hat{\chi}_n(k)=\left[\frac{\sinh(\beta h (n-1))}{\sinh(\beta h n)}\hat{\chi}_{n+1}(k)+\frac{\sinh(\beta h (n+1))}{\sinh(\beta h n)}\hat{\chi}_{n-1}(k)\right]\,.
\label{eq:diffence_equation}
\ee
After a little guess-work we find the following generic solution for $k\neq 0$
\bea
\hat{\chi}_n(k)&=&A(k)\left[\sinh[h\beta(n+1)]e^{-(n-1)|k|\eta} -\sinh[h\beta(n-1)]e^{-(n+1)|k|\eta}\right]\nonumber\\
&+&B(k)\left[\sinh[h\beta(n+1)]e^{+(n-1)|k|\eta} -\sinh[h\beta(n-1)]e^{+(n+1)|k|\eta}\right]\,.
\label{eq:generic_solution}
\eea
Here $A(k)$, $B(k)$ are the two arbitrary constants parametrizing the two-dimensional space of the solutions of \eqref{eq:diffence_equation}. Note that when $k= 0$ the two additive terms in \eqref{eq:generic_solution} are no longer linearly independent. In order to get the generic solution for $k=0$ then one has to choose an appropriate combination of the two which remains independent in the limit $k\to 0$. This is straightforward and yields the following generic solution of \eqref{eq:diffence_equation} for $k=0$
\bea
\hat{\chi}_n(0)&=&C\left\{\sinh[h\beta(n+1)] -\sinh[h\beta(n-1)]\right]\}\nonumber\\
&+&D\left\{\sinh[h\beta(n+1)](n-1) -\sinh[h\beta(n-1)](n+1)\right\}\,,
\label{eq:generic_solution_k0}
\eea
where $C$ and $D$ are again arbitrary constants parametrizing the two-dimensional space of the solutions of \eqref{eq:diffence_equation}.

We can fix the value of the constants from the boundary conditions. Since we require $\delta_n(\lambda)\ll 1$, we can immediately set $B(k)=0$, $D=0$. In order to fix the constants $A(k)$ and $C$, we consider \eqref{eq:TBA} for $n=2$ 
\be
\frac{\sinh(2h\beta)}{\sinh(h\beta)\bar{\eta}_2}\chi_2=s\ast \eta_1+s\ast \frac{\sinh(3h\beta)}{\sinh(h\beta)(1+\bar{\eta}_3)}\chi_3\,,
\label{eq:determine_a}
\ee
Going in Fourier space this equations is readily inverted and gives immediately the constants $A(\kappa)$ and $D$. Collecting all the previous calculations, we can write down the functions $\chi_n$ explicitly. Using \eqref{eq:generic_solution}, \eqref{eq:generic_solution_k0} with the explicit value of the constants $A(k)$ and $D$, we finally arrive at
\bea
\chi_n\simeq G^{\beta,h}_n\ast \eta_1\,,
\eea
where
\bea
\hat{G}^{\beta,h}_n(k)=
\frac{e^{|k|\eta  (1-n)}}{\sinh(2\beta h)} \left\{\sinh [\beta  h (n+1)]-e^{-2 |k|\eta } \sinh [\beta  h (n-1)]\right\}\,.
\label{eq:G_function}
\eea


\subsection{Simplified formulae for the expectation values of densities and currents of $\{\boldsymbol{Q}_m\}_{m=1}^\infty$ and $\boldsymbol{S}^z$}
\label{app:simpformulae}

In this appendix we derive the Formulae \eqref{eq:partial_1}--\eqref{eq:partial_4} for the stationary-state expectation values of densities and currents of the local charges $\{\boldsymbol{Q}_m\}_{m=1}^\infty$ (\emph{cf}. Sec.~\ref{sec:charges}) and of the magnetization $\boldsymbol{S}^z$. We start by defining 
\bea
R^{\zeta}_{n,i}(k)&=&
\left\{
\begin{array}{ll}
	\frak{F}[\rho_{n,\zeta}](k)\,, &\quad i=1\,,\\
	\frak{F}[\rho_{n,\zeta}v_{n,\zeta}](k)\,, & \quad i=2\,,
\end{array}
\right.
\\
R^{h, \zeta}_{n,i}(k)&=&
\left\{
\begin{array}{ll}
	\frak{F}[\rho^{h}_{n,\zeta}](k)\,, &\quad i=1\,,\\
	\frak{F}[\rho^{h}_{n,\zeta}v_{n,\zeta}](k)\,, & \quad i=2\,,
\end{array}
\right.
\eea
where $\frak{F}$ denotes the Fourier transform as in \eqref{eq:convention_FT}. Using these definitions one can rewrite \eqref{eq:BT_decoupled} and \eqref{eq:gapped_vel_decoupled} in Fourier space as
\be
R_{n,i}(k)=\frac{1}{2\cosh(k\eta)}\left[R^h_{n-1,i}(k)+R^h_{n+1,i}(k)\right]-R^h_{n,i}(k)\,,
\label{eq:fourier_transform_bethe}
\ee
where we have dropped the dependence on $\zeta$ and where
\be
R^h_{0,i}(k)=
\left\{
\begin{array}{ll}
	1\,, &\quad i=1\,,\\
	ik\sinh\eta\,, & \quad i=2\,.
\end{array}
\right.
\ee
Using \eqref{eq:fourier_transform_bethe}, one can cast the equations for the densities and currents of the local charges $\{\boldsymbol{Q}_m\}_{m=1}^\infty$ in the simplified form
\bea
\frac{\braket{\boldsymbol{q}_{m+1}}_\zeta}{\sinh^m\left(\eta\right)}=\sum_{k\in\mathbb{Z}}\frac{R^{h,\zeta}_{1,1}(k)-R^{h,\zeta}_{0,1}(k)e^{-|k|\eta}}{2\cosh(k\eta)}(ik)^{m-1}\,,\label{eq:to_derive1}\\
\frac{\braket{\boldsymbol{J}_{m+1}}_\zeta}{\sinh^m\left(\eta\right)}=\sum_{k\in\mathbb{Z}}\frac{R^{h,\zeta}_{1,2}(k)-R^{h,\zeta}_{0,2}(k)e^{-|k|\eta}}{2\cosh(k\eta)}(ik)^{m-1}\,.\label{eq:to_derive2}
\eea
The derivation of Eq.~\eqref{eq:to_derive1} can be found in \cite{BWFD14}, and will not be reported here. Analogously, Eq.~\eqref{eq:to_derive2} is obtained by following the same steps. Exploiting the convolution theorem, we arrive at the final result
\bea
\frac{\braket{\boldsymbol{q}_{m+1}}_\zeta}{\sinh^m\left(\eta\right)}&=&\frac{\pi}{(-2)^{m-1}}\int_{-\pi/2}^{\pi/2}d\lambda s^{(m-1)}(\lambda)\left[\rho^{h}_{1,\zeta}(\lambda)-a_1(\lambda)\right]\,,\label{eq:apppartial_1}\\
\frac{\braket{\boldsymbol{J}_{m+1}}_\zeta}{\sinh^m\left(\eta\right)}&=&\frac{\pi}{(-2)^{m-1}}\int_{-\pi/2}^{\pi/2}d\lambda s^{(m-1)}(\lambda)\left[\rho^{h}_{1,\zeta}(\lambda)v_1(\lambda)+\frac{\sinh\eta}{2}a^\prime_1(\lambda)\right]\,,\label{eq:apppartial_2}
\eea 
where
\bea
s^{(n)}(\lambda)&=&\frac{d^n}{d\lambda^n}s(\lambda)\,,\\
s^{(0)}(\lambda) &=& s(\lambda)\,.
\eea
The expressions for the spin density and current can also be simplified. In this case, a straightforward computation yields
\bea
\braket{\boldsymbol{s}}_\zeta&=&\frac{1}{2}-\sum_{n=1}^{\infty}n\int_{\pi/2}^{\pi/2}d\lambda\rho_{n,\zeta}(\lambda)=\frac{1}{2}-\sum_{n=1}^{\infty}n\hat{\rho}_{n,\zeta}(0)\label{eq:apppartial_3}\nonumber\\
&=&\frac{1}{2}-\sum_{n=1}^{\infty}n\left[\frac{1}{2}\left(\hat{\rho}^{h}_{n+1,\zeta}(0)+\hat{\rho}^{h}_{n-1,\zeta}(0)\right)-\hat{\rho}^{h}_{n,\zeta}(0)\right]\nonumber\\
&=&\lim_{n\to\infty}\left[\frac{1}{2}\hat{\rho}^{h}_{n,\zeta}(0)-\frac{n}{2}\left(\hat{\rho}^{h}_{n+1,\zeta}(0)-\hat{\rho}^{h}_{n,\zeta}(0)\right)\right]=\frac{1}{2}\lim_{n\to\infty}\int_{-\pi/2}^{\pi/2}\rho^{t}_{n,\zeta}(\lambda)\,,
\eea
where we assumed that $\hat{\rho}^{h}_{n,\zeta}-\hat{\rho}^{h}_{n-1,\zeta}$ vanishes faster than $1/n$, we used that $\lim_{n\to\infty}\rho^{h}_{n,\zeta}(\lambda)=\lim_{n\to\infty}\rho^{t}_{n,\zeta}(\lambda)$ and the identity
\be
\sum_{n=1}^{m}n\left[\frac{1}{2}\left(\hat{\rho}^{h}_{n+1,\zeta}(0)+\hat{\rho}^{h}_{n-1,\zeta}(0)\right)-\hat{\rho}^{h}_{n,\zeta}(0)\right]=\frac{m}{2}\left(\hat{\rho}^{h}_{m+1,\zeta}(0)-\hat{\rho}^{h}_{m,\zeta}(0)\right)-\frac{1}{2}\hat{\rho}^{h}_{m,\zeta}(0)\,,
\ee
which can be easily proven by induction on $m$. Analogously, one can derive
\bea
\braket{\boldsymbol{J}_s}_\zeta&=&\frac{1}{2}\lim_{n\to\infty}\int_{-\pi/2}^{\pi/2}\rho^{t}_{n,\zeta}(\lambda)v_{n,\zeta}(\lambda)\,.\label{eq:apppartial_4}
\eea

\end{document}